# Gel-Chemistry-Controlled Heavy-Metal Transport and Retention in Cementitious Gel Nanopores: A Molecular Dynamics Study

Weiqiang Chen[a,b,c], Qiyao He[a,b,c], Kai Gong[a,b,c,*]

[a] Department of Civil and Environmental Engineering, Rice University, Houston, Texas 77005, United States

[b] Rice Advanced Materials Institute, Rice University, Houston, Texas 77005, United States

[c] Ken Kennedy Institute, Rice University, Houston, Texas 77005, United States

* Corresponding author. E-mail: kg51@rice.edu

# Gel-Chemistry-Controlled Heavy-Metal Transport and Retention in Cementitious Gel Nanopores: A Molecular Dynamics Study

## Abstract

Cementitious materials are widely used for hazardous-waste encapsulation, yet the molecular mechanisms governing heavy-metal ion retention across different gel chemistries remain insufficiently resolved. Previous MD studies have generally focused on individual gel systems and pore-averaged transport metrics, limiting direct comparison across gel chemistries. Here, classical molecular dynamics simulations were employed to investigate the adsorption-controlled mobility of representative heavy-metal ions ($Pb^{2+}$, $Ba^{2+}$, and $Cs^{+}$) within nanopores of C–S–H, C–(N)–A–S–H and N–A–S–H gels, representing different cementitious systems. By combining pore-averaged diffusivity, spatially resolved diffusivity and residence-time analysis, ion-density profiles, two-dimensional adsorption maps, radial distribution functions, coordination analysis, and interfacial binding-strength descriptors, this study establishes a comparative atomistic framework linking gel surface chemistry to ion mobility suppression under nanoconfinement, while distinguishing the contributions of local mobility and spatial ion partitioning to pore-averaged transport. Ion mobility is substantially reduced in all gel nanopores relative to bulk solutions, but the extent and mechanism of suppression vary strongly with gel chemistry. C–(N)–A–S–H with higher interfacial Al/Si substitution exhibits the strongest retention, driven by ion accumulation around Al-linked oxygen species via an ion-exchange-like mechanism with charge-balancing $Na^{+}$. C–S–H immobilizes ions primarily through surface hydroxyl oxygens and Ca-mediated linkages, whereas N–A–S–H exhibits more distributed binding environments. $Pb^{2+}$ and $Ba^{2+}$ exhibit broadly similar immobilization mechanisms, whereas $Cs^{+}$ shows more distinct, gel-dependent interactions with silicate and aluminosilicate oxygen sites. A relative total binding strength (rTBS) descriptor is introduced, showing a strong positive correlation with the extent of ion immobilization across gel types, ion species, and pore sizes examined. These results clarify gel-specific and ion-specific mechanisms controlling heavy-metal retention and provide a comparative atomistic framework for linking interfacial binding strength to pore-averaged mobility suppression in idealized cementitious nanopores.

# 1 Introduction

**Background and motivation.** Heavy-metal contamination arising from mining, smelting, fossil-fuel combustion, industrial waste disposal, and nuclear-waste management poses long-term risks to soil and groundwater quality because many heavy metal species are persistent, toxic, bioaccumulative and difficult to degrade [1, 2]. Among these contaminants, Pb, Ba, and Cs are of particular concern. Pb is a potent neurotoxin; soluble Ba compounds can affect cardiovascular and muscular function; and radioactive Cs isotopes (e.g., $^{137}Cs$) present long-term chemical and radiological hazards. These concerns underscore the urgent need for durable, low-cost and scalable immobilization technologies that reduce contaminant mobility and limit environmental exposure.

Solidification/stabilization (S/S) using cementitious binders offers a promising and widely used strategy owing to their low cost, broad availability, and ability to reduce contaminant mobility and leachability over extended timescales via combination of chemical binding with physical encapsulation in a dense, low-permeability matrix [3]. As such, cementitious binders based on ordinary Portland cement (OPC) and alkali-activated materials (AAMs), including geopolymers, have been extensively investigated for heavy-metal immobilization [1, 3-5]. To quantitatively assess research trends on heavy-metal immobilization in OPC- and AAM-based systems, we conducted a large-scale text-mining analysis of the Web of Science Core Collection from 1970 to 2025 using an automated literature-mining pipeline based on large language models [6, 7]. As shown in **Figure 1**a, publications on both material systems have increased markedly over the past several decades, reaching nearly 200 papers in 2024. Notably, although AAM systems emerged more recently, they have grown more rapidly over the past decade and have recently become comparable to OPC-based systems in annual publication output. **Figure 1**b further shows that literature is overwhelmingly dominated by experimental studies (> 98.9%). In contrast, MD-based studies have appeared only recently and account for a very small fraction of the total literature (~0.53%). This quantitative literature mapping provides direct evidence that, although macroscopic immobilization performance has been widely studied, molecular-level simulation studies remain comparatively limited.

The predominance of experimental studies in **Figure 1**b reflects a substantial body of work on macroscopic immobilization performance of cementitious binders and the material factors that control it. For OPC-based systems, immobilization performance can be influenced by binder chemistry [8, 9], hydration products and phase assemblage [10], microstructure and porosity [11], the presence of additives or supplementary cementitious materials [12, 13], and the type and oxidation state of the heavy-metal species [14]. These effects have been summarized in several review articles [3, 5]. For AAM-based systems, immobilization performance similarly depends on precursor chemistry [15], gel composition and structure [16], activator type and alkalinity [17], microstructure and porosity [18], and the type and speciation of heavy metals [19, 20], as summarized in Refs. [4, 15, 21, 22].

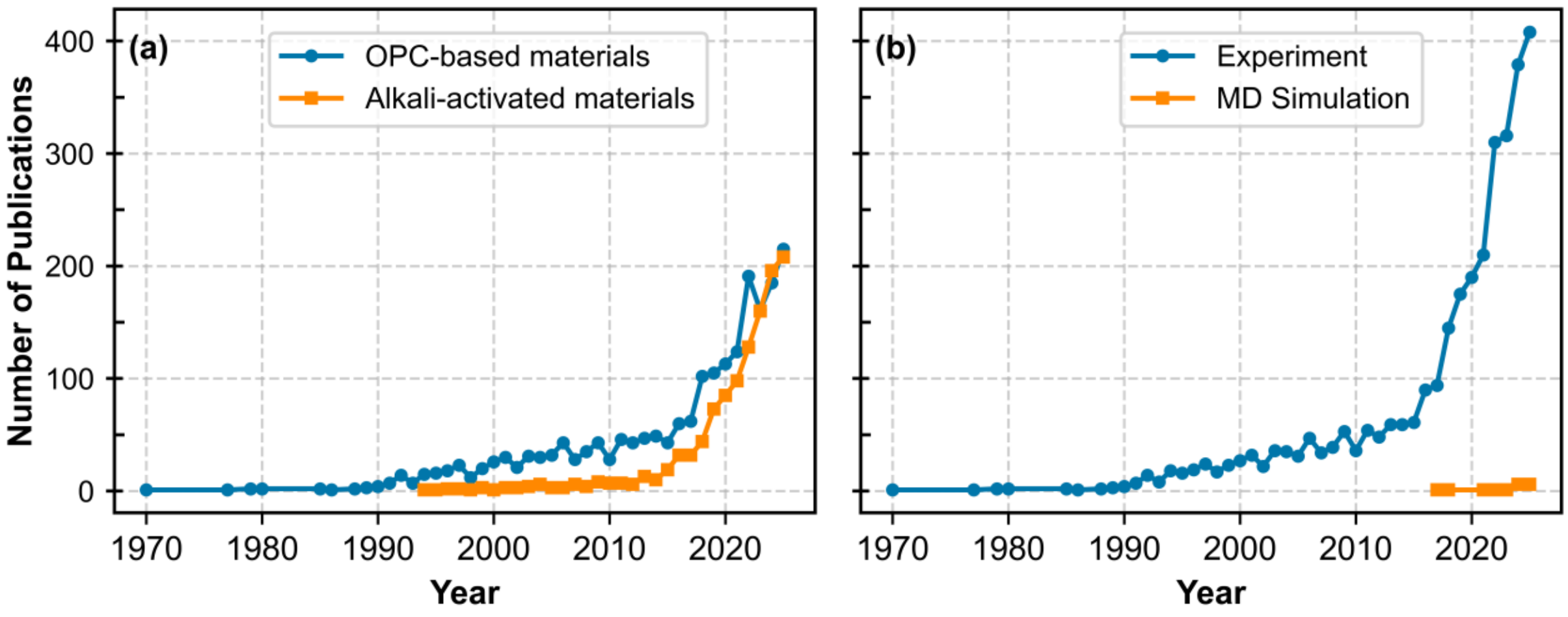

**Figure 1.** Trends in publications on heavy-metal immobilization using cementitious materials from 1970 to 2025: (a) number of publications per year categorized by cementitious material type, including OPC-based systems and alkali-activated materials; and (b) number of publications per year categorized by research method, including experimental studies and those containing MD simulations.

While these experimental investigations have provided valuable insights into bulk immobilization behaviors and the factors influencing heavy-metal retention, elucidating the underlying molecular-scale mechanisms remains challenging. Hydrated cementitious binders are chemically and structurally heterogeneous, containing variable gel compositions, mixed crystalline–amorphous phases, evolving pore-solution chemistries, broad pore size distributions, and complex solid-liquid interfaces [23, 24]. Even simplified model systems, such as synthetic calcium silicate hydrate (C–S–H), display intrinsic variability in Ca/Si ratio, defect density, surface hydroxylation, and surface topology. Consequently, macroscopic measurements such as leaching, uptake capacity, and bulk transport cannot by themselves resolve atomic-scale sorption sites, local coordination environments, or the relative binding characteristics of heavy-metal species. Even when combined with advanced characterization techniques, such as electron microscopy, X-ray diffraction, tomography, solid-state NMR, or X-ray spectroscopy, unambiguous mechanistic attribution remains difficult because multiple immobilization processes often occur simultaneously in chemically heterogeneous and partially amorphous cementitious matrices [3, 25].

**Role of Molecular Dynamics (MD) Simulations.** Molecular dynamics (MD) simulations offer a powerful complementary approach for investigating heavy-metal transport and immobilization at the atomic scale. By explicitly controlling gel composition, pore geometry, solution chemistry, and environmental conditions, MD enables systematic evaluation of the physicochemical factors governing heavy-metal adsorption and transport. At the same time, MD simulations provide continuous spatial and temporal resolution of ion trajectories, hydration structures, interfacial water dynamics, and surface coordination environments, allowing detailed analysis of the molecular events that underpin immobilization. Nevertheless, as shown in **Figure 1**b, the number of MD studies addressing heavy-metal immobilization in OPC- and AAM-based systems remains far smaller than number of experimental investigations.

Most existing MD studies have focused on ionic transport and adsorption within nanopores of individual representative cementitious gel systems, including C–S–H in OPC systems [26, 27], sodium-containing calcium aluminosilicate hydrate (C–(N)–A–S–H) in AAM systems based on high-Ca precursor (e.g., slag) [28, 29], and sodium aluminosilicate hydrate (N–A–S–H) in AAM systems based on low-Ca precursors (e.g., metakaolin and class F fly ash) [4, 30-36]. Similar to light ions, heavy-metal ions generally exhibit reduced mobility under nanoconfinement. For example, previous MD simulations have shown that the average $Cs^+$ diffusivity in C–S–H (Ca/Si $\approx$ 1.1) nanopores is lower than in bulk solution [37], with the suppression intensifying as the pore width narrows [29]. For C–S–H gel in OPC systems and its crystalline analogs, tobermorite (Ca/Si = 1), the literature points to a multifactorial control of interfacial transport and immobilization. Jiang et al. [38] reported that $Cs^+$ undergoes inner-sphere adsorption at bridging and pairing silicate oxygen sites in tobermorite nanopores (Ca/Si = 1), leading to reduced mobility relative to bulk solution. Wang et al. [39] found that, in tobermorite C–S–H nanopores (Ca/Si = 1), increasing the CsCl concentration from 0 to 2 M decreases the overall diffusivities of water and ions, which was attributed to enhanced ion–surface interactions and ion clustering at higher ionic strength, based on radial distribution functions (RDFs) and density profiles. Qiao et at. [40] found that the immobilization capacity of a C–S–H (Ca/Si = 1.3) nanopore follows the order $Pb^{2+}$ > $Zn^{2+}$ > $Cu^{2+}$, which reflects their relative ability to substitute for $Ca^{2+}$, as indicated by hydration-shell stability analysis. Zha et al. [41] reported that, on tobermorite-like C–S–H surfaces (Ca/Si = 1), competitive adsorption favors $Pb^{2+}$ over $Cd^{2+}$ and $Zn^{2+}$ because of its stronger electrostatic interactions with the surface. Kai et al. [42] further reported that the average diffusivities of $H_2O$, $Cs^+$ and $Ca^{2+}$ in C–S–H gels (Ca/Si = 1.1–1.8) follow the order $H_2O$ > $Cs^+$ > $Ca^{2+}$. They also found that increasing interlayer $Cs^+$ content enhances overall diffusivity, which was attributed to (i) interlayer expansion, (ii) weakened gel–solution interfacial interactions, as indicated by potential energy calculations, and (iii) reduced stability of Cs–water coordination relative to Ca–water coordination as indicated by time correlation functions (TCFs) analysis.

For C–(N)–A–S–H and C–A–S–H systems, MD research to date has primarily focused on diffusion of light ions (e.g., $Na^+$, $Ca^{2+}$), with only limited efforts devoted to heavy-metal immobilization. Duque-Redondo et al. [29] demonstrated that Al incorporation into C–S–H (Ca/Si = 1.1) gel to form C–A–S–H gel improves its retention and adsorption capacity for different ions ($Cs^+$, $Ca^{2+}$, and $Na^+$), while decreasing $Cs^+$ diffusivity by ~40%. This retention capacity follows the trend $Na^+$ > $Ca^{2+}$ > $Cs^+$, which has been attributed to their differences in ionic charge and size. Su et al. [28] investigated the leaching of heavy metals from C–A–S–H nanopore and found that the leaching rate follows the order $Pb^{2+}$ > $Cr^{3+}$ > $Cd^{2+}$. This trend is inversely correlated with the calculated adsorption strengths of the ions in the gels, evaluated from potential-energy difference between the combined gel–ion system and the isolated gel and ion components. This simulation-derived trend was also consistent with their experimental heavy-metal leaching measurements.

In the case of N–A–S–H gels, Zhang et al. [35] observed that a higher Al/Si ratio in N–A–S–H gels and elevated temperature favor $Pb^{2+}$ and $Cs^+$ adsorption by enhancing electrostatic attraction and facilitating cation exchange. Hou et al. [31] reported that abundant non-bridging oxygens and strong interfacial H-bonding promote inner-sphere adsorption and longer residence for smaller, higher–charge-density cations, leading to a mobility trend of $Na^+$ < $Cs^+$< $K^+$. Nanoconfinement has also been found to induce anisotropic diffusivity in N–A–S–H nanopore, where the motion of heavy-metal ions such as $Cs^{+}$ [31] and $Cu^{2+}$ [34] perpendicular to the gel surface is slower than motion parallel to it.

Although these MD studies provide valuable atomistic insights into ion adsorption and transport within specific cementitious gels, most investigations have focused on individual gel systems and relied predominantly on pore-averaged transport metrics. Consequently, direct comparison of heavy-metal transport and retention across different gel chemistries remains difficult, while pore-averaged diffusivities do not distinguish the contribution of local ion mobility from spatial partitioning into near-surface adsorption layers. In addition, the dominant binding environments governing heavy-metal retention have not been systematically compared across gel chemistries and ion species. To the best of our knowledge, a common descriptor linking interfacial binding with mobility suppression across different ions and gel systems also remains lacking.

**Research Gap and Objectives.** Our literature review shows that, despite growing interest in molecular-scale modeling of heavy-metal adsorption and transport in cementitious nanopores, systematic comparative MD studies across different binder chemistries remain scarce. Existing simulations typically focus on individual gel systems, most commonly tobermorite-based C–S–H [38, 39, 41], or N–A–S–H [31, 34, 35]. This limits the ability to distinguish general nanoconfinement effects from gel-specific immobilization mechanisms. Furthermore, most prior MD studies rely on pore-averaged diffusivities [31, 34, 35, 38, 39, 41, 42], which provide useful overall mobility information but obscure spatial variations in ion transport near solid–liquid interfaces. Our recent MD simulations [43] of NaCl solutions in a C–S–H (Ca/Si = 1.67) gel pore, for example, revealed that diffusivity near the gel surface is markedly lower than in the pore center, underscoring the need for spatially resolved transport analysis. To the best of our knowledge, no prior MD study has systematically compared spatially resolved heavy-metal ion transport across different cementitious gel nanopores.

This gap is important because pore-averaged diffusivity can be controlled by two coupled factors: the local mobility of ions in different regions of the pore and the fraction of ions partitioned into near-surface adsorption layers. In other words, a gel may exhibit strong overall immobilization not only because ions move slowly near the interface, but also because a larger fraction of ions accumulates in low-mobility interfacial regions. Resolving this distinction is essential for identifying the molecular origins of heavy-metal retention and for developing more accurate nanoscale descriptors that better inform multiscale transport models [44, 45]. Such spatially resolved understanding may also help interpret anomalous transport behaviors observed in other nanoconfined and interfacial systems [46, 47].

This study addresses these limitations through systematic comparative MD simulations of representative heavy-metal ions ($Pb^{2+}$, $Ba^{2+}$, and $Cs^{+}$) confined within cementitious gel nanopores. The simulations resolve both pore-averaged and position-dependent transport and adsorption behavior across five gel nanochannels with a focus on a representative pore width ≈ 4 nm. The simulated systems include (i) a C–S–H (Ca/Si = 1.67) pore representative of OPC-based binders, (ii) C–(N)–A–S–H pores derived from tobermorite with interfacial Al/Si ratios of 0, 0.059, and 0.125, representing controlled model systems relevant to high-Ca AAMs [48, 49], and (iii) an amorphous N–A–S–H pore (Al/Si = 0.33) representing low-Ca AAM systems [49, 50]. These models are used to examine how gel surface properties govern ion partitioning, ion–surface interactions, and mobility suppression under nanoconfinement. Within the tobermorite-derived series, varying the interfacial Al/Si substitution systematically modifies the surface charge distribution and Al-associated binding environments within a common structural framework, whereas the C–S–H and N–A–S–H models provide representative contrasts among OPC-, high-Ca AAM-, and low-Ca AAM-type gel surfaces.

The objectives of this work are threefold. First, we compare the pore-averaged and spatially resolved mobility of $Pb^{2+}$, $Ba^{2+}$, and $Cs^{+}$ in different cementitious gel nanopores, with particular attention to the contribution of near-surface low-mobility layers. Second, we identify the interfacial adsorption environments responsible for heavy-metal retention by analyzing residence-time distributions, ion-density profiles, two-dimensional adsorption maps, partial radial distribution functions (RDFs) and coordination environments. Third, we introduce total binding strength and relative total binding strength descriptors that combine ion–surface coordination numbers and pair interaction energies to identify the dominant adsorption sites and rationalize the relationship between interfacial binding and mobility suppression across gel chemistries, ion species, and selected pore-size conditions. Together, these analyses establish a unified cross-gel framework that advances beyond previous single-system or pore-averaged studies by distinguishing the respective contributions of local mobility and spatial ion partitioning to effective transport and relating both to interfacial binding.

By linking gel surface properties, interfacial binding environments, and nanoconfined diffusivity suppression, this work provides molecular-level insights into adsorption-controlled heavy-metal immobilization in representative cementitious gel pores. More broadly, the comparison of tobermorite-derived calcium silicate and aluminosilicate interfaces with an amorphous sodium aluminosilicate glass-derived gel provides insight into how network chemistry, charge compensation, and interfacial oxygen speciation regulate hazardous-ion retention in hydrated inorganic materials. These findings can aid the interpretation of experimental observations and provide mechanistic insights that may inform future efforts to improve contaminant retention in cementitious binders for industrial and nuclear waste containment. The resulting composition–structure–transport framework may also inform analogous studies of ion transport in nanoporous ceramics and glasses, glass dissolution and reactivity, and hazardous- or radioactive-waste immobilization in glass-based materials.

## 2 Computational Methods

### 2.1 C–S–H/C–(N)–A–S–H/N–A–S–H pore model

To investigate the transport and adsorption behaviors of heavy-metal ions in nanoscale cementitious gel pores, five slit-shaped nanochannels with a nominal pore width of ~4 nm were constructed from three representative gel families, as shown in **Figure 2**. These include: (a) C–(N)–A–S–H gels derived from a tobermorite structure (**Figure 2**a–c), representing binder gels in high-Ca precursor-based AAMs; (b) C–S–H gel with a Ca/Si ratio of 1.67 (**Figure 2**d), representing the binder gel in ordinary Portland cement (OPC) systems [51]; and (c) an amorphous N–A–S–H model (**Figure 2**e), representing binder gels in low-Ca AAM systems. Together these models encompass gel pores for both conventional and emerging alternative cementitious materials. The selected pore width (~4 nm) lies within the experimentally measured pore size distribution of cementitious gel pores (0.5–10 nm) [52, 53]. In addition, pore-size effects were examined for a representative ion–gel system using pore widths of 1, 2, 4, and 8 nm. These slit-pore models are idealized representations designed to isolate surface-controlled adsorption and transport mechanisms under well-defined gel surface chemistries. They do not capture the full pore-network connectivity, phase heterogeneity, or evolving pore-solution chemistry of real cementitious binders; these limitations are discussed in **Section 4**.

For the C–(N)–A–S–H gel pores (**Figure 2**a–c), the initial atomic model of the unit cell was from Hamid's tobermorite structure, with a basal spacing of 11 Å and a Ca/Si ratio of ~0.67–1 depending on the occupancy of additional Ca sites and the associated protonation state [54]. After construction and orthogonalization of the supercell, the structure was cleaved through the interlayer region along the [001] direction to create a 4 nm-wide nanochannel. Surface Ca atoms exposed by cleavage were removed where necessary, and the resulting non-bridging oxygen atoms of the surface silicate tetrahedra were protonated to terminate dangling bonds and form silanol groups. Following these surface modifications, the final gel nanopore model had an overall Ca/Si ratio of 0.89, which is close to that observed in blast furnace slag-based AAM gels (Ca/Si ≈ 0.8–1) [55, 56]. Then, surface bridging Si atoms in the "dreierketten" silicate chains were uniformly replaced with Al atoms according to the specified Al/Si ratio (ranging from 0 to 0.125), and charge-balancing $Na^+$ ions were added to maintain electroneutrality of the negatively charged $[Al(O_{1/2})_4]^{-1}$ tetrahedra. Note that the Al/Si aliovalent substitution was constrained to the bridging silicate tetrahedra protruding into the gel pore, rather than at the pairing ones, as supported by NMR studies [57-59] and *ab initio* calculations [60]. The range of Al/Si ratios explored in the C–(N)–A–S–H gel aligns with the range observed in experimental studies, 0–0.26 [16, 61-63]. For efficient model construction, Al/Si substitution was first introduced in a single unit cell, which was subsequently expanded into a supercell to obtain the final structural model shown in **Figure 2**a–c. This construction produces a periodically repeated distribution of interfacial Al sites. This idealized surface substitution scheme was designed to isolate the direct influence of interfacial Al substitution on ion adsorption and transport, while preserving a common tobermorite-derived framework. Accordingly, the C–(N)–A–S–H models should be regarded as controlled interfacial model systems rather than representations of the heterogeneous Al distributions expected in real C–(N)–A–S–H gels. Consequently, the observed trends should therefore be interpreted as the coupled effects of increasing interfacial Al-site density, associated framework charge, Al-linked oxygen environments, and charge-balancing $Na^+$ within a common tobermorite-derived framework. Al substitution at subsurface or other tetrahedral sites was not considered, although such distributions may alter the surface structure and electrostatic environment of real C–(N)–A–S–H gels. The effects of alternative Al distributions warrant investigation in future studies. The charge-balancing $Na^+$ ions were not fixed during the simulations, and were allowed to migrate during equilibration and production, enabling possible exchange with heavy-metal ions. Their redistribution and exchange behaviors are analyzed in **Section 3**.

For the C–S–H nanopore model (**Figure 2**d), the molecular model of the C–S–H gel was adopted from a widely used structure developed by Mohamed et al. [64], which has a Ca/Si ratio of 1.67, representing the poorly crystalline binder phase in OPC system [51]. This structure was generated from a defective 14 Å tobermorite structure based on a building block approach and optimized through classical MD and density functional theory (DFT) calculations [64]. A ~4 nm nanopore was created by expanding the interlayer spacing while preserving the intralayer atomic configuration. This approach of creating nanochannels is commonly used in previous MD studies [26, 43].

Although both C–(N)–A–S–H and C–S–H nanopore model families are derived from tobermorite-based structures, they differ markedly in composition and chain topology. The C–(N)–A–S–H models retain the continuous, relatively intact dreierketten silicate or aluminosilicate chains inherited from 11 Å tobermorite. The C–(N)–A–S–H (0) model contains no Al substitution or charge-balancing $Na^+$, whereas the Al-containing models are generated by progressively substituting bridging Si with Al, followed by the introduction of charge-balancing $Na^+$ ions, without interrupting the aluminosilicate chains. In contrast, the C–S–H model is based on defective 14 Å tobermorite, in which partial removal of bridging silicate tetrahedra produces shorter and discontinuous silicate chains. The removal of silicate tetrahedra, together with the incorporation of additional $Ca^{2+}$ ions, increases the Ca/Si ratio to 1.67. The Al/Si and Ca/(Si+Al) ratios are provided in **Figure 2**. The detailed $Q^n$ distributions and mean chain lengths (MCL) are reported in **Table S2** of the **Supplementary Material**. The C–S–H model has an MCL of approximately 3.31, consistent with previously reported values for similar Ca/Si ratios [65]. However, the absence of $Q^1$ chain-end species in the idealized C–(N)–A–S–H models result in a formally infinite MCL according to the conventional $Q^1$–$Q^2$-based expression, reflecting their continuous chain topology rather than an infinitely extended physical chain.

The N–A–S–H gel nanopore (**Figure 2**e) was generated from an amorphous sodium aluminosilicate (N–A–S) glass model, followed by the insertion of hydroxyl (OH) groups and water ($H_2O$) molecules to represent hydrated, amorphous N–A–S–H gel. This approach is commonly adopted due to the structural similarity between N–A–S glass and N–A–S–H gel [30]. The initial N–A–S glass structure was generated following the widely adopted "melt-and-quench" method in MD simulations [66-68] using the Pedone force field [69]. More details on the force field and simulation are given in **Section S1** of the **Supplementary Material**. Briefly, a system containing 5200 atoms (1200 $SiO_2$, 200 $Al_2O_3$, and 200 $Na_2O$ units) was first equilibrated at 5000 K for 1 ns to erase memory of the initial configuration. The melt was then quenched to 2000 K over 1.5 ns, equilibrated at 2000 K for 1 ns, further quenched to 300 K over 1.7 ns, and finally equilibrated at 300 K for 1 ns. The MD simulations were performed in the isothermal–isobaric NPT ensemble using the Martyna-Tobias-Klein (MTK) thermostat and a time step of 1 fs. The resulting amorphous N–A–S structure was validated by comparing its pair distribution function (PDF) with experimental X-ray PDF data for a sodium silicate-activated metakaolin geopolymer binder [70]. The level of agreement was quantified using the weighted residual factor, $R_w$, as defined in the PDFgui [71], yielding $R_w$ = 0.33 **(Figure S1)**. This value is comparable to those reported in previous studies of amorphous structure refinement as summarized in Ref. [72], indicating that the constructed model reasonably reproduces the main real-space structural correlations of the experimental geopolymer binder within the fitted range. Accordingly, the model structure is used here as a representative hydrated amorphous aluminosilicate environment resembling N–A–S–H gel, rather than as a uniquely defined N–A–S–H configuration. Further discussion of the structural attributes is given in **Section S2** of the **Supplementary Material**.

The established dry amorphous N–A–S glass model was then cleaved and separated to create a 4 nm-wide slit-shaped nanopore. The connectivity of the exposed aluminosilicate network was then identified using cutoff distances of 2.2 and 2.5 Å for Si–O and Al–O pairs, respectively. Hydroxyl groups were introduced to terminate undercoordinated surface Si and Al sites, while exposed non-bridging oxygen atoms were protonated. The accessible void space within the hydroxylated N–A–S framework was subsequently identified through isosurface analysis. In addition to the newly created nanochannel, internal voids with sufficient volume to accommodate at least one water molecule were identified within the amorphous N–A–S skeleton. During the subsequent solution-packing step described below, the nanochannel was filled with the prescribed aqueous electrolyte in the same manner as the other gel models, while additional water molecules were inserted into the accessible internal voids at an initial target density of approximately 1.0 g $cm^{-3}$. Ions were initially placed only within the nanochannel and were not inserted into the internal voids during model construction. This treatment differs from that used for the C–S–H and C–(N)–A–S–H models, in which water molecules and ions were introduced only into the nanochannel. Similar non-GCMC hydration strategies have been adopted in previous MD studies of cementitious gel models, in which a prescribed number of water molecules was directly inserted using structure-generation codes, PACKMOL, or related packing algorithms, followed by MD relaxation [65, 73, 74].

Once the nanochannels were constructed, they were filled with aqueous salt solutions using the PACKMOL software [75]. Three electrolyte compositions were considered: 1 M $PbCl_2$ (corresponding to a $H_2O/Pb^{2+}$ ratio of ~55.5:1), 1 M $BaCl_2$ ($H_2O/Ba^{2+}$ ratio of ~55.5:1), and 2 M CsCl ($H_2O/Cs^+$ ratio of ~27.8:1). These concentrations were selected such that the total ionic charges contributed by dissolved cations were equivalent across all cases. The relatively high concentrations also ensure sufficient statistical sampling in the MD simulations and are consistent with previous MD studies of heavy-metal ions in nanopores of cementitious gels [33, 39, 40] and clays [76]. The electrolyte compositions are therefore treated as controlled reference states for comparing ion–surface interactions across gel models, rather than as direct representations of equilibrium cement pore solutions.

The number of water molecules and ion pairs in each system was initially estimated from the channel volume by assuming a target solution density of $\sim 1.2 \mathrm{g/cm^3}$. This value was used only as a starting point, since the subsequent NPT equilibration at 300 K and 1 atm allowed the pore space to relax and the final solution density to be refined. After system equilibration at 300 K and 1 atm, the bulk solution densities of $PbCl_2$, $BaCl_2$, and CsCl were determined by averaging over configurations sampled at each timestep (8,000,000 in total) within the central 1.0 nm region of C–(N)–A–S–H (Al/Si = 0) and C–S–H nanopores (where mid-channel properties are expected to approximate bulk conditions) were $\sim 1.268 \mathrm{\ g/cm^3}$, $\sim 1.160 - 1.172 \mathrm{\ g/cm^3}$, and $\sim 1.240 - 1.274 \mathrm{\ g/cm^3}$, respectively. These values are in close agreement with reported experimental densities for 1.0 M $BaCl_2$ (~$1.176\ g/cm^3$ at 20 °C), estimated by linear interpolation of tabulated data in ref. [77], and for 2.0 M CsCl (~$1.240\ g/cm^3$ at 25 °C), estimated from interpolation and conversion between molality ($m$, mol/kg) and molarity ($M$, mol/L) using $M = \frac{m\rho}{1+\frac{mM_s}{1000}}$, where $M_s$ is the solute molar mass (g/mol) and $\rho$ is the mass density of the solution (g/cm$^3$, obtained from ref. [78]). The cross-sectional views and dimensions of the final models for various gel nanopores are shown in **Figure 2**, with each model consisting of ~13,000 atoms.

Although the solubility of $PbCl_2$ in pure water at ambient temperature is limited, the 1 M $PbCl_2$ solution modeled here is treated as a simplified chloride-rich reference electrolyte for comparative analysis, rather than as a realistic representation of cement pore solution chemistry. The imposed composition enables comparison of dissolved Pb-containing species across the model systems and allows us to isolate adsorption-controlled transport under chloride-rich conditions. Such simplified model systems are commonly used in molecular simulations to enable systematic comparison of ion behavior across systems and concentrations [33, 37, 39, 40, 76]. Precipitation and solubility equilibria are outside the scope of the present non-reactive MD framework. To verify the internal consistency of the simulated dissolved state over the MD timescale, Pb–Cl and Pb–O coordination statistics and the Pb–Pb radial distribution function were monitored; no persistent Pb-rich aggregation was observed over the simulated timescale.

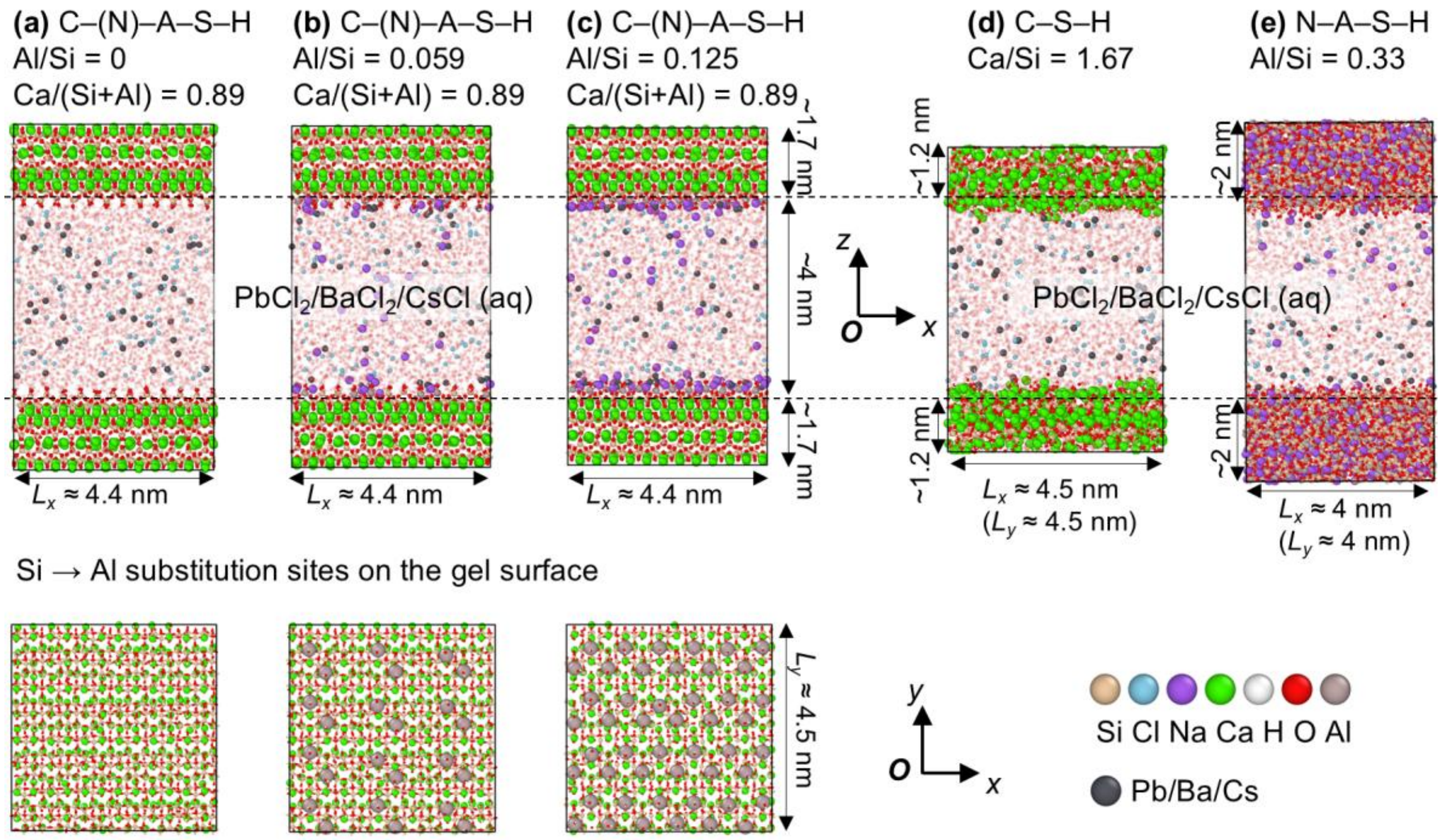

**Figure 2.** Cross-sectional views of the model structures of the gel nanochannels: (a–c) C–(N)–A–S–H with Al/Si ratios of 0, 0.059, and 0.125; (d) C–S–H with a Ca/Si ratio of 1.67, constructed using atomic coordinates reported in ref. [64]; and (e) N–A–S–H with an Al/Si ratio of 0.33. Each constructed ~4 nm-wide gel nanochannel is filled with an aqueous solution of 1 M $PbCl_2$ or $BaCl_2$ or 2 M CsCl. The coordinate origin ($z$ = 0 nm) corresponds to the center of the nanochannel. Water molecules are shown with partial transparency to enhance visualization.

## 2.2 Simulation details

Based on the saturated nanopore models in **Section 2.1**, classical MD simulations were performed to investigate the immobilization mechanisms of heavy-metal ions in C–S–H, C–(N)–A–S–H, and N–A–S–H nanopores. All simulations were carried out using the Large-scale Atomic/Molecular Massively Parallel Simulator (LAMMPS) [79] with visualizations generated with the Open Visualization Tool (OVITO) [80]. Periodic boundary conditions were applied in all directions, and the equations of motion were integrated with the velocity-Verlet algorithm using a timestep of 1 fs. After energy minimization of the liquid-filled nanochannel via the Polak–Ribière conjugate-gradient method, the system was equilibrated for 8 ns in the isothermal–isobaric (NPT) ensemble at 300 K and 1 atm. Convergence of the potential energy (**Figure S2**) confirmed that the system had reached equilibrium. Subsequently, production trajectories were generated over 8 ns in the canonical (NVT) ensemble at 300 K, with configurations recorded every 8 ps. Both the NPT and NVT stages employed Nosé–Hoover coupling [81, 82], with thermostat and barostat relaxation times of 0.1 ps and 1 ps, respectively.

**Table 1.** Summary of simulation cases performed in this study.

| Simulation case | Gel type | Solution type | Numbers of water molecules and ions in the nanopore |
|---|---|---|---|
| Case 1 | C–(N)–A–S–H (Al/Si = 0) | 1 M $PbCl_2$ | $47Pb+94Cl+2689H_2O+0Na$ |
| Case 2 | C–(N)–A–S–H (Al/Si = 0.059) | 1 M $PbCl_2$ | $47Pb+94Cl+2689H_2O+48Na$ |
| Case 3 | C–(N)–A–S–H (Al/Si = 0.125) | 1 M $PbCl_2$ | $47Pb+94Cl+2689H_2O+96Na$ |
| Case 4 | C–S–H (Ca/Si = 1.67) | 1 M $PbCl_2$ | $48Pb+96Cl+2641H_2O+0Na$ |
| Case 5 | N–A–S–H (Al/Si = 0.33) | 1 M $PbCl_2$ | $38Pb+76Cl+2090H_2O+400Na$ |
| Case 6 | C–(N)–A–S–H (Al/Si = 0) | 1 M $BaCl_2$ | $47Ba+94Cl+2689H_2O+0Na$ |
| Case 7 | C–(N)–A–S–H (Al/Si = 0.059) | 1 M $BaCl_2$ | $47Ba+94Cl+2689H_2O+48Na$ |
| Case 8 | C–(N)–A–S–H (Al/Si = 0.125) | 1 M $BaCl_2$ | $47Ba+94Cl+2689H_2O+96Na$ |
| Case 9 | C–S–H (Ca/Si = 1.67) | 1 M $BaCl_2$ | $48Ba+96Cl+2641H_2O+0Na$ |
| Case 10 | N–A–S–H (Al/Si = 0.33) | 1 M $BaCl_2$ | $38Ba+76Cl+2090H_2O+400Na$ |
| Case 11 | C–(N)–A–S–H (Al/Si = 0) | 2 M CsCl | $94Cs+94Cl+2689H_2O+0Na$ |
| Case 12 | C–(N)–A–S–H (Al/Si = 0.059) | 2 M CsCl | $94Cs+94Cl+2689H_2O+48Na$ |
| Case 13 | C–(N)–A–S–H (Al/Si = 0.125) | 2 M CsCl | $94Cs+94Cl+2689H_2O+96Na$ |
| Case 14 | C–S–H (Ca/Si = 1.67) | 2 M CsCl | $96Cs+96Cl+2641H_2O+0Na$ |
| Case 15 | N–A–S–H (Al/Si = 0.33) | 2 M CsCl | $76Cs+76Cl+2090H_2O+400Na$ |
| Case 16 | Bulk | 1 M $PbCl_2$ | $37Pb+74Cl+2028H_2O$ |
| Case 17 | Bulk | 1 M $BaCl_2$ | $37Ba+74Cl+2028H_2O$ |
| Case 18 | Bulk | 2 M CsCl | $74Cs+74Cl+2028H_2O$ |

Fifteen primary nanochannel cases were considered, covering three representative contaminant cations, $Pb^{2+}$, $Ba^{2+}$, and $Cs^{+}$, across five gel surface chemistries (**Table 1**), which also provides the final numbers of water molecules and ions added to each nanopore. Each case was simulated in triplicate using randomized initial configurations to improve statistical robustness. In addition, three bulk reference systems, 1 M $PbCl_2$, 1 M $BaCl_2$, and 2 M CsCl, were simulated in a 4 × 4 × 4 $nm^3$ periodic cell under identical simulation conditions. Comparison between nanoconfined and bulk systems enables elucidation of nanoconfinement effects on heavy-metal adsorption and transport and provides reference data for validation against published bulk solution data. To examine pore-size effects, additional simulations were performed for a representative system ($Cs^{+}$ in C–(N)–A–S–H with Al/Si = 0.125) using pore widths of 1, 2, 4, and 8 nm.

All MD simulations employed the ClayFF force field [83]. The potential parameters and partial charges follow refs. [83-87], with full details given in the **Supplementary Material** (**Section S5; Tables S3–S5**). The Lorentz–Berthelot mixing rule [88, 89] was used to obtain cross Lennard–Jones parameters, consistent with the ClayFF framework. Long-range electrostatics were treated with the particle-particle-particle-mesh (PPPM) method [90], with a target accuracy of $10^{-5}$, using a real-space cutoff of 1.2 nm for both Coulombic and short-range van der Waals interactions; tail corrections were applied to the van der Waals energy. ClayFF has been shown to reproduce structural and dynamical properties of hydrated cementitious systems with good accuracy, while offering substantially lower computational cost compared to reactive force fields (e.g., ReaxFF [91]) and DFT methods [26]. For these reasons, ClayFF has been widely used for simulating ionic transport and interfacial processes in cementitious nanopores [31, 33, 39, 43, 92-96]. In the present context, the use of reactive force fields (e.g., ReaxFF) is constrained by the lack of a unified and validated parameterization capable of consistently describing all relevant elements and interactions across both solid and aqueous phases; combining multiple parameter sets may introduce additional uncertainties. Within the non-reactive ClayFF framework, the simulations are intended to capture physical adsorption, hydration, ion pairing, coordination environments, and nanoconfined mobility suppression, but not precipitation, chemical incorporation, or bond-breaking/bond-forming reactions. These implications are discussed further in **Section 4**.

# 3 Results & Discussion

The Results and Discussion are organized to progressively link transport behavior to interfacial mechanisms. We first compare bulk and nanoconfined diffusivities to quantify overall mobility suppression and then use spatially resolved diffusivity and residence-time analyses to identify near-surface low-mobility regions. Ion-density profiles and adsorption maps are then used to determine how ions partition between the pore center and interfacial regions. Finally, radial distribution function (RDF), coordination, and binding-strength analyses are used to identify the surface environments responsible for adsorption-controlled immobilization and to rationalize the relationship between interfacial binding and mobility suppression across gel chemistries, ion species, and selected pore-size conditions.

## 3.1 Diffusion of Heavy-Metal Ions in Bulk and Confined Nanopores

### *3.1.1 Calculation of average self-diffusion coefficients*

The mobility of heavy-metal ions is closely linked to their environmental transport, toxicity and bioavailability. Ion and water mobility were quantified using the mean-square displacement (MSD), a standard metric for diffusion analysis in MD simulations [96-100]. The MSD describes the average squared displacement of particles from their initial positions and was calculated from the MD trajectories using **Equation** (1):

$$\mathrm{MSD}(i, t_0, \tau) = (\boldsymbol{r}_i(\tau + t_0) - \boldsymbol{r}_i(t_0))^2, \tag{1}$$

where, $\boldsymbol{r}_i(t_0)$ and $\boldsymbol{r}_i(\tau + t_0)$ denote the position vectors of particle $i$ at times $t_0$ and $\tau + t_0$, respectively. The self-diffusion coefficient ($D$) of ions and water molecules was then calculated from the Einstein's diffusion equation [101, 102],

$$D = \lim_{\tau \to \infty} \frac{\langle \mathrm{MSD}(i, t_0, \tau) \rangle_{i, t_0}}{2d\tau}, \tag{2}$$

where, $\langle \cdot \rangle_{i, t_0}$ denotes the ensemble average over all particles ($i$) and all-time origins ($t_0$), and $d$ is the dimensionality of the diffusion process ($d = 3$ in this work). In many prior studies (e.g., refs. [76, 98, 99, 103]), diffusion coefficients are estimated using MSD trajectories initiated from a single time origin ($t_0 = 0$):

$$D = \lim_{\tau \to \infty} \frac{\langle \mathrm{MSD}(i, t_0 = 0, \tau) \rangle_{i}}{2d\tau}, \tag{3}$$

In contrast, **Equation** (2) considers all available time origins ($t_0 \in$ 0–8 ns) and lag times ($\tau \in$ 0–8 ns). Nevertheless, the number of valid time origins decreases with increasing $\tau$: many trajectories contribute at short lag times, whereas at $\tau = 8$ ns only a single trajectory ($t_0 = 0$) remains, resulting in poor sampling of the long-time MSD. To improve statistical robustness, the diffusion coefficient was computed by ensemble-averaging instantaneous estimates over all particle-time $(i, t_0, \tau)$ combinations:

$$D = \langle \frac{\mathrm{MSD}(i, t_0, \tau)}{2d\tau} \rangle_{i, t_0, \tau}, \tag{4}$$

This formulation, introduced in our recent work [43], improves the statistical estimation of MSD by incorporating all available displacement events, which is particularly beneficial when long-time sampling is limited for spatially resolved diffusivity in heterogeneous environments. As a result, it yields smoother and more linear MSD profiles, especially in spatially resolved analyses of heterogeneous systems where conventional estimators suffer from sparse long-lag statistics. Accordingly, this approach is adopted here as a statistically robust and consistent method for comparative diffusivity analysis. For each simulation case, three independent production runs with distinct initial configurations were performed, and the reported diffusion coefficients represent the mean values, with standard deviations shown as error bars.

**Figure 3** shows representative MSD profiles of $Ba^{2+}$ ions in a 1 M bulk $BaCl_2$ solution over the 8-ns NVT production stage, with each curve corresponding to one of the ensemble-averaging schemes in **Equations** (2)–(4). As expected, MSD increases approximately linearly with time, reflecting the random, uncorrelated motion of ions in solution. The MSD calculated using **Equation** (2) exhibits improved linearity compared with **Equation** (3) due to enhanced statistical averaging. However, at large lag times, the scarcity of valid time origins reduces the reliability of the MSD obtained from **Equation** (2). The estimator defined in **Equation** (4), which is adopted in this work, yields result consistent with the other two methods while providing improved statistical robustness.

To validate both the force field and the diffusivity calculation method (**Equation** (4)), **Table 2** compares the calculated bulk self-diffusion coefficients of $Pb^{2+}$ (1 M), $Ba^{2+}$ (1 M), and $Cs^{+}$ (2 M) at 300 K, with literature values from experiments and MD simulations under similar conditions. For reference, infinite-dilution (~0 M) experimental values are also included, where higher diffusivities are expected due to reduced ionic interactions [104]. As anticipated, the computed diffusivities fall below infinite-dilution limits yet remain in close agreement with prior experimental [104-108] and MD [109] data, demonstrating the reliability of present methodology.

In addition, **Table 2** shows that the bulk diffusivities follow the trend $Pb^{2+} \approx Ba^{2+} < Cs^{+}$, consistent with literature reports. To rationalize this behavior, the ionic properties of $Pb^{2+}$, $Ba^{2+}$, $Cs^{+}$, and $Na^{+}$ were compared (**Table 3**), including ionic charge ($z$), ionic radius ($r_\mathrm{I}$), hydration shell thickness ($\Delta r$), hydration radius ($r + \Delta r$), hydration number, charge density ($z/r_\mathrm{I}^3$), and hydration free energies ($\Delta G_\mathrm{hyd}$) [110, 111]. $Pb^{2+}$ and $Ba^{2+}$ exhibit higher charge densities and more negative hydration free energies than $Cs^{+}$, indicative of stronger ion–water interactions. Consequently, $Pb^{2+}$ and $Ba^{2+}$ form more structured, tightly bound hydration shells, which increase their effective hydrodynamic (Stokes) radii and suppress diffusion, consistent with the Stokes–Einstein relation linking diffusivity inversely to the hydration radius [111]. In contrast, $Cs^{+}$ has a lower charge density and weaker hydration, allowing faster diffusion in aqueous solution. For comparison, $Na^{+}$ displays a higher charge density and a more negative hydration free energy than $Cs^{+}$, which is consistent with the well-established classification of $Cs^+$ as a structure-breaker and $Na^+$ as a structure-former in aqueous solutions [104]. Accordingly, $Na^+$ generally exhibits slower bulk diffusivity (e.g., $1.07 \times 10^{-9} \mathrm{m}^2/\mathrm{s}$ [43]) than $Cs^+$ under similar conditions.

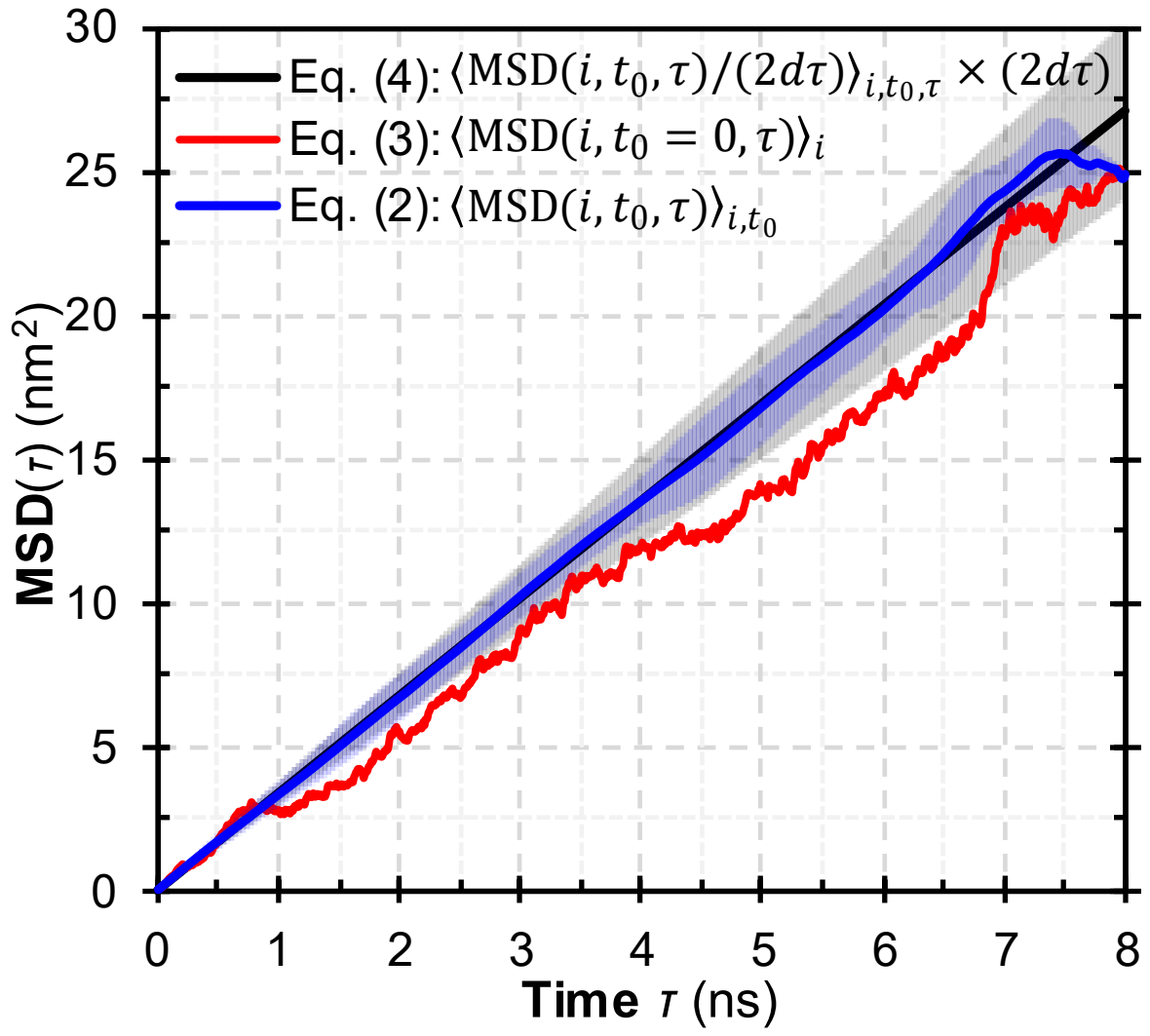

**Figure 3.** Typical MSD curves for $Ba^{2+}$ in a 1 M $BaCl_2$ bulk solution at 300 K, obtained from an 8-ns MD production run and analyzed using three different ensemble-averaging methods corresponding to **Equations** (2), (3), and (4) in the text, respectively. The error bars represent the standard deviations from ensemble-averaging calculations.

**Table 2.** Comparison of the resulting average diffusion coefficients for $Pb^{2+}$, $Ba^{2+}$, and $Cs^+$ ions in the $PbCl_2$/$BaCl_2$/CsCl bulk solution at 300 K with literature values from MD simulations [109] and experimental measurements [104-108]. The calculated values reported in this work are accompanied by standard deviations from three independent simulations using different initial configurations.

| Heavy metal ion species | Source | Approach | Concentration | Value |
|---|---|---|---|---|
| $Pb^{2+}$ in $PbCl_2$ (aq) | This work | MD simulation | ~1 M | 0.57±0.01 |
| | Ref. [105] | Experiment | Infinite dilution | 0.94 |
| | Ref. [107] | Experiment | Infinite dilution | 0.95 |
| $Ba^{2+}$ in $BaCl_2$ (aq) | This work | MD simulation | ~1 M | 0.56±0.01 |
| | Ref. [109] | MD simulation | Infinite dilution | 0.64±0.03 |
| | Ref. [107] | Experiment | Infinite dilution | 0.85 |
| | Ref. [106] | Experiment | ~1 M | 1.18 |
| | Ref. [108] | Experiment | Infinite dilution | 0.85 |
| $Cs^+$ in CsCl (aq) | This work | MD simulation | ~2 M | 1.84±0.09 |
| | Ref. [109] | MD simulation | Infinite dilution | 1.45±0.05 |
| | Ref. [104] | Experiment | ~2 M | 1.80 |
| | Ref. [105] | Experiment | Infinite dilution | 2.17±0.05 |
| | Ref. [107] | Experiment | Infinite dilution | 2.07 |
| | Ref. [108] | Experiment | Infinite dilution | 2.06 |

**Table 3.** Ionic charge ($z$, in $e$), radii ($r_{\mathrm{I}}$, in Å), hydration shell thickness ($\Delta r$), hydration radius ($r + \Delta r$), hydration number, estimated charge density ($z/r_{\mathrm{I}}^3$, in $e/\text{Å}^3$), and hydration free energies ($\Delta G_{\mathrm{hyd}}$, in kJ/mol) of the three heavy metal ions and the sodium ion covered in this work.

| Ion type | Ionic charge, $z$ ($e$) | Ionic radius, $r_{\mathrm{I}}$ (Å) [110] | Width of hydration shell, $\Delta r$, (Å) [110] | Hydration radius, $r + \Delta r$, (Å) | Hydration number [110] | Charge density, $z/r_{\mathrm{I}}^3$ ($e/\text{Å}^3$) | Hydration free energy, $\Delta G_{\mathrm{hyd}}$ (kJ/mol) [111] |
|---|---|---|---|---|---|---|---|
| **$Pb^{2+}$** | 2 | 1.18 | 1.43 | 2.61 | 6.1 | 1.22 | −1510 |
| **$Ba^{2+}$** | 2 | 1.36 | 1.18 | 2.54 | 5.3 | 0.80 | −1270 |
| **$Cs^{+}$** | 1 | 1.70 | 0.49 | 2.19 | 2.1 | 0.20 | −266 |
| **$Na^{+}$** | 1 | 1.02 | 1.16 | 2.18 | 3.5 | 0.94 | −383 |

Using the same diffusivity calculation method (**Equation** (4)), we calculated the average self-diffusion coefficients of $Pb^{2+}$, $Ba^{2+}$, and $Cs^+$ ions confined within different gel nanochannels (**Figure 4**). The corresponding results for $Cl^-$ ions and water molecules are provided in **Figure S3** of the **Supplementary Material**. To further quantify the confinement effect, the extent of diffusivity suppression (hereafter referred to as extent of immobilization) for $Pb^{2+}$, $Ba^{2+}$, and $Cs^+$ ions relative to their bulk diffusivities (**Figure 4**) was estimated using **Equation** (5):

$$S = \frac{D_{\mathrm{Bulk}} - D_{\mathrm{Nanoconfinement}}}{D_{\mathrm{Bulk}}}, \quad (5)$$

where $D_{\mathrm{Bulk}}$ and $D_{\mathrm{Nanoconfinement}}$ denote diffusion coefficients in bulk solution, and under nanoconfinement, respectively.

The results demonstrate that confinement within the 4 nm-wide gel pores leads to pronounced reductions in diffusivity for all species (**Figure 4** and **Figure S3**), consistent with previous studies of ionic transport in nanopores [37, 43, 92, 98-100]. All studied gels significantly hinder the transport of heavy metal ions, although the extent of immobilization varies with gel chemistry (**Figure 4**). In the C–S–H pore (representing OPC systems), diffusivities of $Pb^{2+}$, $Ba^{2+}$, and $Cs^{+}$ are reduced by ~59%, ~57%, and ~47%, respectively, relative to their bulk values. In C–(N)–A–S–H systems, the immobilization effect increases significantly with Al/Si ratio, rising from ~22–29% at Al/Si = 0 to ~55–67% at Al/Si = 0.125, where the extent of immobilization surpasses that observed in C–S–H for all three ions studied (**Figure 4** and **Figure S3**). Similar trends have been reported in montmorillonite nanopores, where tetrahedral Al substitution enhances the retention of heavy-metal ions such as $Cd^{2+}$, $Pb^{2+}$, and $Zn^{2+}$ [76]. In contrast, the N–A–S–H pore exhibits slightly weaker suppression than C–S–H for $Pb^{2+}$ and $Ba^{2+}$ but slightly stronger suppression for $Cs^{+}$. Despite these gel-specific differences, the relative diffusion trend $Pb^{2+} \approx Ba^{2+} < Cs^{+}$ persists across all systems, suggesting that ion-specific hydration effects remain influential under nanoconfinement, as will be discussed further in **Section 3.2.3**.

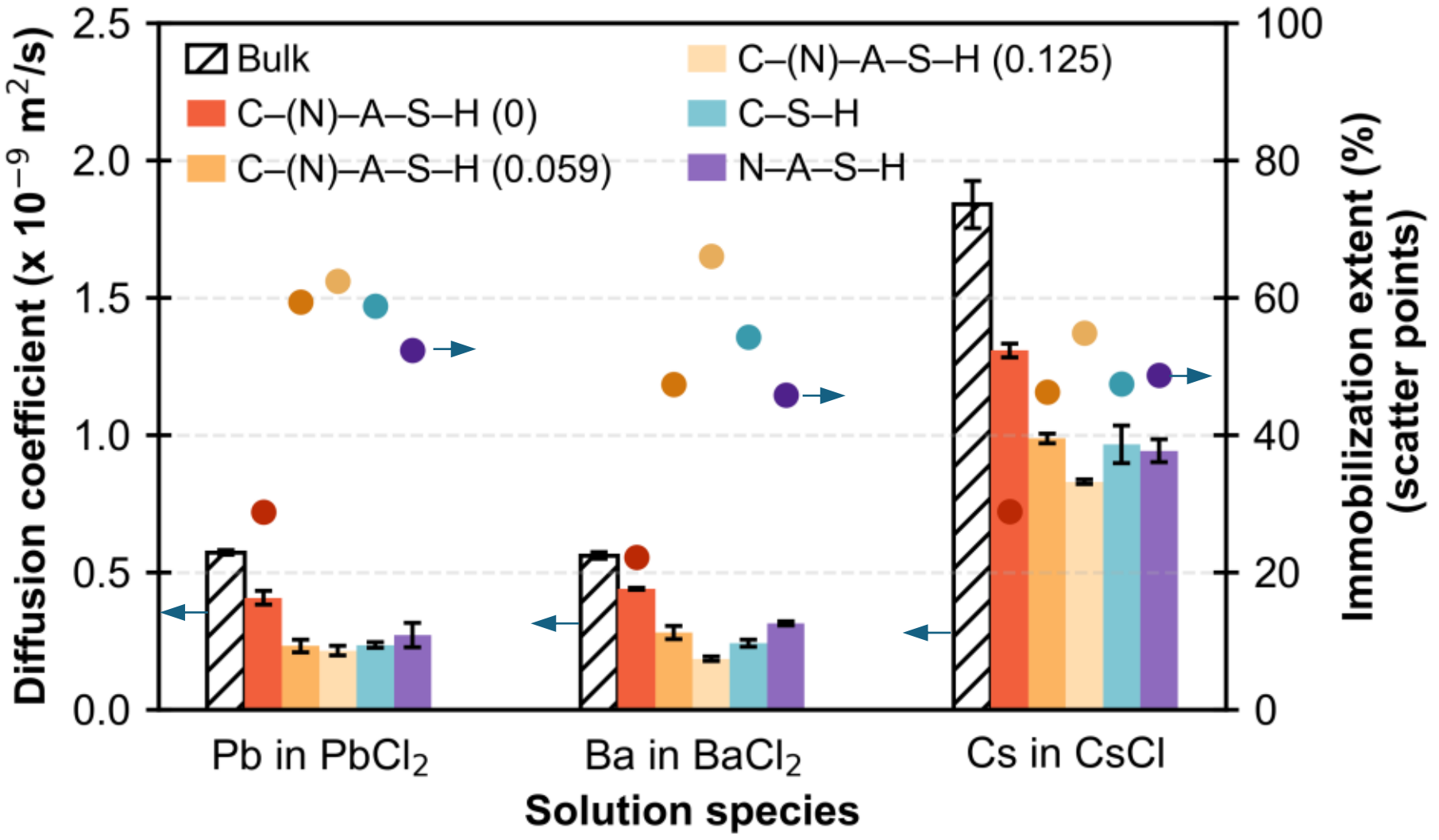

**Figure 4.** Impact of nanoconfinement within 4 nm-wide gel pore channels on the average self-diffusion coefficients of $Pb^{2+}$, $Ba^{2+}$, and $Cs^{+}$ ions, and the corresponding extent of immobilization. **Bar plots (left y-axis)** show the average self-diffusion coefficients, with error bars representing standard deviations from three independent simulations. **Scatter points (right y-axis)** represent the extent of immobilization, defined as the relative reduction in diffusivity under nanoconfinement compared with the corresponding bulk values.

### *3.1.2 Spatially resolved diffusivity and residence time*

Unlike most prior cementitious-gel MD studies, which report pore-averaged transport metrics, the present analysis resolves position-dependent mobility across different gel chemistries. To elucidate the origin of the reduced heavy-metal ion mobility under nanoconfinement, spatially resolved diffusivity profiles were computed for all solution species across the gel nanochannel (see **Section S7** of **Supplementary Material** for details). Diffusivities were resolved along the $z$-direction, normal to the pore surface, with the channel mid-plane defined as $z = 0$ nm to enable consistent comparison across gel chemistries. A bin width of 1 Å is adopted to construct the spatially resolved diffusivity profiles, representing a balance between spatial resolution and statistical reliability: smaller bins yield insufficient sampling for stable averaging and lead to pronounced fluctuations, whereas larger bins smooth near-surface gradients, obscure interfacial features, and increase the computational overhead of trajectory analysis, particularly for rougher N–A–S–H surfaces. A representative example is shown in **Figure S5**e, demonstrating that although minor statistical variations exist, the resulting profiles remain consistent in their overall distributions and qualitative trends. The resulting profiles for $Pb^{2+}$, $Ba^{2+}$, and $Cs^{+}$, spanning from the pore surface to channel mid-plane, are shown in **Figure 5**a, 5c, and 5e, respectively, with bulk diffusion coefficients indicated by horizontal dashed lines. Complete profiles across the entire pore channel for all species, including $Cl^{-}$, $Na^{+}$ (if present), and water, are given in **Figures S4–S6** of the **Supplementary Material**.

In all cases, diffusivity exhibits a characteristic parabolic distribution across the pore channel. The mobility of $Pb^{2+}$, $Ba^{2+}$, and $Cs^{+}$ approaches bulk-like values at the central region of the pore ($z \approx 0$ nm) and decreases progressively towards the pore surfaces, approaching near-zero values in the interfacial region (e.g., $z \approx \pm 2.0$ nm). These results suggest that the reduction in pore-averaged diffusivity (**Figure 4**) arises primarily from strongly hindered transport near the solid-liquid interface, underscoring the critical role of interfacial interactions in suppressing ionic mobility. Similar near-surface retardation is observed for $Cl^{-}$, $Na^{+}$ (if present), and water molecules (**Figure S4–S6** of the **Supplementary Material)**. Such interfacial slowing of molecular transport has been previously reported in MD studies of nanoconfined systems, including water near silica surfaces [112], water in silica nanopores [100, 113], and ionic transport in C–S–H [43, 114] and silicon nitride nanopores [115].

The extent of near-surface immobilization varies considerably with gel chemistry. Across all heavy-metal ions studied, the diffusivity profiles follow the trend: C–(N)–A–S–H (Al/Si = 0) > N–A–S–H > C–S–H, with the differences most pronounced within ~1 nm of the pore surface. This trend is consistent with the pore-averaged results in **Figure 4**. Increasing the Al/Si ratio in C–(N)–A–S–H progressively suppresses ionic mobility, especially near the interface. However, even at Al/Si ratio of 0.125, the local diffusivity remains less suppressed than in C–S–H across much of the pore channel. This appears inconsistent with the pore-averaged results in **Figure 4**, where C–(N)–A–S–H (Al/Si = 0.125) exhibits the highest overall extent of immobilization for all three heavy metals. This apparent discrepancy indicates that pore-averaged results are governed not only by local mobility but also by the spatial distribution of ions within the pore, which will be examined in the next section through number density profiles.

To further assess which species in the nanoconfined solution are most strongly immobilized, spatially resolved diffusivity ratios were calculated across the pore channel using the data from **Figure 5**a, **5**c, and **5**e and **Figure S4–S6**. For the $PbCl_2$ system, the diffusivity ratios $H_2O/Pb^{2+}$, $H_2O/Cl^-$, and $Cl^-/Pb^{2+}$ in the C–(N)–A–S–H (Al/Si = 0.059) nanopore are presented in **Figure 5**b, with bulk solution ratios indicated by horizontal dashed lines. The diffusivity-ratio profiles were obtained by averaging the corresponding values from the two symmetric halves of the nanopore ($-2.5$ nm < $z$ < 0 nm and 0 nm < $z$ < 2.5 nm, with the latter reflected about $z = 0$), and the error bars represent the standard deviation of the two symmetry-related values. Near the gel surface, the diffusivity in the denominator can approach zero, causing its uncertainty to be strongly amplified in the resulting ratio and leading to large variations. These near-surface ratios should therefore be interpreted with caution. Near the pore surface, both $H_2O/Pb^{2+}$ and $H_2O/Cl^-$ diffusivity ratios increase markedly relative to their bulk values, suggesting that ionic mobility is suppressed more strongly than that of water due to stronger ion-surface interactions. Toward the pore center, however, the $H_2O/Pb^{2+}$ ratio falls slightly below the bulk value, indicating that $Pb^{2+}$ mobility becomes relatively enhanced compared to water as interfacial effects diminish. The above-bulk $Cl^-/Pb^{2+}$ diffusivity ratio near the surface indicates that $Pb^{2+}$ mobility is more strongly suppressed than that of $Cl^-$, reflecting stronger adsorption of $Pb^{2+}$ onto the gel surface. Similar trends are generally observed for $BaCl_2$ (**Figure 5**d) and CsCl systems (**Figure 5**f), across all gel chemistries (**Figure S7–S9** of the **Supplementary Material**). An exception is observed for CsCl, where the $Cl^-/Cs^+$ ratio is below the bulk value near the surface (**Figure 5**f and **Figure S9**), indicating that $Cl^-$ is more strongly immobilized than $Cs^+$.

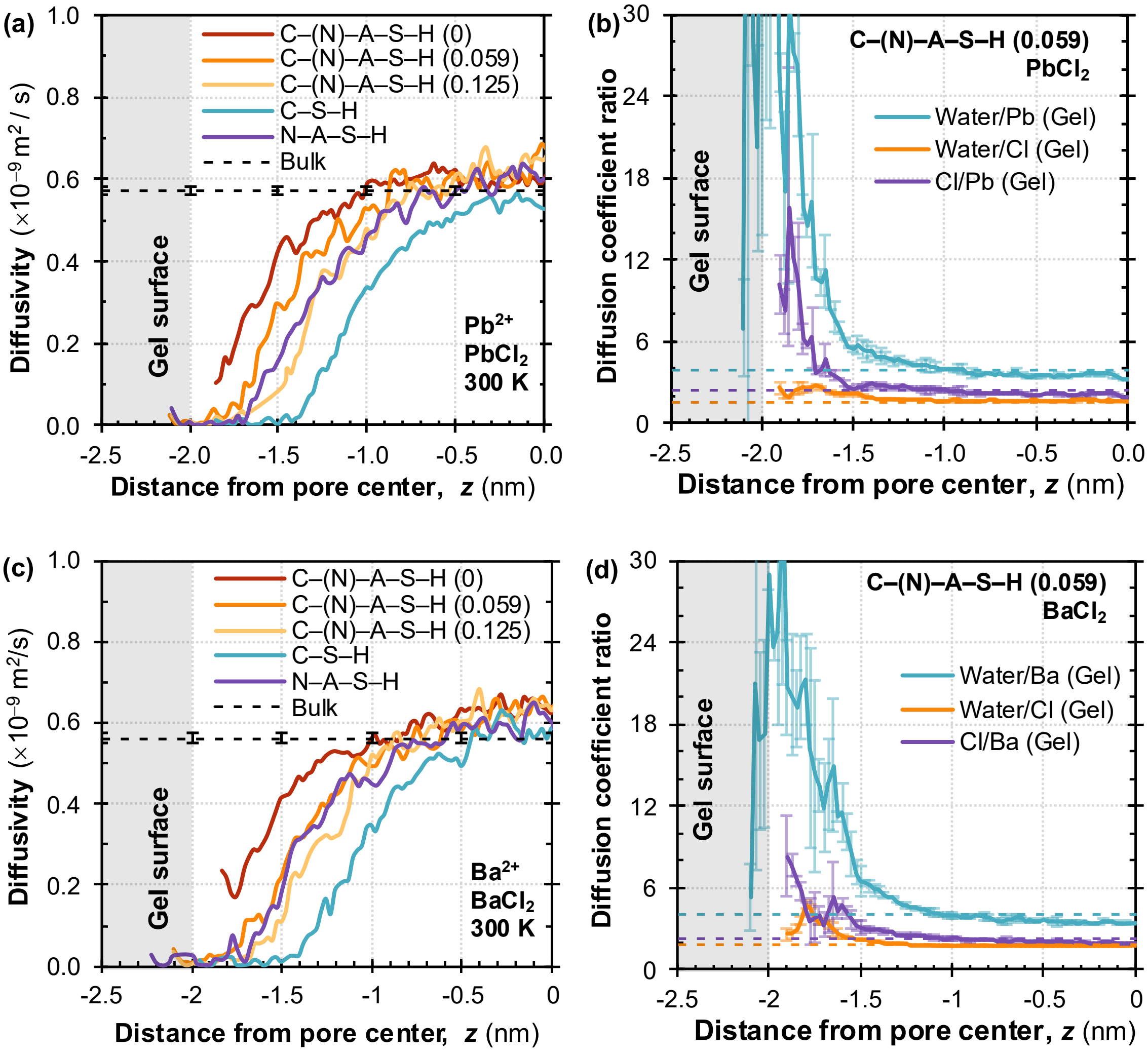

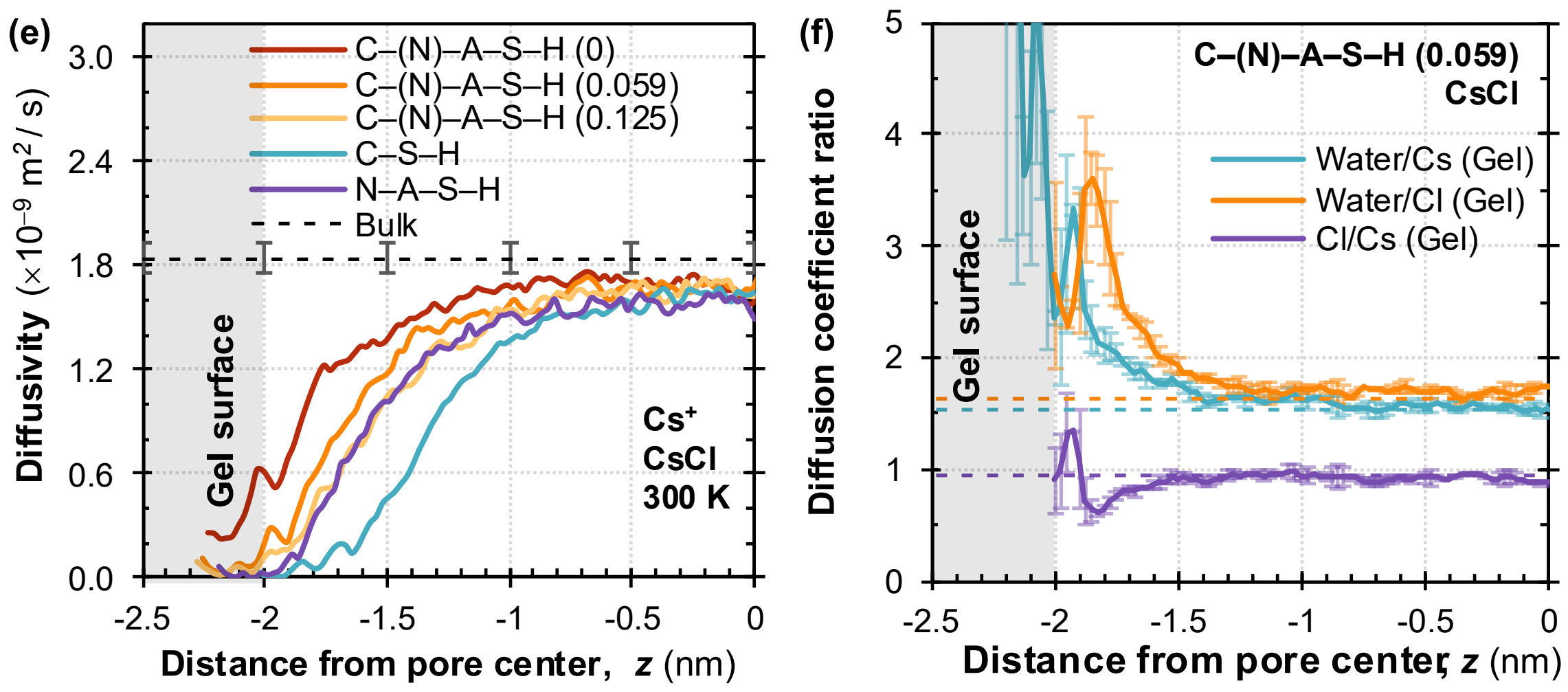

**Figure 5**. (a, c, e) Spatial profiles of the self-diffusion coefficient ($D$, in $10^{-9}$ $m^2/s$) for $Pb^{2+}$, $Ba^{2+}$, and $Cs^+$ across different gel nanopores filled with (a) 1 M $PbCl_2$, (c) 1 M $BaCl_2$, and (e) 2 M CsCl solutions at 300 K, plotted from channel center ($z$ = 0 nm) toward the pore surface ($z \approx -2$ nm). (b, d, f) Corresponding spatial profiles of the diffusivity ratios for (b) $PbCl_2$, (d) $BaCl_2$, and (f) CsCl solutions confined within a C–(N)–A–S–H (Al/Si =0.059) gel nanopore, where the error bars are estimated by averaging the profiles over the two symmetric halves of the nanopore ($-2.5$ nm < $z$ < 0 nm and 0 nm < $z$ < 2.5 nm), with the latter reflected about $z$ = 0 nm. Horizontal dashed lines indicate the corresponding bulk solution values at 300 K.

In addition to diffusivity, spatially resolved residence time (RT) analysis was performed as an independent dynamical descriptor to verify whether regions of low apparent diffusivity correspond to prolonged ion retention near the gel surface. RT is defined here as the average duration for which a species remains within a given spatial region or bound to specific sites. Because longer residence generally corresponds to slower molecular exchange, RT provides a complementary measure of position-dependent mobility under nanoconfinement. The same binning scheme along the $z$-direction (bin size of 1 Å) used to generate the diffusivity profiles in **Figure 5** was applied to compute RT distributions for all species across the different gel chemistries. To our knowledge, such spatially resolved RT profiles introduced here have not been applied to cementitious nanopores. This analysis therefore introduces a complementary descriptor for quantifying position-dependent dynamical confinement and enables a systematic comparison across different cementitious nanopores.

The results for $Ba^{2+}$ ions in the $BaCl_2$ solution are presented in **Figure 6**a, which shows that $Ba^{2+}$ ions exhibit significantly longer RT near the pore surface but gradually recover to the bulk values toward the channel center. This near-surface RT enhancement mirrors the parabolic diffusivity profiles observed in **Figure 5**, further indicating that mobility suppression arises primarily from strong interfacial interactions. Notably, peaks and valleys in the RT profiles correspond to favorable and unfavorable adsorption sites for $Ba^{2+}$, respectively, which align with variations in $Ba^{2+}$ number density, as discussed in the next section.

The extent of RT enhancement near the surface also varies systematically with gel chemistry. For all heavy-metal ions studied here (e.g., $Ba^{2+}$ in **Figure 6**a), near-surface RT generally follows the trend: C–S–H > N–A–S–H > C–(N)–A–S–H (Al/Si = 0). Within the C–(N)–A–S–H series, increasing the Al/Si ratio leads to progressively larger RT near the pore surface. The corresponding RT profiles for $Cl^-$, $Na^+$, and water, as well as for $Pb^{2+}$ and $Cs^+$ systems, are presented in **Figures S10–S12** of the **Supplementary Material** and show similar position-dependent trends.

To assess the consistency between the two mobility descriptors, **Figure 6**b compares the spatially resolved RT and diffusion coefficient ($D$) across all systems. A clear inverse correlation between $D$ and RT is observed, consistent with theoretical master curve described in **Section S10** in the **Supplementary Material**. This agreement indicates that the near-surface reduction in apparent diffusivity is not an artifact of the diffusivity estimator alone but reflects longer-lived interfacial residence events. Together, the spatially resolved diffusivity and RT analysis provides a coherent framework for characterizing heterogeneous ion transport in cementitious nanopores.

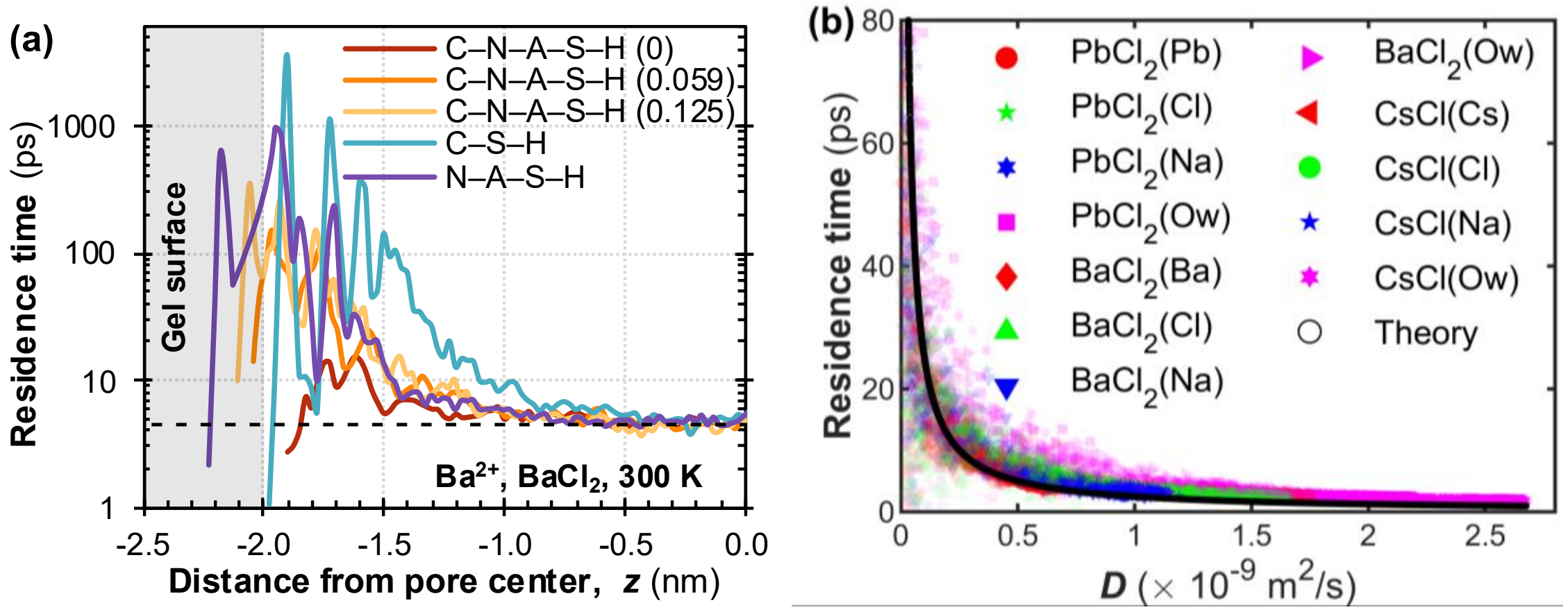

**Figure 6.** (a) Spatial profiles of residence time (RT, ps) for $Ba^{2+}$ ions across different gel nanopores filled with 1 M $BaCl_2$ solution at 300 K, plotted from channel center ($z$ = 0 nm) toward the pore surface ($z \approx -2$ nm). (b) Correlation between diffusion coefficient ($D$) and *RT* across all systems, derived from 54 spatially resolved diffusivity profiles and the corresponding 54 RT distributions. Higher symbol transparency indicates positions closer to the gel surface.

## 3.2 Interfacial Mechanisms of Heavy Metal Immobilization

### *3.2.1 Number density profiles of ions and water*

To further interpret the spatially resolved diffusivity ($D$) and residence time (RT) distributions of nanoconfined heavy-metal ions (**Figure 5–6**, **Figures S4–S6, and Figures S10–S12**), we analyzed the number density profiles of the confined $PbCl_2$, $BaCl_2$, and CsCl solutions across the five gel nanopores. Representative profiles for water oxygen (Ow), water hydrogen (Hw), $Ba^{2+}$, and $Cl^-$ in the $BaCl_2$ system are shown in **Figure 7**, from the pore center ($z \approx 0$ nm) towards the pore surface ($z \approx -2$ nm). The corresponding bulk solution densities are indicated by horizontal dashed lines for comparison. Overall, the number density profiles are strongly modulated near the solid–liquid interface for all gel chemistries, while converging toward bulk-like values at the pore center. The oscillatory water density profiles (Ow and Hw) in C–(N)–A–S–H (**Figure 7**a–c) and C–S–H (**Figure 7**d) reveal pronounced structural ordering near the surface, with a characteristic layer spacing of ~0.3 nm, consistent with previous studies [43, 116-118]. In contrast, water layering is less pronounced in N–A–S–H (**Figure 7**e), likely reflecting its amorphous nature and higher surface roughness.

The $Ba^{2+}$ and $Cl^{-}$ density profiles also exhibit layered structures near the surfaces of all gel pores (**Figure 7**a–e), with densities significantly higher than that in the pore center and bulk solution, confirming strong interfacial adsorption. The peak positions and intensities, however, vary with surface composition. In C–(N)–A–S–H series, increasing the Al/Si ratio from 0 (**Figure 7**a) to 0.059 (**Figure 7**b) and 0.125 (**Figure 7**c) progressively enhances the $Ba^{2+}$ density peak at the surface, reflecting greater $Ba^{2+}$ accumulation. $Ba^{2+}$ peak also shifts closer to the gel surface, with $Ba^{2+}$ preceding $Cl^{-}$ at Al/Si ratios of 0.059–0.125, which is consistent with preferential cation adsorption onto negatively charged Al-tetrahedral sites. Correspondingly, the $Na^{+}$ profiles in **Figure 7**b–c reveal substantial leaching of surface-bound $Na^{+}$ ions that originally balanced the negatively charged Al tetrahedra. This behavior suggests a competitive adsorption process in which $Ba^{2+}$ uptake at negatively charged sites is accompanied by $Na^{+}$ release, consistent with cation-exchange-like mechanisms that maintains surface charge balance.

In C–S–H (**Figure 7**d) and N–A–S–H (**Figure 7**e), $Ba^{2+}$ and $Cl^{-}$ densities are also elevated near the surface relative to the pore center and bulk solution, but remain substantially lower than in C–(N)–A–S–H with Al/Si = 0.059–0.125 (**Figure 7**b–c). A direct comparison of $Ba^{2+}$ number density and diffusivity profiles in the C–(N)–A–S–H (Al/Si = 0.125) and C–S–H pores (**Figure 7**f) help resolve the apparent discrepancy between local and pore-averaged diffusivity trends. In C–(N)–A–S–H (Al/Si = 0.125), $Ba^{2+}$ exhibits a narrower near-surface "immobile layers" (yellow-shaded regions in **Figure 7**c, corresponding to the first one or two $Ba^{2+}$ adsorption layers) where $Ba^{2+}$ diffusivity approaches zero within the simulation timescale, but mobile $Ba^{2+}$ away from surface shows higher local diffusivity than in C–S–H case. Despite the narrower near-surface immobile region, the substantially larger fraction of $Ba^{2+}$ partitioned into this region in C–(N)–A–S–H (Al/Si = 0.125) dominates the pore-averaged response, leading to lower overall diffusivity and a higher extent of immobilization than in C–S–H (**Figure 4**). These results indicate that pore-averaged diffusivity is governed by both local mobility and ion partitioning: a gel can exhibit strong overall mobility suppression not only because ions move slowly near the interface, but also because a larger fraction of ions accumulates in the immobile regions at the interface.

A notable distinctive feature of the C–S–H system is that the $Cl^{-}$ and $Ba^{2+}$ adsorption layers appear at similar position, unlike the ordering observed in Al-substituted C–(N)–A–S–H and N–A–S–H. This difference may arise from stronger Ca–Cl interactions at the C–S–H surface, together with the absence of Al-associated cation-binding sites that are present in Al-substituted C–(N)–A–S–H and N–A–S–H gels. The atomistic origin of these differences will be examined further in the next section. The corresponding density profiles for $PbCl_2$ and CsCl systems are provided in **Figures S13–15** of the **Supplementary Material**, which show broadly similar trends as $BaCl_2$. One distinct feature is observed for CsCl in C–(N)–A–S–H (Al/Si = 0), where the first $Cs^+$ density peak lies closer to the gel surface than those of $Cl^-$ and other heavy-metal ions (**Figure S15**a). This behavior is attributed to inner-sphere association of $Cs^+$ ions with bridging and pairing silicate oxygen sites, which are absent for $Pb^{2+}$ and $Ba^{2+}$ in C–(N)–A–S–H (Al/Si = 0), supported by the coordination-number analysis discussed later. The formation of inner-sphere $Cs^+$ adsorption in tobermorite nanopores is consistent with a previous MD study [38].

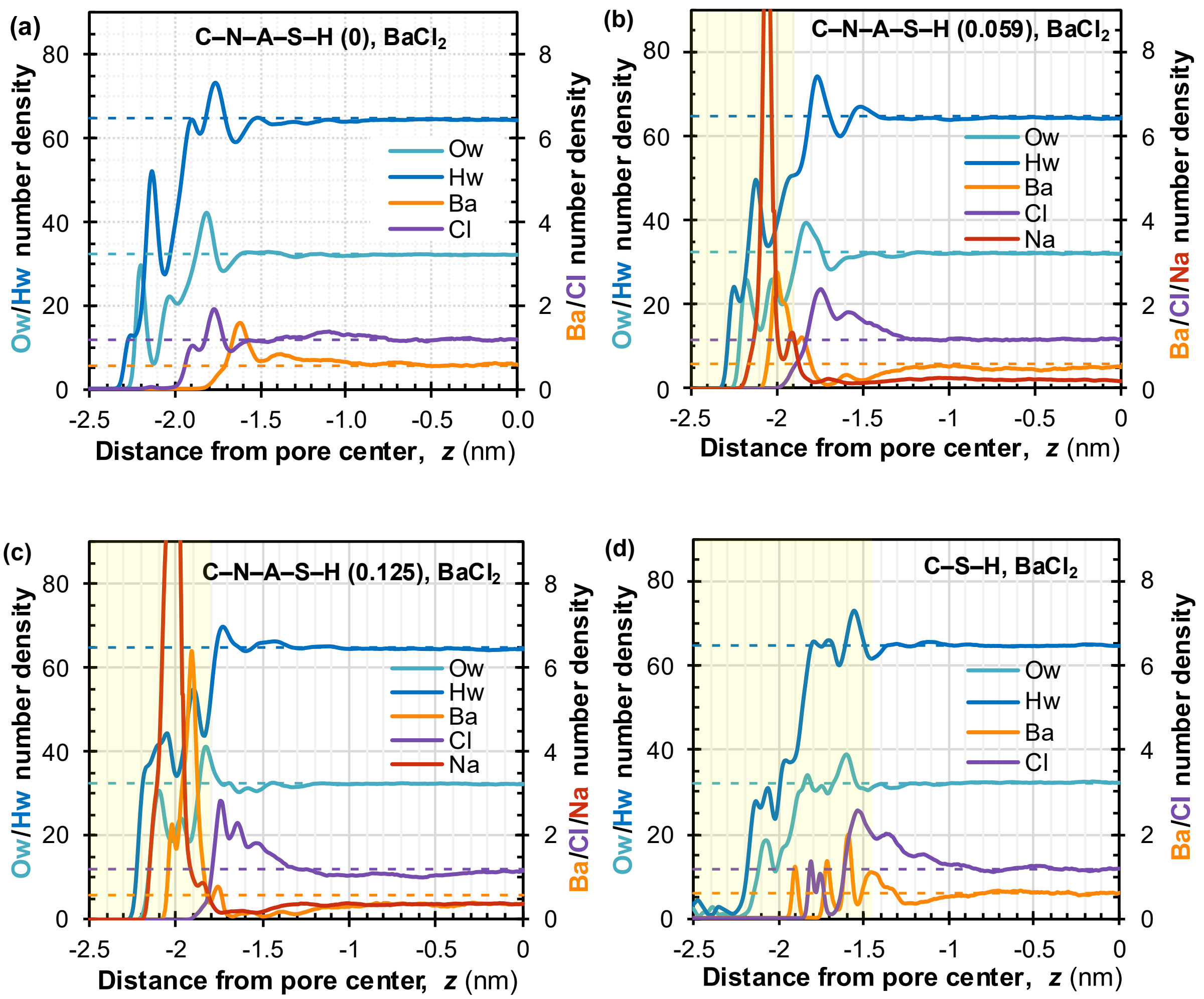

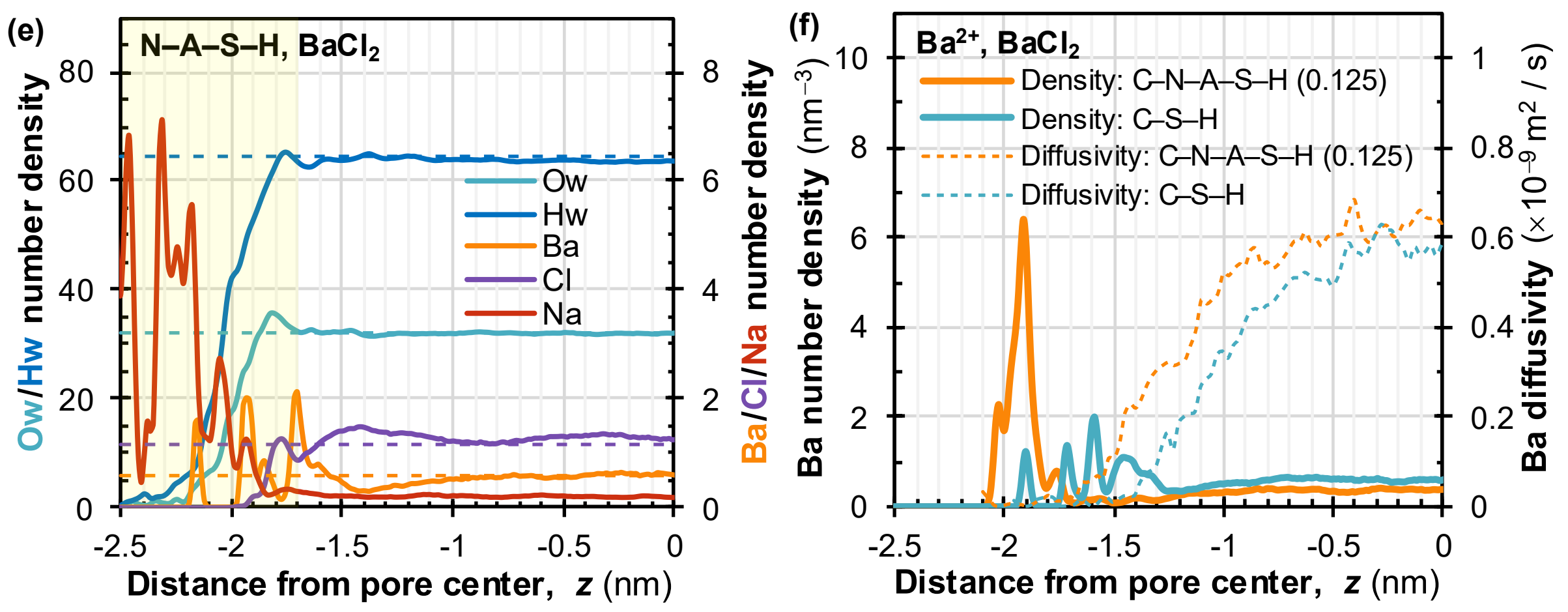

**Figure 7.** Number density profiles (in $nm^{-3}$) of ions and water molecules along the *z*-direction (perpendicular to the pore surface) for 1 M $BaCl_2$ solution inside different gel nanochannels: C–(N)–A–S–H with an Al/Si ratio of (a) 0, (b) 0.059, and (c) 0.125; (d) C–S–H; and (e) N–A–S–H. Profiles include water hydrogen (Hw), water oxygen (Ow), barium ions ($Ba^{2+}$), sodium ions ($Na^+$ if present), and chloride ions ($Cl^-$). The position $z = 0$ nm corresponds to the center of the nanochannel. Horizontal dashed lines in (a)–(e) represent the respective bulk solution densities for comparison. The yellow-shaded regions denote the immobile layer of $Ba^{2+}$, which was first identified from the near-zero-mobility regions in the *z*-directional diffusivity profiles shown in **Figure 5**c and further refined using the corresponding peak/valley positions in the one-dimensional number-density profiles, as detailed in **Section S11** of the **Supplementary Material**. (f) Comparison of $Ba^{2+}$ number density and diffusivity profiles for C–(N)–A–S–H with an Al/Si ratio of 0.125 and C–S–H, illustrating the adsorption layer near the pore surface.

For quantitative comparison, we further computed the amount and fraction of immobilized heavy-metal ions in each system by integrating the number density profiles within the immobile regions (yellow-shaded areas in **Figure 7** and **Figure S13–15**), defined as regions where diffusivity approaches zero. The corresponding *z*-ranges, number of immobilized ions, and fractions are summarized in **Table 4**. We note that this quantity should be interpreted as immobilized fraction within the MD-timescale, rather than as a direct measure of permanent immobilization under macroscopic service conditions. No immobilized $Pb^{2+}$, $Ba^{2+}$, and $Cs^+$ ions are observed in C–(N)–A–S–H (Al/Si = 0). However, as the Al/Si ratio increases, the immobilized fraction increases substantially, reaching approximately 52%, 54% and 25% for $Pb^{2+}$, $Ba^{2+}$, and $Cs^+$ ions, respectively, at Al/Si = 0.125. These values are considerably higher than those in C–S–H and N–A–S–H, demonstrating the strong influence of Al substitution on heavy-metal immobilization.

Together with the diffusivity distribution profiles in **Figure 5**a, 5c, and 5e, these results explain why the C–(N)–A–S–H (Al/Si = 0.125) gel exhibits the lowest pore-averaged diffusivity among the simulated gel systems (**Figure 4**).

For Cs ions, the immobilized fraction is consistently lower than that of $Pb^{2+}$ and $Ba^{2+}$ across most pore systems (except for C–(N)–A–S–H with Al/Si = 0, where the immobilized fraction is zero in all cases), reflecting weaker or less persistent interactions with pore surfaces. This difference is particularly evident in C–S–H (Ca/Si = 1.67), where the trapped $Cs^{+}$ fraction is much lower than in Al-bearing gel systems. This weaker $Cs^{+}$ retention in C–S–H is consistent with experimental observations showing that Cs exhibits a much higher leaching percentage (41.4%) from OPC paste (hydrated over 90 days) than divalent ions such as $Sr^{2+}$, $Co^{2+}$, and $Eu^{2+}$ (<1.5%) [14]. Although the present simulations isolate adsorption-controlled mobility in idealized nanopores, this qualitative consistency supports the relevance of gel-specific interfacial binding environments for interpreting contaminant retention trends.

**Table 4.** Summary of the *z*-position of the immobile region (yellow-shaded regions in **Figure 7**), along with the number and percentage of immobilized heavy-metal ions within these regions for each gel nanochannel. The corresponding number and percentage of $Na^{+}$ ions leached from the original pore surface is also given.

| # | Gel type | Solution type | *Z*-position of the immobile region (nm) | Number and percentage of immobilized heavy metal ions | Leached Na ion number (percentage) |
|---|---|---|---|---|---|
| 1 | C–(N)–A–S–H (Al/Si = 0) | $PbCl_2$ | – | (0.00)/47 = (0.00) % | 0 |
| 2 | C–(N)–A–S–H (Al/Si = 0.059) | $PbCl_2$ | $\|z\| \geq 1.90$ | (16.25±0.00)/47 = (34.57±0.00) % | 27.53±0.00; (57.36±0.01)% |
| 3 | C–(N)–A–S–H (Al/Si = 0.125) | $PbCl_2$ | $\|z\| \geq 1.80$ | (24.44±0.12)/47 = (52.01±0.25) % | 60.81±0.03; (63.34±0.03)% |
| 4 | C–S–H (Ca/Si = 1.67) | $PbCl_2$ | $\|z\| \geq 1.40$ | (14.14±0.07)/48 = (29.46±0.15) % | 0 |
| 5 | N–A–S–H (Al/Si = 0.33) | $PbCl_2$ | $\|z\| \geq 1.70$ | (10.80±0.05)/38 = (28.42±0.14) % | 18.27±0.03; (4.57±0.01)% |
| 6 | C–(N)–A–S–H (Al/Si = 0) | $BaCl_2$ | – | (0.00)/47 = (0.00) % | 0 |
| 7 | C–(N)–A–S–H (Al/Si = 0.059) | $BaCl_2$ | $\|z\| \geq 1.90$ | (14.06±0.03)/47 | 18.62±0.06; |
| | | | | = (29.92±0.05) % | (38.79±0.12)% |
| 8 | C–(N)–A–S–H (Al/Si = 0.125) | $BaCl_2$ | $\|z\| \geq 1.80$ | (25.17±0.04)/47 = (53.54±0.08) % | 47.41±0.57; (49.39±0.60)% |
| 9 | C–S–H (Ca/Si = 1.67) | $BaCl_2$ | $\|z\| \geq 1.45$ | (14.64±0.07)/48 = (30.50±0.15) % | 0 |
| 10 | N–A–S–H (Al/Si = 0.33) | $BaCl_2$ | $\|z\| \geq 1.70$ | (7.70±0.00)/38 = (20.27±0.01) % | 19.89±0.00; (4.97±0.00)% |
| 11 | C–(N)–A–S–H (Al/Si = 0) | CsCl | – | (0.00)/94 = (0.00) % | 0 |
| 12 | C–(N)–A–S–H (Al/Si = 0.059) | CsCl | $\|z\| \geq 1.99$ | (13.65±0.13)/94 = (14.52±0.14) % | 15.10±0.12; (31.45±0.25)% |
| 13 | C–(N)–A–S–H (Al/Si = 0.125) | CsCl | $\|z\| \geq 1.98$ | (23.89±0.04)/94 = (25.41±0.04) % | 17.68±0.02; (18.42±0.02)% |
| 14 | C–S–H (Ca/Si = 1.67) | CsCl | $\|z\| \geq 1.85$ | (1.61±0.00)/96 = (1.67±0.00) % | 0 |
| 15 | N–A–S–H (Al/Si = 0.33) | CsCl | $\|z\| \geq 1.92$ | (12.48±0.23)/76 = (16.42±0.30) % | 15.59±0.21; (3.90±0.05)% |

Using the same procedure, we also quantified the total amount of $Na^+$ released from the surface of N–A–S–H and C–(N)–A–S–H pores (**Table 4**). As shown in **Figure 8**, the total charge of adsorbed heavy-metal ions within the immobile regions correlates strongly with the total charge of leached $Na^+$, with a slope close to unity for linear regression. This near-charge-balanced relationship indicates that heavy-metal adsorption in these Al-bearing gels is closely coupled with displacement of charge-compensating $Na^+$ from negatively charged tetrahedral Al sites. Within the present non-reactive MD framework, this behavior is consistent with an ion-exchange-like interfacial adsorption mechanism and identifies $Na^+$ leaching as a quantitative signature of adsorption-driven charge compensation in these systems. The specific surface environments responsible for this behavior are further analyzed in **Section 3.2.3**.

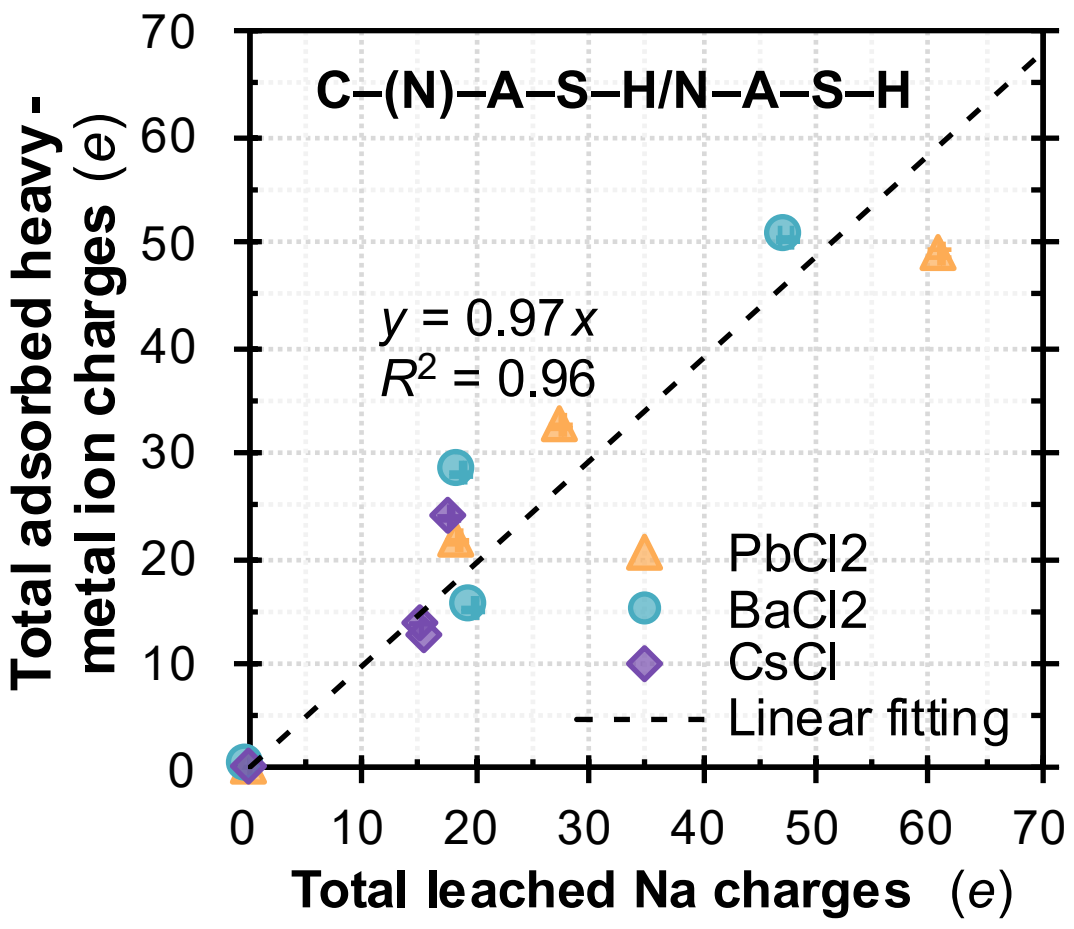

**Figure 8.** Correlation between the total charge of adsorbed heavy-metal ions within the immobile regions (yellow-shaded regions in **Figure 7**) and the total charge of the leached $Na^+$ ions from the C–(N)–A–S–H and N–A–S–H gel surfaces. Error bars represent standard deviations from three independent simulations.

### *3.2.2 Two-dimensional absorption patterns on nanopore surfaces*

To link the observed mobility suppression to specific interfacial environments, we mapped the two-dimensional adsorption density of $Ba^{2+}$, $Cl^-$, and $Na^+$ ions on the different pore surfaces by integrating over the immobile regions (yellow-shaded zones in **Figure 7**). The resulting 2D density maps (**Figure 9**a–c) reveal non-uniform spatial distributions of adsorbed species and distinct patterns across gel chemistries. For C–(N)–A–S–H with Al/Si = 0, adsorption of $Ba^{2+}$ (**Figure 9**a) and $Cl^-$ (**Figure 9**b) is relatively diffuse, with only weak alignment to surface silicate chain trenches, reflecting limited affinity for specific surface sites. In contrast, increasing the Al/Si ratio leads to pronounced site-specific adsorption: $Ba^{2+}$ ions and $Na^+$ ions preferentially accumulate near Al/Si substitution sites (**Figure 9**a and **9**c). This behavior is in line with previous MD studies of montmorillonite, where divalent $Zn^{2+}$/$Cd^{2+}$/$Pb^{2+}$ ions preferentially adsorbed at tetrahedral Al substitution sites [76]. These results indicate that immobilization is governed by localized ion–surface interactions, with specific atomic-scale binding environments examined further by radial distribution function (RDF) analysis in **Section 3.2.3**. Notably, $Ba^{2+}$ adsorption is spatially heterogeneous even among Al/Si sites: certain regions exhibit substantially higher $Ba^{2+}$ densities, likely facilitated via localized $Ba^{2+}$/$Na^+$ ion exchange, as indicated by the corresponding reduction in $Na^+$ densities at those positions (**Figure 9**c). In contrast, the spatial distributions of $Cl^-$ ions are more diffuse and largely follows the arrangement of preabsorbed $Ba^{2+}$ and $Na^+$ ions.

For C–S–H and N–A–S–H, localized regions of elevated $Ba^{2+}$ and $Cl^-$ density are also observed, but without clear ordering or well-defined adsorption motifs, consistent with their more structurally disordered surfaces. The specific binding environments underlying these adsorption hotspots will be analyzed through partial RDF analysis in **Section 3.2.3**. The corresponding adsorption density maps for the $PbCl_2$ and CsCl systems are given in **Figure S16–18** in the **Supplementary Material**, and show broadly similar features to those observed for $BaCl_2$.

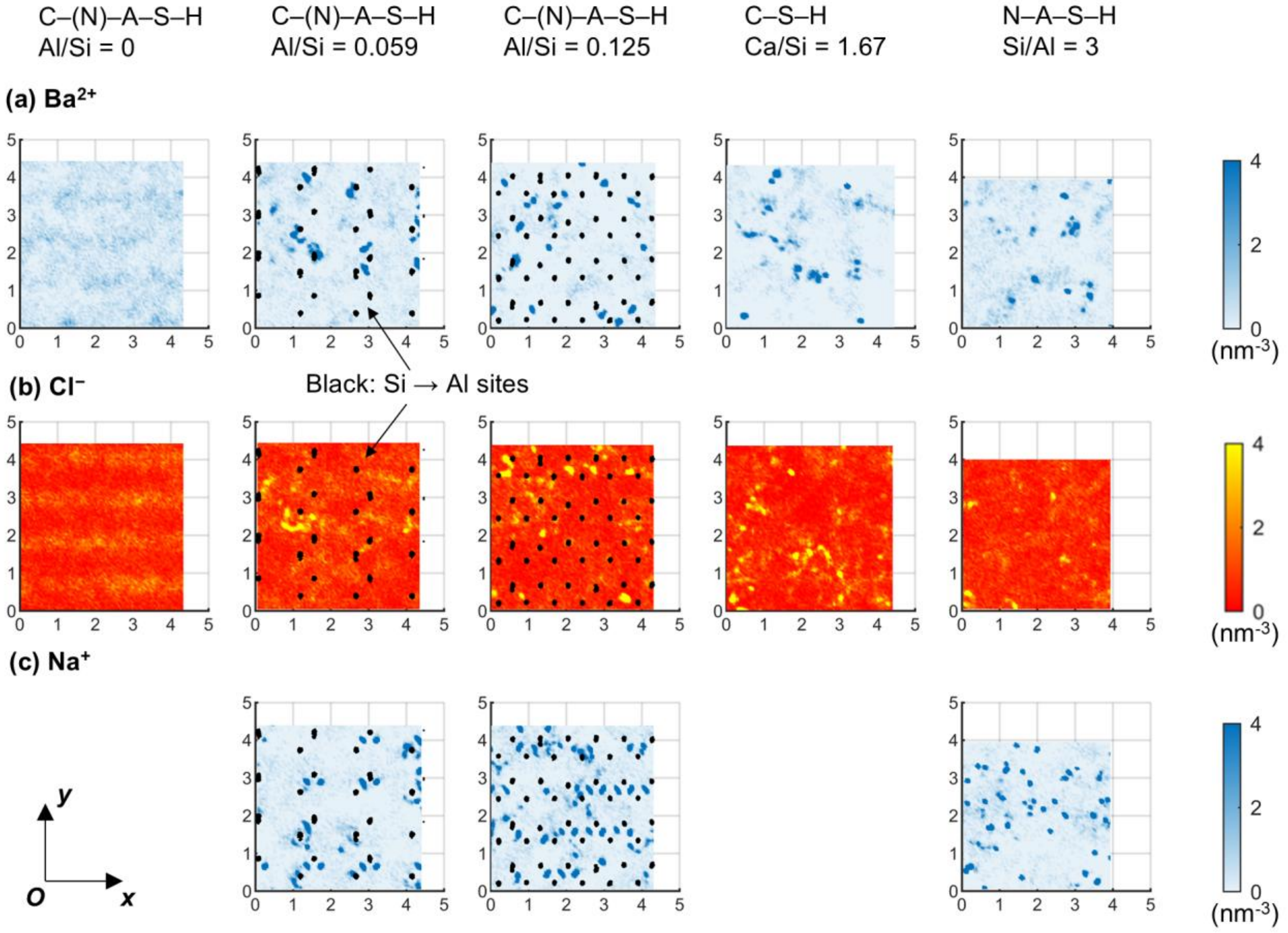

**Figure 9.** Two-dimensional adsorption density maps of (a) $Ba^{2+}$, (b) $Cl^-$, and (c) $Na^+$ ions (present in some of the original gels to charge-balance alumina tetrahedra) on various gel nanopore surfaces for the nanoconfined $BaCl_2$ solutions. The maps were obtained by temporal and spatial averaging of ion coordinates within the surface-immobile layers (yellow-shaded

regions in **Figure 7**). $Na^+$ ions originate from the original gel structure and act as charge-balancing species for Al tetrahedra.

### *3.2.3 Atomistic ion-surface interaction from radial distribution functions (RDFs) analysis*

To identify the local interfacial environments responsible for the adsorption and immobilization behaviors described above, we conducted a detailed local structural analysis of both bulk and nanoconfined solutions across different gel nanopores. Partial radial distribution functions (RDFs) were computed for relevant atom-atom pairs to probe preferred binding environments at the solid–liquid interface. **Figure 10** presents representative partial RDFs for selected Ba–X and Cl–X pairs in the $BaCl_2$ system, while the complete sets of partial RDFs are provided in **Figure S20** of the **Supplementary Material**. To focus on adsorption-relevant interactions, i.e., those that are energetically favorable and structurally significant, only ion-surface atom pairs exhibiting attractive interactions and pronounced first-neighbor RDF peaks are highlighted here. The positions of the first RDF maxima, corresponding to the most probable nearest-neighbor distances, $r_{0,\mathrm{Ba-j}}$, for Ba–j atom pairs, are summarized in **Table 5**. Based on these interatomic distances, the corresponding Ba–j pair interaction energies ($E_{\mathrm{Ba-j}}$) were computed following our recent study [43], with calculation details provided in **Section S5** of the **Supplementary Material**; the results are summarized in **Table 6**. Coordination numbers ($\mathrm{CN}_{\mathrm{Ba-j}}$) were obtained by integrating the partial Ba–j RDFs up to their first minima and are summarized in **Table 7**. Detailed procedures for the calculation of partial RDFs and CNs are described in Ref. [66].

For C–(N)–A–S–H with Al/Si = 0 (**Figure 10**a and **Figure S20**a), no distinct RDF peaks are observed between $Ba^{2+}$ and surface atoms, indicating negligible direct ion-surface binding. Instead, pronounced Ba–Ow (water oxygen) in **Figure S20**a and Ba–Cl peaks in **Figure 10**a reflect the formation of a well-defined hydration shell around $Ba^{2+}$ and contact ion pairing with $Cl^-$, respectively, consistent with bulk solution behavior (**Figure S20**k). The presence of a clear Cl–Hh peak (**Figure 10**b) indicates that $Cl^-$ ions interact with surface hydroxyl hydrogens (Hh), explaining why $Cl^-$ density peak occurs closer to the surface than $Ba^{2+}$ in **Figure 7**a. By contrast, introducing Al/Si substitutions into C–(N)–A–S–H (**Figure 10**c and e) generates distinct RDF peaks between $Ba^{2+}$ and surface oxygens associated with the substitution sites, revealing strong site-specific binding. Among these interactions, Ba–Ohs (surface hydroxyl oxygens connected to tetrahedral Al) plays a dominant role. Although the Ba–Obts (bridging oxygens between Si and Al) interactions exhibit higher individual bond strengths than Ba–Ohs (**Table 6**), their coordination numbers are substantially lower (**Table 7**), leading to a reduced contribution to the overall $Ba^{2+}$ binding compared to Ba–Ohs (as quantified in **Table 8**). Together, these results indicate that $Ba^{2+}$ immobilization in Al-containing C–(N)–A–S–H is dominated by binding with Ohs sites, with secondary contributions from Obts sites. For $Cl^{-}$ ions (**Figure 10**d and f), adsorption occurs primarily through coordination with surface hydroxyl hydrogen atoms (Cl–Hh, with a bond energy of $-4.36 \times 10^{-19}$J) and electrostatic association with pre-adsorbed $Ba^{2+}$ and $Na^{+}$. These interactions are comparatively weak, consistent with the higher-than-bulk $Cl^{-}/Ba^{2+}$ diffusivity ratios near the pore surface (**Figure 5**d), which confirms that $Ba^{2+}$ mobility is more strongly suppressed than that of $Cl^{-}$.

For C–S–H, the partial RDFs of Ba–X pairs (**Figure 10**g and **Figure S20g**) show that $Ba^{2+}$ ions are immobilized primarily via coordination with surface hydroxyl oxygens (Oh). Notably, this Ba–Oh interaction exhibit the shortest bond distance (~2.62 Å, **Table 5**), and the highest individual bond energy among all observed Ba–O interactions ($-24.44 \times 10^{-19}$J, **Table 6**). Yet, the relatively low Ba–Oh coordination number (~0.34, **Table 7**) limits its overall contribution to $Ba^{2+}$ retention compared with Al-containing C–(N)–A–S–H systems (**Table 8**). In addition, $Ba^{2+}$ retention in C–S–H includes a secondary contribution from Ca–Ow–Ba linkages, where surface-associated $Ca^{2+}$ remains strongly coordinated with water oxygen and indirectly stabilizes $Ba^{2+}$ near the interface. The Cl–X RDFs for C–S–H (**Figure 10**h) indicate that $Cl^{-}$ adsorption is mainly associated with surface Ca species (Cl–Ca) and electrostatic interactions with pre-adsorbed $Ba^{2+}$ (Cl–Ba).

In N–A–S–H (**Figure 10**i), $Ba^{2+}$ interactions are primarily associated with bridging oxygens linked to Al tetrahedra (Obts), surface hydroxyl oxygens bonded to Si (Oh), and surface hydroxyl oxygens connected to tetrahedral Al (Ohs) sites. Although Ba–Ohs interactions are individually stronger than Ba–Oh interactions (**Table 6**), their low coordination (**Table 7**) limits their overall contribution relative to other surface sites. This behavior contrasts with Al-containing C–(N)–A–S–H, where Ba–Ohs interactions dominate the total binding strength with negligible contribution from Ba–Oh (**Table 8**). For $Cl^{-}$ ions (**Figure 10**j), immobilization arises primarily from interactions with surface hydroxyl hydrogen atoms (Hh), together with electrostatic association with pre-adsorbed $Ba^{2+}$ and $Na^{+}$ ions. The cation-mediated contribution to $Cl^{-}$ retention is consistent with a previous MD study [31], which suggested that $Cl^{-}$ exhibits weak intrinsic adsorption on N–A–S–H surfaces and is retained primarily through interfacial cation–anion pairing.

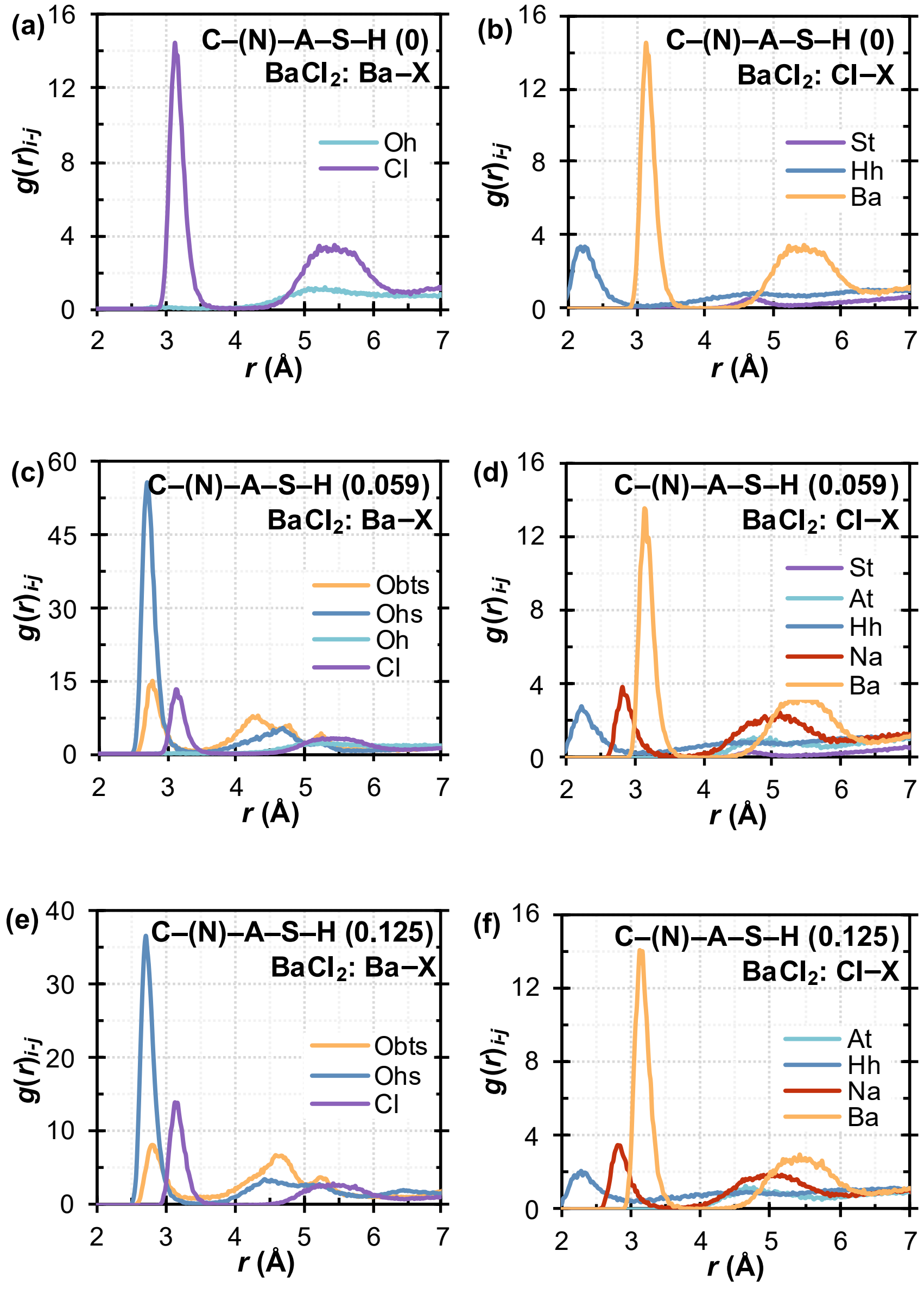

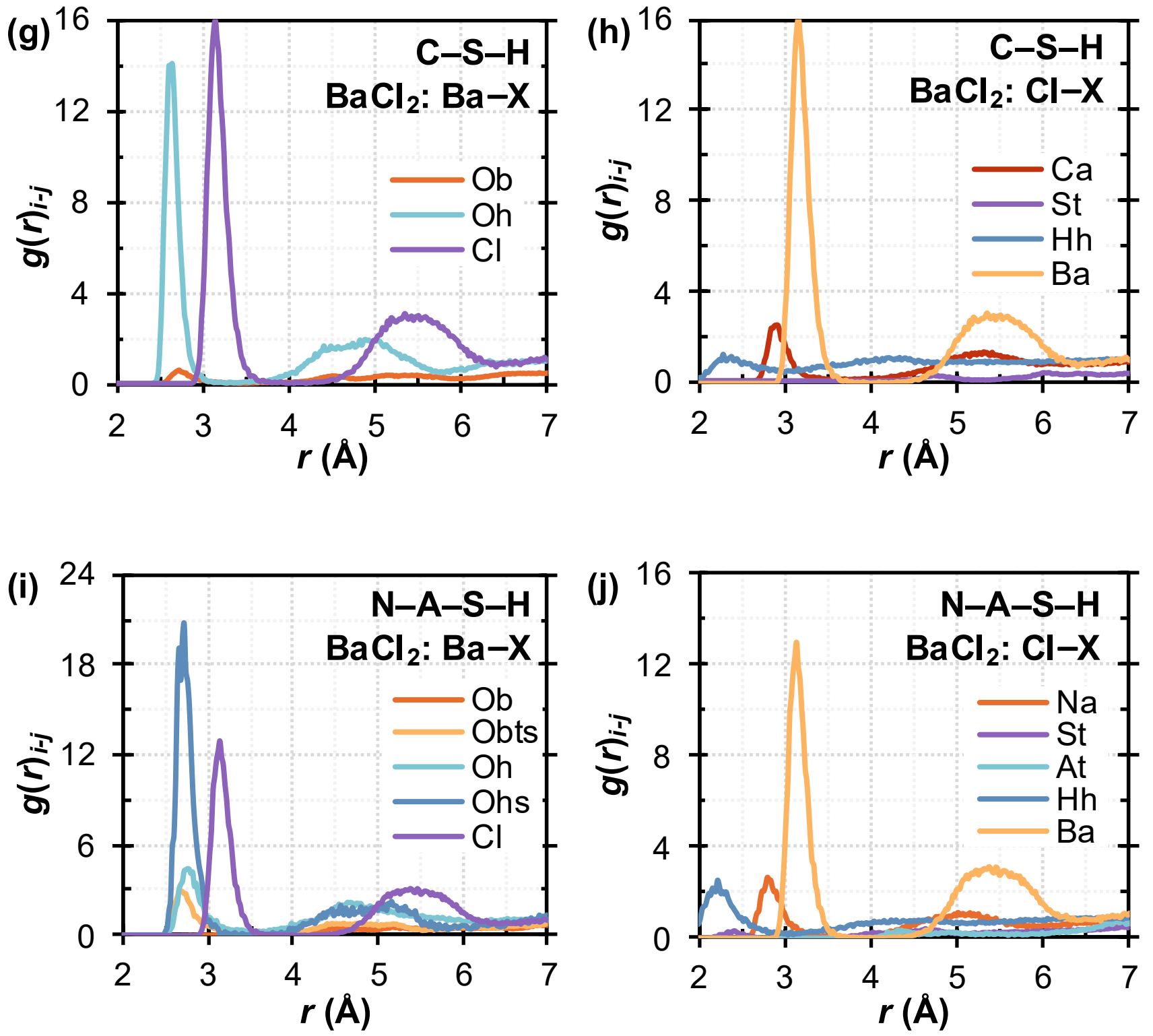

**Figure 10.** (a, c, e, g, i) Partial radial distribution functions (RDFs) of Ba–X pairs in nanoconfined $BaCl_2$ solution within (a) C–(N)–A–S–H gel with a Al/Si ratio of 0, (c) 0.059, (e) 0.125, (g) C–S–H gel, and (i) N–A–S–H gel, showing nearest-neighbor interactions between $Ba^{2+}$ and surface atoms or $Cl^-$ ions. (b, d, f, h, j) Corresponding partial RDFs for Cl–X pairs. Atom labels: St = tetrahedral Si atoms; At = tetrahedral Al atoms; Ca = interlayer $Ca^{2+}$ ions; Ow = oxygen atoms in water molecules; Hw = hydrogen atoms in water molecules; Ob = bridging oxygen atoms connected to tetrahedral Si; Obts = bridging oxygen atoms connecting tetrahedral Si and Al; Oh = oxygen atoms in surface hydroxyl groups; Ohs = hydroxyl oxygen atoms bonded to Al tetrahedral sites; Hh = hydrogen atoms in surface hydroxyl groups.

**Table 5.** Summary of the nearest interatomic distances, $\boldsymbol{r}_{0,\mathrm{Ba-j}}$, for Ba–j atom pairs, in the first coordination shell of $Ba^{2+}$ ions for bulk and nanoconfined $BaCl_2$ solutions within various gel nanopores. These distances were derived from the first-peak positions of the partial radial distribution functions (RDF) shown in **Figure 10** and **Figure S20**. Atom labels: Ob = bridging oxygen atoms connected to tetrahedral Si atoms; Obts = bridging oxygen atoms connecting tetrahedral Si and Al; Oh = oxygen atoms in surface hydroxyl groups; Ohs = hydroxyl oxygen atoms bonded to Al tetrahedral sites. The corresponding results for all other Ba–X pairs, together with those for bulk and nanoconfined $PbCl_2$ and CsCl solutions, are presented in **Section S14** of the **Supplementary Material**. The standard deviations reported in the table were calculated based on three independent simulations.

| Gel type | Interatomic distance: $r_{0,\mathrm{Ba-j}}$ (Å) | | | |
|---|---|---|---|---|
| | **Obts** | **Ob** | **Oh** | **Ohs** |
| Bulk | − | − | − | − |
| C–(N)–A–S–H (Al/Si = 0) | − | − | − | − |
| C–(N)–A–S–H (Al/Si = 0.059) | 2.78±0.01 | 2.83±0.02 | − | 2.72±0.01 |
| C–(N)–A–S–H (Al/Si = 0.125) | 2.80±0.01 | 2.78±0.02 | − | 2.71±0.00 |
| C–S–H (Ca/Si = 1.67) | − | 2.72±0.01 | 2.62±0.02 | − |
| N–A–S–H (Al/Si = 0.33) | 2.68±0.01 | (2.77±0.04) | 2.78±0.01 | 2.70±0.02 |

**Table 6.** Summary of nearest-neighbor interatomic interacting strength (bond energies) for Ba–j pairs ($E_{\mathrm{Ba-j}}$) in the first coordination shell of $Ba^{2+}$ ions for bulk and nanoconfined $BaCl_2$ solutions within various gel nanopores. Bond energies were calculated from the total potential energy function using the corresponding Ba–j bond lengths listed in **Table 5**, following the method described in **Section S5** of the **Supplementary Material**. More negative bond energy values indicate stronger attractive interactions. The complete results for all other Ba–X pairs, together with those for bulk and nanoconfined $PbCl_2$ and CsCl solutions, are presented in **Section S14** of the **Supplementary Material**. The standard deviations reported in the table were calculated based on three independent simulations.

| Gel type | Bond energy: $E_{\mathrm{Ba-j}}$ ($10^{-19}$ J) | | | |
|---|---|---|---|---|
| | **Obts** | **Ob** | **Oh** | **Ohs** |
| Bulk | − | − | − | − |
| C–(N)–A–S–H (Al/Si = 0) | − | − | − | − |
| C–(N)–A–S–H (Al/Si = 0.059) | −19.15±0.05 | −16.91±0.08 | − | −17.98±0.05 |
| C–(N)–A–S–H (Al/Si = 0.125) | −19.00±0.05 | −17.17±0.12 | − | −18.01±0.00 |
| C–S–H (Ca/Si = 1.67) | − | −19.36±0.05 | −24.44±0.12 | − |
| N–A–S–H (Al/Si = 0.33) | −19.65±0.05 | (−17.20±0.21) | −15.48±0.04 | −18.08±0.12 |

**Table 7.** Summary of the coordination numbers ($\mathrm{CN_{Ba-j}}$) for Ba–j pairs in the first coordination shell of $Ba^{2+}$ ions for bulk and nanoconfined $BaCl_2$ solutions within various gel nanopores. Coordination numbers were obtained by integrating the corresponding partial radial distribution functions (RDFs) shown in **Figure 10** and **Figure S20** up to the cutoff radius ($r_c$) listed in the table. The complete results for all other Ba–X pairs, together with those for bulk and nanoconfined $PbCl_2$ and CsCl solutions, are presented in **Section S14** in the **Supplementary Material**. The standard deviations reported in the table were calculated based on three independent simulations.

| Gel type | Coordination number $\mathrm{CN_{Ba-j}}$ at $r_c$ = 3.19Å | | | |
|---|---|---|---|---|
| | **Obts** | **Ob** | **Oh** | **Ohs** |
| Bulk | − | − | − | − |
| C–(N)–A–S–H (Al/Si = 0) | − | − | − | − |
| C–(N)–A–S–H (Al/Si = 0.059) | 0.19±0.01 | 0.01±0.01 | − | 0.66±0.01 |
| C–(N)–A–S–H (Al/Si = 0.125) | 0.29±0.01 | 0.07±0.01 | − | 1.06±0.04 |
| C–S–H (Ca/Si = 1.67) | – | 0.09±0.00 | 0.34±0.01 | – |
| N–A–S–H (Al/Si = 0.33) | 0.25±0.02 | 0.01±0.00 | 0.22±0.05 | 0.09±0.01 |

**Table 8.** Total binding strength (TBS) and individual Ba–j contribution in the first coordination shell of $Ba^{2+}$ ions for bulk and nanoconfined $BaCl_2$ solutions within various gel nanopores. Each TBS component ($-\mathrm{CN}_{\mathrm{Ba-j}}E_{\mathrm{Ba-j}}$) was obtained by multiplying the Ba–j pair interaction energy (**Table 6**) by the corresponding coordination number (**Table 7**). Since attractive Ba–j interactions have negative energies, the leading negative sign converts them into positive binding-strength magnitudes for easier comparison. The total TBS values ($-\sum(\mathrm{CN}_{\mathrm{Ba-j}}E_{\mathrm{Ba-j}})$, where $j$ = Obts, Ob, Oh, Ohs) represent the collective contribution of all Ba–surface interactions. The corresponding results for all other Ba–X pairs, together with those for bulk and nanoconfined $PbCl_2$ and CsCl solutions, are presented in **Section S14** in the **Supplementary Material**. The standard deviations reported in the table were calculated based on three independent simulations.

| **Gel type** | **Individual components of TBS**: $-\mathrm{CN}_{\mathrm{Ba-j}}E_{\mathrm{Ba-j}}$ ($10^{-19}$ J) | | | | **Total binding strength** (TBS): $-\sum(\mathrm{CN}_{\mathrm{Ba-j}}E_{\mathrm{Ba-j}})$ | **Diffusion coefficient**: $D$ ($10^{-9}\mathrm{m}^2/\mathrm{s}$) | **Immobilization extent (%)** |
|---|---|---|---|---|---|---|---|
| | **Obts** | **Ob** | **Oh** | **Ohs** | | | |
| Bulk | 0.00 ±0.00 | 0.00 ±0.00 | 0.00 ±0.00 | 0.00 ±0.00 | 0.00±0.00 | 0.56±0.01 | 0.00±0.00 |
| C–(N)–A–S–H (Al/Si = 0) | 0.00 ±0.00 | 0.00 ±0.00 | 0.00 ±0.00 | 0.00 ±0.00 | 0.00±0.00 | 0.44±0.00 | 21.52±0.75 |
| C–(N)–A–S–H (Al/Si = 0.059) | 3.65 ±0.19 | 0.13 ±0.14 | 0.00 ±0.00 | 11.82 ±0.11 | 15.60±0.34 | 0.28±0.02 | 49.92±4.36 |
| C–(N)–A–S–H (Al/Si = 0.125) | 5.52 ±0.18 | 1.13 ±0.13 | 0.00 ±0.00 | 19.12 ±0.74 | 25.77±0.81 | 0.19±0.01 | 67.01±1.51 |
| C–S–H (Ca/Si = 1.67) | 0.00 ±0.00 | 1.74 ±0.02 | 8.37 ±0.32 | 0.00 ±0.00 | 14.21±0.13* | 0.24±0.01 | 56.82±2.31 |
| N–A–S–H (Al/Si = 0.33) | 4.84 ±0.39 | 0.10 ±0.05 | 3.40 ±0.78 | 1.57 ±0.16 | 9.91±1.31 | 0.31±0.01 | 44.05±1.31 |

*The secondary Ca–Ow–Ba interaction for the C–S–H gel was also included in the calculation of the TBS due to its relatively high strength. The calculated amount of such interactions per $Ba^{2+}$ ion within the gel system is provided in **Section S14** of the **Supplementary Material**. According to our previous study [43] and the present results, the Ca–Ow bond ($-15.14 \times 10^{-19}$J) is stronger than other interactions, such as Ba–Ow ($-13.37 \times 10^{-19}$J) and Na–Ow ($-7.84 \times 10^{-19}$J). In addition, interlayer $Ca^{2+}$ ions remain strongly bound to the gel surface throughout the simulation process. Therefore, the secondary immobilization pathway of $Ba^{2+}$ ions via the Ba–Ow–Ca linkage should be taken into consideration for the C–S–H gel, whereas other gels do not exhibit comparably strong secondary interactions.

The complete RDFs and coordination-number results for $PbCl_2$, $BaCl_2$, and CsCl systems are provided in **Figures S19–21** and **Tables S7–S21** of the **Supplementary Material.** The results for $PbCl_2$ systems are broadly consistent with those observed for the $BaCl_2$, indicating that these two divalent cations interact with similar surface oxygen environments across the gel systems. To quantify this similarity, **Figure 11**a compares the coordination numbers of $Pb^{2+}$ and $Ba^{2+}$ with the corresponding surface oxygen species across all gel chemistries. The strong linear correlation, with $R^2 \approx 0.97$, indicates that $Pb^{2+}$ and $Ba^{2+}$ exhibit similar coordination-site preferences, despite differences in ionic radius and hydration structure. This similarity suggests that their immobilization mechanisms are governed by the same dominant interfacial oxygen environments, particularly Al-associated Ohs and Obts sites in Al-containing C–(N)–A–S–H, Oh sites in C–S–H, and distributed Obts/Oh/Ohs sites in N–A–S–H.

In contrast, the comparison between $Cs^+$ and $Ba^{2+}$ coordination numbers (**Figure 11**b) shows a weaker overall correlation ($R^2 \approx 0.59$), with $Cs^+$ not following the same coordination-site hierarchy as the divalent cations across all surface oxygen environments. The lower correlation arises mainly from differences in coordination with Ob and Oh sites. However, this difference is surface-dependent rather than uniformly enhanced for $Cs^+$. In the Al-free tobermorite-derived pore (C–(N)–A–S–H (Al/Si=0) and N–A–S–H, $Cs^+$ shows appreciable coordination with silicate-related Ob and/or Oh sites, consistent with its weaker hydration and greater tendency to form inner-sphere-like associations with silicate oxygen environments. In C–S–H, by contrast, Cs–Ob and Cs–Oh coordination are both low, consistent with the small $Cs^+$ fraction retained in the near-surface immobile region (**Table 4**). For Al-associated Ohs and Obts sites, the $Cs^+$–$Ba^{2+}$ correlation is stronger (**Figure 11**b), suggesting that these negatively charged Al-related environments act as important binding sites for both monovalent and divalent cations. Taken together, the coordination-number comparisons show that $Pb^{2+}$ and $Ba^{2+}$ share broadly similar immobilization mechanisms across the simulated gel pores, whereas $Cs^+$ exhibits more gel-dependent coordination behavior. Thus, heavy-metal retention depends on both gel surface chemistry and ion-specific hydration and coordination characteristics.

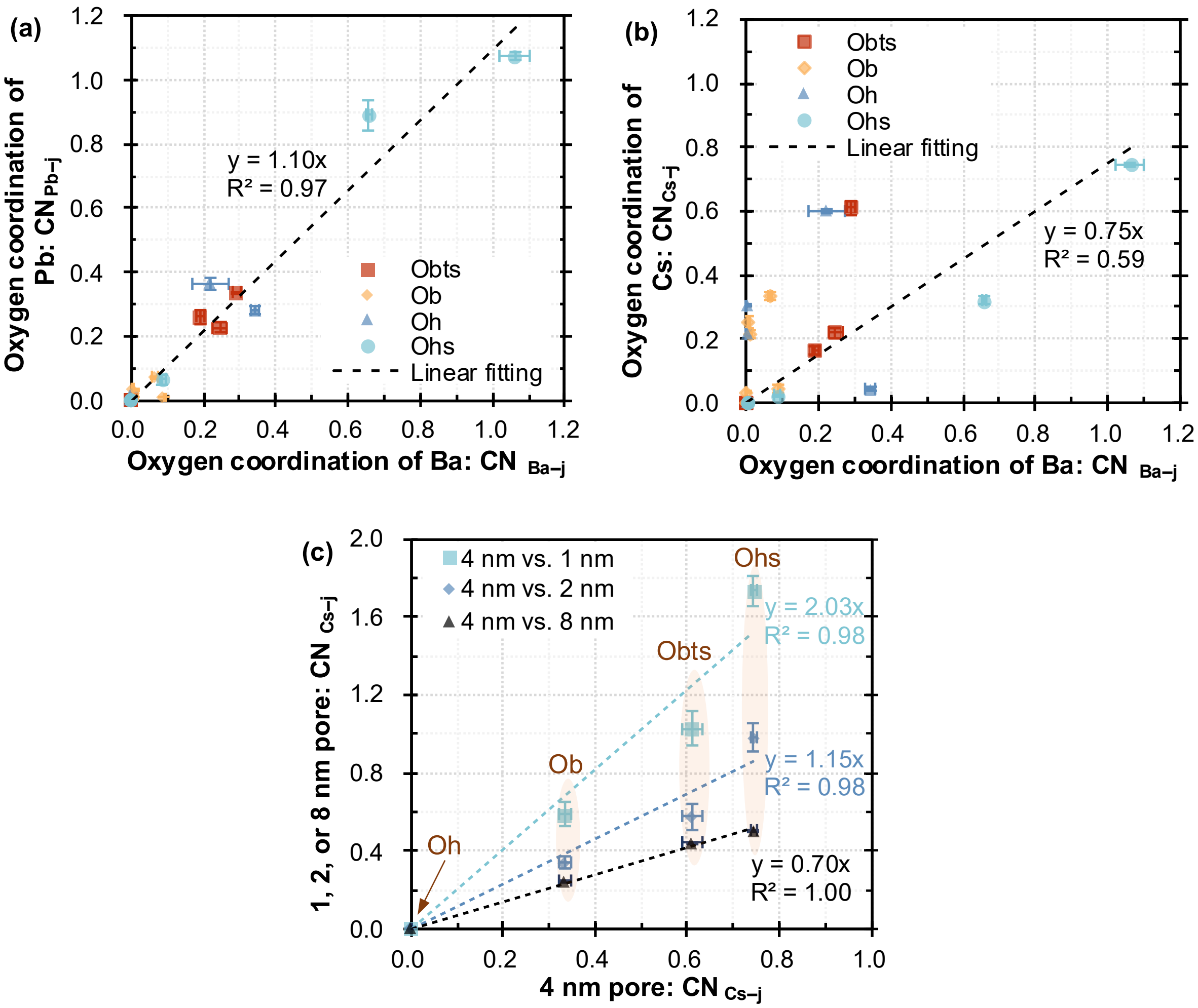

**Figure 11.** Correlations of individual coordination-number (CN) components for ion-surface interactions: (a) $Ba^{2+}$ versus $Pb^{2+}$, (b) $Ba^{2+}$ versus $Cs^{+}$, and (c) $Cs^{+}$ in C–(N)–A–S–H (Al/Si = 0.125) with varying pore sizes (4 nm versus 1, 2 and 8 nm). In panels (a) and (b), each data point represents the same surface-oxygen CN component evaluated for two different cation species within the same gel pore. In panel (c), each data point represents the same CN component evaluated at two different pore sizes. Error bars indicate standard deviations from three independent simulations.

To assess whether the local immobilization mechanism depends on pore size, additional simulations were performed for a representative C–(N)–A–S–H system (Al/Si = 0.125) and $Cs^{+}$ over a pore width range of ~1–8 nm (**Section S15**), where the corresponding partial RDF and coordination analyses show that the dominant Cs–surface oxygen coordination environments remain consistent across the pore sizes examined. As shown in **Figure 11**c, Cs–O coordination numbers obtained for the 1, 2, and 8 nm pores correlate strongly with those from the 4 nm reference pore, with $R^2$ values of 0.97–1.00. This suggests that the relative hierarchy of $Cs^+$ binding sites is largely preserved as pore width changes. Although decreasing pore size increases the relative contribution of interfacial regions and therefore enhances the overall extent of immobilization (**Figures S22–S24**), it does not substantially alter the local surface coordination environments responsible for $Cs^+$ retention. These results indicate that, for the representative system examined here, pore size primarily modulates the proportion of ions affected by interfacial confinement, while the underlying adsorption-controlled immobilization mechanism remains governed by the same local ion–surface interactions.

### *3.2.4 Binding strength descriptor and correlation with diffusivity*

The RDF-derived structural attributes (bond distances, coordination numbers, and interaction energies) provide a quantitative description of specific ion-surface interactions. Building on these, inspired by our recent work on structural descriptor development for amorphous systems [43, 66, 67, 119], we calculated a total binding strength (TBS) descriptor to quantify the collective effect of these interactions. The $TBS_Y$ of each heavy-metal ion (Y = $Pb^{2+}$, $Ba^{2+}$ and $Cs^+$) with surface atoms (j) exhibiting attractive interactions is defined as:

$$\mathrm{TBS_Y} = -\sum(\mathrm{CN_{Y-j}}E_{\mathrm{Y-j}}), \tag{6}$$

where $\mathrm{CN}_{\mathrm{Y}-j}$ is the coordination number between cation Y and surface atom type $j$, and $E_{\mathrm{Y}-j}$ is the corresponding pair interaction energy. Because attractive pair interaction energies are negative in the ClayFF potential, TBS is defined here as the positive magnitude of the summed attractive ion–surface interaction contributions. Thus, larger TBS values indicate stronger effective interfacial binding. TBS should not be interpreted as an absolute binding free energy; rather, it is a force-field-consistent descriptor that combines the abundance and strength of attractive ion–surface associations within the present simulation framework.

The TBS values along with individual TBS contributions for $Ba^{2+}$ are summarized in **Table 8**, with corresponding results for $Pb^{2+}$ and $Cs^{+}$ given in **Tables S7–S21** of the **Supplementary Material**. The $Ba^{2+}$ results (**Table 8**) show that TBS varies substantially with gel surface chemistry. Al-containing C–(N)–A–S–H systems exhibit larger TBS values than C–S–H and N–A–S–H, indicating stronger overall interfacial binding. Within the C–(N)–A–S–H series, TBS increases substantially with Al/Si ratio, showing that tetrahedral Al substitution enhances surface binding. For $Ba^{2+}$ in Al-substituted C–(N)–A–S–H, TBS is dominated by contributions from Ba–Ohs interactions (~70%) with secondary contributions from Ba–Obts interactions (~20%). In contrast, TBS in C–S–H arises mainly from by Ba–Oh interactions (~60%), with a secondary contribution (~30%) from Ca–Ow–Ba interactions arising from strong Ca–Ow binding that facilitate indirect $Ba^{2+}$ immobilization. In N–A–S–H, the binding environment is more distributed, with notable contributions from Ba–Obts, Ba–Oh, and Ohs interactions.

To relate surface binding to transport behavior, we examined the relationship between TBS and pore-averaged mobility suppression (**Figure 12**). The extent of immobilization, $S$, defined from the reduction in diffusivity due to nanoconfinement relative to corresponding bulk solution diffusivity, is plotted against the relative total binding strength, $rTBS_Y$, defined as:

$$\mathrm{rTBS_Y} = \mathrm{TBS_Y} / |\mathrm{CN_{Y-Ow}} E_{\mathrm{Y-OW}} + \mathrm{CN_{Y-Cl}} E_{\mathrm{Y-Cl}}|, \tag{7}$$

where the denominator represents the combined first-shell ion–water and ion–chloride interaction descriptor of the heavy-metal ion (Y = $Pb^{2+}$, $Ba^{2+}$ and $Cs^{+}$) in the corresponding bulk solution and serves as an ion-specific reference interaction scale. This normalization accounts for differences in hydration and ion pairing among the studied ions, which would otherwise obscure direct comparison of their absolute surface-binding strengths. Consequently, rTBS enables comparison of the relative surface-binding effects across different ions and facilitates the correlation between binding strength and ion diffusivity.

As shown in **Figure 12**a, a positive correlation is observed between $rTBS_Y$ and the extent of immobilization across the ion–gel systems studied here, indicating that stronger relative ion-surface binding corresponds to greater suppression of pore-averaged ion mobility. Because individual adsorbed ions can coordinate simultaneously with multiple surface oxygen species, a simple summation of all pairwise interaction contributions may overestimate the effective constraint on ion mobility. Hence, we further refined the analysis by correcting for multiple ion–surface atom interactions associated with the same adsorbed ion (see **Table S11, S16, S21** of **Supplementary Material** for details). The corrected data, presented in **Figure 12**b, show a substantially improved linear correlation ($R^2$ = 0.90 compared with 0.76 before correction). This improvement indicates that the refined TBS descriptor better captures the effective constraint imposed by interfacial interactions. Notably, the pore-size data for $Cs^+$ ions confined in C–(N)–A–S–H (Al/Si = 0.125) also generally follow the same trends in **Figure 12**, suggesting that the descriptor rationalizes both local binding effects and the increased contribution of interfacial regions under stronger confinement.

Overall, the TBS/rTBS analysis provides a compact, descriptor-based link between atomistic binding environments and pore-averaged mobility suppression in cementitious gel nanopores. The observed trends across ion–gel systems studied here indicate that, within the present non-reactive MD framework, the extent of heavy-metal ion immobilization within cementitious gel pores is primary governed by interfacial ion–surface interactions, which can be systematically captured using the TBS framework. However, because the absolute TBS/rTBS values depend on the underlying force field, charge assignment and interaction definitions, these descriptors should be viewed as a mechanistic tool for rationalizing relative mobility trends, rather than as a directly transferable predictor of macroscopic leaching behavior.

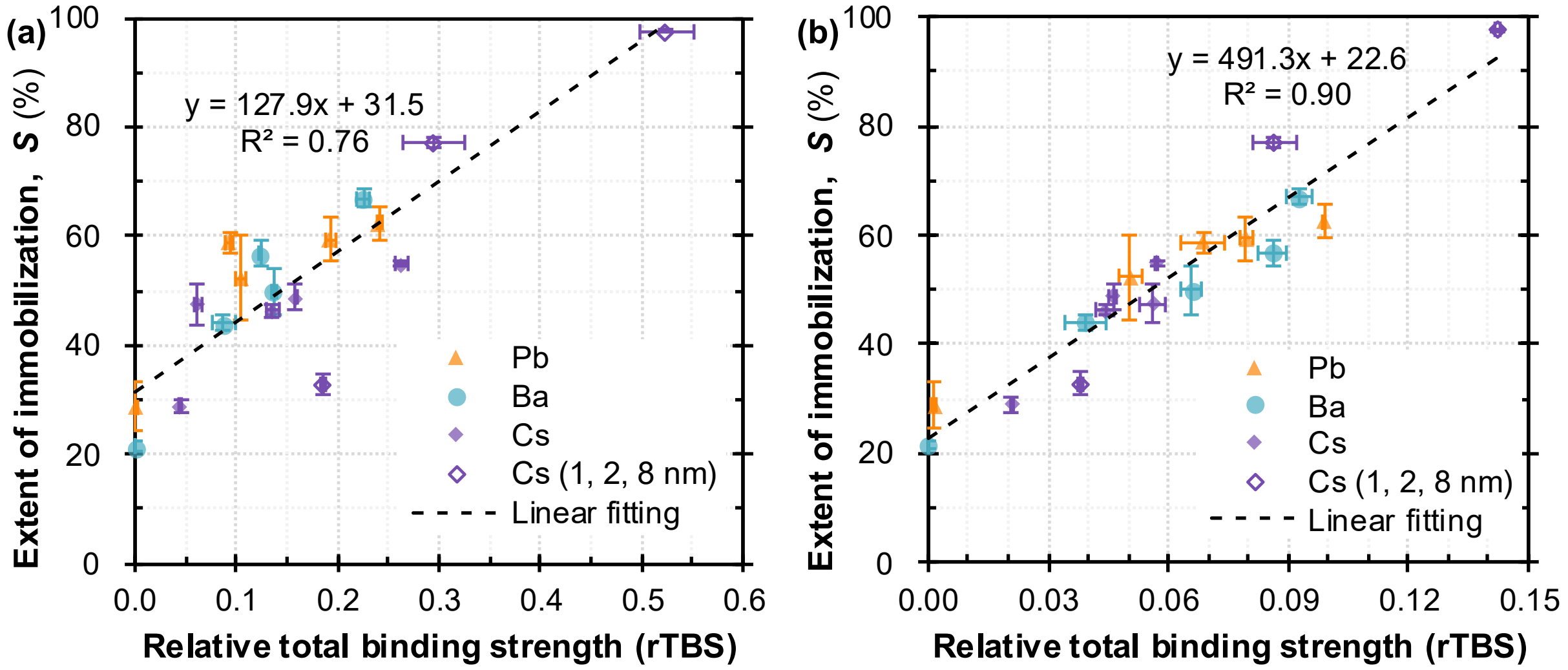

**Figure 12.** Correlation between the extent of immobilization ($S$) and relative total binding strength (rTBS) for heavy-metal ions confined within different gel nanopores, with error bars representing standard deviations from three independent simulations. (a) Based on TBS values reported in **Table 8**, **Table S10**, **Table S15**, and **Table S20**; (b) based on corrected TBS values (**Table S11**, **Table S16**, and **Table S21),** which account for multiple ion–surface atom interactions per ion. rTBS is defined as $\mathrm{rTBS_Y} = \mathrm{TBS_Y}/|\mathrm{CN_{Y-Ow}}E_{\mathrm{Y-OW}} + \mathrm{CN_{Y-Cl}}E_{\mathrm{Y-Cl}}|$, where the denominator represents the total solvation–pairing strength of the heavy-metal ion (Y = $Pb^{2+}$, $Ba^{2+}$, and $Cs^{+}$) in bulk conditions.

# 4 Broader Impact and Limitations

## 4.1 Broader Impact

This work advances the molecular understanding of adsorption-controlled heavy-metal retention in cementitious gel nanopores by establishing connections among gel chemistry, interfacial binding, spatial ion partitioning, and nanoconfined mobility. A central implication is that pore-averaged ion mobility is governed not only by the local mobility of ions within different pore regions, but also by how ions partition among these regions. Consequently, strong pore-averaged mobility suppression can arise from both reduced ion mobility near the interface and preferential accumulation within low-mobility interfacial regions. Heavy-metal retention in gel nanopores should therefore be viewed as a coupled consequence of local coordination chemistry, surface-site accessibility, interfacial charge, spatial ion redistribution, and nanoconfinement.

This perspective provides a molecular basis for considering gel-surface chemistry as a design variable for contaminant retention. In the idealized systems examined here, changes in Al-associated interfacial environments influence oxygen coordination, framework charge compensation, ion exchange, and the spatial distribution of adsorbed ions. These relationships suggest that the density, accessibility, and chemical identity of interfacial binding sites may be more informative for adsorption-controlled retention than bulk composition alone. The different responses of $Pb^{2+}$, $Ba^{2+}$, and $Cs^{+}$ further indicate that optimization of binder chemistry should account for the coordination preferences and solution-phase interactions of the target contaminant rather than assume a universal retention mechanism.

The findings are particularly relevant to the development of cementitious and alkali-activated materials for hazardous- and radioactive-waste containment. C–S–H, C–(N)–A–S–H, and N–A–S–H provide representative hydrated calcium silicate and aluminosilicate environments in which network chemistry, charge compensation, and interfacial oxygen speciation govern ion–surface interactions. The identified composition–structure–transport relationships provide mechanistic guidance for interpreting how changes in gel chemistry may influence contaminant partitioning and mobility and may inform the development of low-carbon binders that combine reduced environmental impact with effective contaminant retention and long-term durability. The spatially resolved diffusivity and residence-time analyses, together with the comparative binding-strength descriptors, also provide tools for connecting atomistic coordination environments with mobility suppression. With appropriate validation and upscaling, these relationships may provide mechanistic constraints or physically informed constitutive inputs for mesoscale and continuum models of contaminant transport and durability.

Beyond cementitious materials, the conceptual relationships among nanoscale structure, interfacial binding, ion partitioning, and mobility may be relevant to other nanoconfined systems in which adsorption and transport are strongly coupled. These include nanoporous solids such as ceramics [120], glasses [100], silicates [121], clays [122], zeolites [32], and metal–organic frameworks [123], as well as systems associated with mineral weathering [124, 125], cell-membrane signal transduction [126], water desalination and purification [127, 128], membrane-based ion separation and resource recovery [129], battery energy storage [130], and subsurface waste containment [122, 131]. By linking nanoscale structure, interfacial binding, and ion dynamics, this study contributes to a broader understanding of ion transport in chemically heterogeneous nanoconfined environments. More generally, analogous composition–structure–property frameworks have been used to relate MD-derived structural descriptors to glass dissolution and reactivity [66, 67, 132], to predict other glass properties [133], and to guide the formulation of nuclear-waste glasses [134, 135].

## 4.2 Limitations

Despite the insights obtained, this study and force field–based MD simulations more generally, has several limitations that should be considered when interpreting the results. First, real cementitious gels are intrinsically heterogeneous, with variable surface compositions, hierarchical pore networks, and dynamically evolving pore solution chemistries [136]. The gel models adopted here simplify this complexity into idealized nanochannel systems, enabling isolation of fundamental interfacial adsorption and transport mechanisms under controlled conditions. Most comparisons across gel chemistries were performed at a representative pore width of ~4 nm to enable consistent evaluation of gel surface chemistry effects. To assess pore size effects, additional simulations were performed for a representative C–(N)–A–S–H system (Al/Si = 0.125) and CsCl over a pore size range of ~1–8 nm, as described in **Section S15**. Within this nanometer-scale regime, pore size primarily modulates the relative proportion of interfacial and bulk-like regions, while the dominant local immobilization mechanism remains qualitatively unchanged across the pore sizes examined.

Second, the solution compositions considered here, including high-concentration chloride solutions such as 1 M $PbCl_2$, are intended as controlled reference states for comparative analysis rather than representations of equilibrium pore solutions in hydrated cement systems. In real cementitious systems, ion speciation, solubility limits, precipitation reactions and interactions with secondary phases may significantly alter ion distributions and retention behavior. More broadly, heavy-metal immobilization in cementitious materials typically involves multiple coupled mechanisms, including adsorption, chemical precipitation, structural incorporation, and physical encapsulation within evolving hydration/gel products [3]. The present MD simulations isolate the interfacial adsorption and nanoconfinement-controlled transport component within representative, idealized gel pores and therefore do not capture retention pathways involving precipitation, incorporation into gel structures, and interaction with secondary phases. This distinction should be considered when extrapolating present results to real cementitious binder. Future studies should systematically examine how variations in pore structure, gel chemistry, solution chemistry, external conditions, and secondary phases alters transport and immobilization behavior.

Third, the simulations employed a classical, non-reactive force field (CLAYFF), which is well-suited for capturing physical adsorption, hydration structure, ion pairing, surface coordination, and transport, but does not describe chemically reactive immobilization pathways. Capturing such processes would require reactive force fields (e.g., ReaxFF [91]) or *ab initio* methods (e.g., density functional theory), albeit at higher computational cost. In the present case, the use of ReaxFF is limited not primarily by system size, but by the lack of a single validated parameterization capable of consistently describing all relevant elements and interactions considered here (Ca, Si, Al, O, H, Na, Cl, Pb, Ba, and Cs) across both solid and aqueous phases. Hybridizing multiple ReaxFF parameter sets would introduce inconsistencies and additional uncertainties. Therefore, CLAYFF was adopted to provide a robust and internally consistent description within the scope of this work. Future studies should explore reactive force fields when suitable parameterizations become available.

Within this framework, the TBS descriptor should be interpreted as model-dependent whose numerical values depend on the underlying force field, charge assignment and interaction definitions. In CLAYFF, the use of partial charges for the solid framework and formal charges for ions may enhance the electrostatic contributions, particularly for multivalent cations. The central contribution of TBS is therefore methodological: it provides a consistent descriptor-based framework for relating ion–surface interaction strength into mobility suppression. Similar descriptors can be defined within other modeling frameworks or higher-level methods to enable systematic comparisons and mechanistic interpretation. Accordingly, TBS is used here to rationalize relative trends across gels within a single, consistent modeling framework, while broader transferability will benefit from benchmarking against alternative force fields and available experimental constraints.

Fourth, it should be noted that the Fickian diffusion analysis adopted in this work assumes a linear relationship between mean-square displacement and time. However, previous studies have shown that diffusion in the immediate vicinity of gel surfaces can become non-Fickian because of strong adsorption, nanoscale confinement, and intermittent trapping [65, 137]. Our previous study of NaCl solutions confined within C–S–H gel nanopores also identified subdiffusive dynamics of interfacial water and ions [43]. Therefore, the spatially resolved diffusion coefficients reported here should be intepretated as apparent local-mobility metrics obtained under the Fickian approximation. They are intended primarily for relative comparison across spatial regions, ion species and gel chemistries, rather than as generalized anomalous-diffusion coefficient.

Finally, MD simulations are intrinsically limited by accessible length and time scales. The nanometer-scale pores and nanosecond simulation times considered here enable detailed resolution of molecular-scale mechanisms but remain far shorter than the spatial and time scales relevant for cementitious materials in service. Within this scope, the production trajectories are sufficient for resolving the comparative adsorption-controlled mobility trends analyzed here, as supported by the trajectory-length convergence test in the **Supplementary Material** (**Figure S5**f), while the bin-size sensitivity analysis further confirms the robustness of the spatially resolved diffusivity profiles (**Figure S5**e). However, the simulations do not capture slower adsorption–desorption exchange or long-term release behavior beyond the probed timescale. Bridging this gap will require multiscale modeling strategies that incorporate atomistic insights into mesoscale and continuum frameworks, along with systematic experimental validation. Accordingly, the connection established in this study should be interpreted as descriptor-based within the atomistic framework, linking binding strength, coordination environment and relative mobility suppression, rather than a direct, quantitative prediction of macroscopic diffusivity or leaching behavior.

## 5 Conclusions

Classical molecular dynamics simulations were used to compare the transport and adsorption-controlled retention of $Pb^{2+}$, $Ba^{2+}$, and $Cs^{+}$ in representative C–S–H, C–(N)–A–S–H, and N–A–S–H gel nanopores. By combining pore-averaged and spatially resolved diffusivity, residence-time analysis, ion-density distributions, adsorption maps, coordination analysis, and interfacial binding descriptors, this study establishes a comparative atomistic framework for linking gel chemistry and interfacial structure to ion mobility suppression under nanoconfinement.

Ion mobility was suppressed in all investigated gel nanopores relative to the corresponding bulk solutions, but the magnitude of suppression depended strongly on gel chemistry. Increasing Al/Si substitution in C–(N)–A–S–H series progressively enhanced cation retention, with the Al/Si =

0.125 model exhibiting the strongest mobility suppression for all three ions. Spatially resolved diffusivity and residence-time analyses further showed that pore-averaged transport is jointly controlled by the local mobility of ions within different pore regions and their spatial partitioning into low-mobility interfacial layers. Strong overall retention therefore arises not only from slower motion near the surface, but also from preferential accumulation within these regions.

The enhanced adsorption in Al-containing C–(N)–A–S–H was accompanied by the displacement of charge-balancing $Na^+$ ions, supporting an ion-exchange-like mechanism associated with negatively charged tetrahedral Al sites. $Pb^{2+}$ and $Ba^{2+}$ exhibited broadly similar coordination mechanisms, with their enhanced retention primarily associated with hydroxyl oxygens linked to tetrahedral Al (Ohs) and, secondarily, Al-associated bridging oxygens (Obts). In C–S–H, their interfacial binding was dominated by surface hydroxyl oxygens (Oh), whereas the N–A–S–H surface provided more distributed binding environments. $Cs^+$ displayed a more distinct and gel-dependent mechanism, with important contributions from inner-sphere interactions with bridging and hydroxyl oxygens. Additional simulations of $Cs^+$ in C–(N)–A–S–H with Al/Si = 0.125 over pore widths of approximately 1–8 nm showed that pore size influenced the magnitude of pore-averaged mobility suppression, while the dominant local binding environments remained largely unchanged.

The relative total binding strength (rTBS) descriptor showed a strong positive correlation with mobility suppression across the investigated ion-gel systems, providing a comparative measure that links collective ion–surface interactions to adsorption-controlled transport. Because its absolute value depends on the force-field parameters, rTBS should be interpreted as a relative mechanistic descriptor rather than an absolute binding free energy or a direct predictor of macroscopic leaching. Overall, this work links gel chemistry, interfacial structure, and ion transport in cementitious nanopores through a spatially resolved atomistic framework. The results identify tetrahedral-Al-associated oxygen environments as key structural features governing enhanced cation retention in Al-bearing gels and demonstrate pore-averaged transport depends on both local ion mobility and interfacial partitioning. Because the present study is based on idealized nanopore models and non-reactive simulations, extension to realistic cementitious waste forms will require consideration of reactive precipitation and structural incorporation, heterogeneous and evolving pore structures and realistic solution chemistries, secondary-phase formation, and multiscale transport and leaching processes.

# 6 Appendix A. Supplementary Material

Additional results can be found in the supplementary material, including: Potential parameters for the sodium aluminosilicate (N–A–S) glass; Structural characteristics of the generated N–A–S glass; $Q^n$ species and mean silicate chain length (MCL) of different gel models; Temporal evolution of the system's potential energy during the relaxation stage; ClayFF force field parameters; Average diffusion coefficient of interlayer solution species; Spatial distribution of water and ion diffusivity inside different gel nanopores; Spatial evolution of the diffusion coefficient ratio for water/cation, water/anion, and anion/cation; Spatial evolution of the residence time (RT); Theoretical calculations of correlation between diffusivity and residence time; One-dimensional number density distribution of different solution species across the gel nanopore; Two-dimensional adsorption pattern of ions on the gel surface; Partial radial distribution function (RDF) for different ion species; Summary of nearest interatomic distances, interaction strengths, coordination numbers, and total binding strengths within the first coordination shell of heavy metal ions; Effect of pore size on heavy-metal immobilization behaviors and mechanisms; and Correlation analysis of TBS components across cation types and pore sizes.

# 7 Use of AI-Assisted Technologies

The authors utilized ChatGPT (GPT-5) to assist with language refinement during the manuscript preparation. All content has been carefully reviewed and revised as needed by the authors to ensure accuracy and clarity, and the authors take full responsibility for the final submitted work.

# 8 Declaration of Competing Interest

The authors declare that they have no known competing financial interests or personal relationships that could have appeared to influence the work reported in this paper.

## 9 Acknowledgments

This work was supported by the Department of Civil and Environmental Engineering at Rice University and the Gulf Research Program's Early-Career Research Fellowship. Access to computational resources is in part supported by the Big-Data Private-Cloud Research Cyberinfrastructure MRI-award funded by NSF under grant CNS-1338099 and by Rice University's Center for Research Computing (CRC). The in-house text-mining pipeline powered by large language models was originally developed by Dr. Zhanzhao Li and Dr. Kengran Yang with support from OpenAI Researcher Access Program.

## 10 References

1. Chen L, Nakamura K, Hama T. Review on stabilization/solidification methods and mechanism of heavy metals based on OPC-based binders. J Environ Manag. 2023;332:117362. https://doi.org/10.1016/j.jenvman.2023.117362.

2. Ma B, Provis JL, Wang D, Kosakowski G. The essential role of cement-based materials in a radioactive waste repository. npj Mater Sustain. 2024;2(1):21. https://doi.org/10.1038/s44296-024-00025-9.

3. Chen Q, Tyrer M, Hills CD, Yang X, Carey P. Immobilisation of heavy metal in cement-based solidification/stabilisation: A review. Waste Manag. 2009;29(1):390–403. https://doi.org/10.1016/j.wasman.2008.01.019.

4. Liang K, Wang XQ, Chow CL, Lau D. A review of geopolymer and its adsorption capacity with molecular insights: A promising adsorbent of heavy metal ions. J Environ Manag. 2022;322:116066. https://doi.org/10.1016/j.jenvman.2022.116066.

5. Barbhuiya S, Das BB, Qureshi T, Adak D. Cement-based solidification of nuclear waste: mechanisms, formulations and regulatory considerations. J Environ Manag. 2024;356:120712. https://doi.org/10.1016/j.jenvman.2024.120712.

6. Li Z, Yang K, He Q, Gong K. Large language model-enabled automated data extraction for concrete materials informatics. arXiv preprint arXiv:260422938. 2026. https://doi.org/10.48550/arXiv.2604.22938.

7. Li Z, He Q, Chen W, Yang K, Gong K. Automating literature screening and knowledge synthesis with large language models. Digital Discovery. 2026. submitted.

8. Missana T, García-Gutiérrez M, Mingarro M, Alonso U. Analysis of barium retention mechanisms on calcium silicate hydrate phases. Cem Concr Res. 2017;93:8–16. https://doi.org/10.1016/j.cemconres.2016.12.004.

9. Liu Q, Feng P, Shao L, Chen C, Liu X, Ma Y, Zhang L, Geng G. Quantifying the immobilization mechanisms of heavy metals by Calcium Silicate Hydrate (CSH): The case of $Cu^{2+}$. Cem Concr Res. 2024;186:107695. https://doi.org/10.1016/j.cemconres.2024.107695.

10. Liu Y, Dalconi MC, Bellotto MP, Valentini L, Molinari S, Yuan X, Wang D, Hu W, Chen Q, Fernandez-Martinez A. Pb-induced retardation of early hydration of Portland cement: Insights from in-situ XRD and implications for substitution with industrial by-products. Cem Concr Res. 2025;193:107867. https://doi.org/10.1016/j.cemconres.2025.107867.

11. Giergiczny Z, Król A. Immobilization of heavy metals (Pb, Cu, Cr, Zn, Cd, Mn) in the mineral additions containing concrete composites. J Hazard Mater. 2008;160(2-3):247–55. https://doi.org/10.1016/j.jhazmat.2008.03.007.

12. Katsioti M, Katsiotis N, Rouni G, Bakirtzis D, Loizidou M. The effect of bentonite/cement mortar for the stabilization/solidification of sewage sludge containing heavy metals. Cem Concr Compos. 2008;30(10):1013–19. https://doi.org/10.1016/j.cemconcomp.2008.03.001.

13. Liu M, Xia Y, Zhao Y, Cao Z. Immobilization of Cu (II), Ni (II) and Zn (II) in silica fume blended Portland cement: Role of silica fume. Constr Build Mater. 2022;341:127772. https://doi.org/10.1016/j.conbuildmat.2022.127772.

14. Goo J-Y, Kim B-J, Kang M, Jeong J, Jo HY, Kwon J-S. Leaching behavior of cesium, strontium, cobalt, and europium from immobilized cement matrix. Appl Sci. 2021;11(18):8418. https://doi.org/10.3390/app11188418.

15. Arokiasamy P, Abdullah MMAB, Abd Rahim SZ, Sadique M, Ming LY, Salleh MAAM, Zainol MRRMA, Ghazali CMR. Diverse material based geopolymer towards heavy metals removal: A review. J Mater Res Technol. 2023;22:126–56. https://doi.org/10.1016/j.jmrt.2022.11.100.

16. Zhang X, Wang B, Chang J. Adsorption behavior and solidification mechanism of Pb (II) on synthetic CASH gels with different Ca/Si and Al/Si ratios in high alkaline conditions. Chem Eng J. 2024;493:152344. https://doi.org/10.1016/j.cej.2024.152344.

17. Zheng L, Wang W, Shi Y. The effects of alkaline dosage and Si/Al ratio on the immobilization of heavy metals in municipal solid waste incineration fly ash-based geopolymer. Chemosphere. 2010;79(6):665–71. https://doi.org/10.1016/j.chemosphere.2010.02.018.

18. Hoe-Woon T, Cheng-Yong H, Yun-Ming L, Qi-Hwa N, Wei-Hao L, Loong FK, Pakawanit P, Supramanian DA, Jia-Ni L, Yu-Xin Y. Elucidating the Interplay between Pore Microstructure and Heavy Metal Leaching of Rubber Sludge in Fly Ash Geopolymers. J Environ Chem Eng. 2025:117698. https://doi.org/10.1016/j.jece.2025.117698.

19. El-eswed BI. Chemical evaluation of immobilization of wastes containing Pb, Cd, Cu and Zn in alkali-activated materials: A critical review. J Environ Chem Eng. 2020;8(5):104194. https://doi.org/10.1016/j.jece.2020.104194.

20. Zhang J, Provis JL, Feng D, van Deventer JS. Geopolymers for immobilization of $Cr^{6+}$, $Cd^{2+}$, and $Pb^{2+}$. J Hazard Mater. 2008;157(2-3):587–98. https://doi.org/10.1016/j.jhazmat.2008.01.053.

21. Genua F, Lancellotti I, Leonelli C. Geopolymer-based stabilization of heavy metals, the role of chemical agents in encapsulation and adsorption. Polymers. 2025;17(5):670. https://doi.org/10.3390/polym17050670.

22. Vance ER, Perera DS. 18 - Geopolymers for nuclear waste immobilisation. In: Provis JL, Van Deventer JSJ, eds. Geopolymers: structures, processing, properties and industrial applications. Cambridge, UK: Woodhead Publishing Limited; 2009.

23. Scrivener K, Ouzia A, Juilland P, Mohamed AK. Advances in understanding cement hydration mechanisms. Cem Concr Res. 2019;124:105823. https://doi.org/10.1016/j.cemconres.2019.105823.

24. Provis JL, Palomo A, Shi C. Advances in understanding alkali-activated materials. Cem Concr Res. 2015;78:110–25. https://doi.org/10.1016/j.cemconres.2015.04.013.

25. Guo B, Liu B, Yang J, Zhang S. The mechanisms of heavy metal immobilization by cementitious material treatments and thermal treatments: A review. J Environ Manag. 2017;193:410–22. https://doi.org/10.1016/j.jenvman.2017.02.026.

26. Duque-Redondo E, Bonnaud PA, Manzano H. A comprehensive review of CSH empirical and computational models, their applications, and practical aspects. Cem Concr Res. 2022;156:106784. https://doi.org/10.1016/j.cemconres.2022.106784.

27. Sun M, Hu K, Zhang Y, Wei G, Rong H. Molecular simulation and experimental investigation on enhancement of microbe cement in municipal solid waste incineration fly ash. Front Mater. 2022;9:1013580. https://doi.org/10.3389/fmats.2022.1013580.

28. Su L, Wu S, Yu F, Zhu W, Zhang X, Liang B. Utilizing municipal solid waste incineration fly ash for mine goaf filling: Preparation, optimization, and heavy metal leaching study. Environ Res. 2025;266:120594. https://doi.org/10.1016/j.envres.2024.120594.

29. Duque-Redondo E, Yamada K, Dolado JS, Manzano H. Microscopic mechanism of radionuclide Cs retention in Al containing CSH nanopores. Comput Mater Sci. 2021;190:110312. https://doi.org/10.1016/j.commatsci.2021.110312.

30. Xu L-Y, Alrefaei Y, Wang Y-S, Dai J-G. Recent advances in molecular dynamics simulation of the NASH geopolymer system: Modeling, structural analysis, and dynamics. Constr Build Mater. 2021;276:122196. https://doi.org/10.1016/j.conbuildmat.2020.122196.

31. Hou D, Zhang J, Pan W, Zhang Y, Zhang Z. Nanoscale mechanism of ions immobilized by the geopolymer: A molecular dynamics study. J Nucl Mater. 2020;528:151841. https://doi.org/10.1016/j.jnucmat.2019.151841.

32. Duque-Redondo E, Yamada K, Masoero E, Prieto JB, Manzano H. Adsorption and migration of Cs and Na ions in geopolymers and zeolites. Mater Today Commun. 2023;36:106496. https://doi.org/10.1016/j.mtcomm.2023.106496.

33. Zhang W, Li J-s, Huang X, Chen Z, Lang L, Huang K. Unraveling the cation adsorption of geopolymer binder: A molecular dynamics study. Chemosphere. 2023;335:139118. https://doi.org/10.1016/j.chemosphere.2023.139118.

34. Liang K, Yang G, Wang XQ, Chow CL, Lau D. Development of effective porous geopolymer adsorbent with high strength for copper (II) ion removal. J Clean Prod. 2024;449:141752. https://doi.org/10.1016/j.jclepro.2024.141752.

35. Zhang W, Lang L, Chen X, Huang K, Li J-S. Adsorption mechanism of $Cs^{+}$ and $Pb^{2+}$ on the NASH surface under the different Si/Al ratio and temperature. J Mol Liq. 2024;401:124558. https://doi.org/10.1016/j.molliq.2024.124558.

36. Wang R, Ye J, Wang J, Peng X. Adsorption and diffusion mechanism of cesium and chloride ions in channel of geopolymer with different Si/Al ratios: molecular dynamics simulation. J Radioanal Nucl Chem. 2023;332(9):3597–607. https://doi.org/10.1007/s10967-023-09046-5.

37. Duque-Redondo E, Kazuo Y, López-Arbeloa I, Manzano H. Cs-137 immobilization in CSH gel nanopores. Phys Chem Chem Phys. 2018;20(14):9289–97. https://doi.org/10.1039/c8cp00654g.

38. Jiang J, Wang P, Hou D. The mechanism of cesium ions immobilization in the nanometer channel of calcium silicate hydrate: a molecular dynamics study. Phys Chem Chem Phys. 2017;19(41):27974–86. https://doi.org/10.1039/c7cp05437h.

39. Wang P, Zhang Q, Wang M, Yin B, Hou D, Zhang Y. Atomistic insights into cesium chloride solution transport through the ultra-confined calcium–silicate–hydrate channel. Phys Chem Chem Phys. 2019;21(22):11892–902. https://doi.org/10.1039/c8cp07676f.

40. Qiao G, Hou D, Li W, Yin B, Zhang Y, Wang P. Molecular insights into migration of heavy metal ion in calcium silicate hydrate (CSH) surface and intra-CSH (Ca/Si=1.3). Constr Build Mater. 2023;365:130097. https://doi.org/10.1016/j.conbuildmat.2022.130097.

41. Zha F, Tai H, Wang Q, Kang B, Xu L, Chu C. Molecular insight into the transport of multicomponent heavy metals in calcium silicate hydrate and its mechanical behavior. J Rock Mech Geotech Eng. 2025;17(8):5135–45. https://doi.org/10.1016/j.jrmge.2025.05.002.

42. Kai M, Zhang L, Liew K. Atomistic insights into structure evolution and mechanical property of calcium silicate hydrates influenced by nuclear waste caesium. J Hazard Mater. 2021;411:125033. https://doi.org/10.1016/j.jhazmat.2020.125033.

43. Chen W, Gong K. Insights into ionic diffusion in C–S–H gel pore from molecular dynamics simulations: spatial distributions, energy barriers, and structural descriptor. J Phys Chem B. 2025;129(40):10550–67. https://doi.org/10.1021/acs.jpcb.5c04534.

44. Yang Y, Patel RA, Churakov SV, Prasianakis NI, Kosakowski G, Wang M. Multiscale modeling of ion diffusion in cement paste: electrical double layer effects. Cem Concr Compos. 2019;96:55–65. https://doi.org/10.1016/j.cemconcomp.2018.11.008.

45. Liu C, Zhang M. Multiscale modelling of ionic diffusivity in unsaturated concrete accounting for its hierarchical microstructure. Cem Concr Res. 2022;156:106766. https://doi.org/10.1016/j.cemconres.2022.106766.

46. Duan C, Majumdar A. Anomalous ion transport in 2-nm hydrophilic nanochannels. Nat Nanotechnol. 2010;5(12):848–52. https://doi.org/10.1038/nnano.2010.233.

47. Wang M, Hou Y, Yu L, Hou X. Anomalies of ionic/molecular transport in nano and sub-nano confinement. Nano Lett. 2020;20(10):6937–46. https://doi.org/10.1021/acs.nanolett.0c02999.

48. Gong K, White CE. Impact of chemical variability of ground granulated blast-furnace slag on the phase formation in alkali-activated slag pastes. Cem Concr Res. 2016;89:310–19. https://doi.org/10.1016/j.cemconres.2016.09.003.

49. Li C, Sun H, Li L. A review: The comparison between alkali-activated slag (Si+Ca) and metakaolin (Si+Al) cements. Cem Concr Res. 2010;40(9):1341–49. https://doi.org/10.1016/j.cemconres.2010.03.020.

50. Provis JL. Alkali-activated materials. Cem Concr Res. 2018;114:40–48. https://doi.org/10.1016/j.cemconres.2017.02.009.

51. Richardson IG. The nature of CSH in hardened cements. Cem Concr Res. 1999;29(8):1131–47. https://doi.org/10.1016/S0008-8846(99)00168-4.

52. Mehta PK, Monteiro P. Concrete: microstructure, properties, and materials. 4th ed. New York: McGraw-Hill Education; 2014.

53. Jiang Z-L, Pan Y-J, Lu J-F, Wang Y-C. Pore structure characterization of cement paste by different experimental methods and its influence on permeability evaluation. Cem Concr Res. 2022;159:106892. https://doi.org/10.1016/j.cemconres.2022.106892.

54. Hamid SA. The crystal structure of the 11Å natural tobermorite $Ca_{2.25}[Si_3O_{7.5}(OH)_{1.5}]\cdot 1H_2O$. Z Kristallogr Cryst Mater. 1981;154(1-4):189–98. https://doi.org/10.1524/zkri.1981.154.14.189.

55. Le Saoût G, Ben Haha M, Winnefeld F, Lothenbach B. Hydration degree of alkali-activated slags: a 29 Si NMR study. J Am Ceram Soc. 2011;94(12):4541–47. https://doi.org/10.1111/j.1551-2916.2011.04828.x.

56. Haha MB, Le Saout G, Winnefeld F, Lothenbach B. Influence of activator type on hydration kinetics, hydrate assemblage and microstructural development of alkali activated blast-furnace slags. Cem Concr Res. 2011;41(3):301–10. https://doi.org/10.1016/j.cemconres.2010.11.016.

57. Faucon P, Charpentier T, Nonat A, Petit J. Triple-quantum two-dimensional $^{27}$Al magic angle nuclear magnetic resonance study of the aluminum incorporation in calcium silicate hydrates. J Am Chem Soc. 1998;120(46):12075–82. https://doi.org/10.1021/ja9806940.

58. Wang S-D, Scrivener KL. $^{29}$Si and $^{27}$Al NMR study of alkali-activated slag. Cem Concr Res. 2003;33(5):769–74. https://doi.org/10.1016/S0008-8846(02)01044-X.

59. Richardson IG, Brough AR, Brydson R, Groves GW, Dobson CM. Location of aluminum in substituted calcium silicate hydrate (C-S-H) gels as determined by $^{29}$Si and $^{27}$Al NMR and EELS. J Am Ceram Soc. 1993;76(9):2285–88. https://doi.org/10.1111/j.1151-2916.1993.tb07765.x.

60. Manzano H, Dolado JS, Ayuela A. Aluminum incorporation to dreierketten silicate chains. J Phys Chem B. 2009;113(9):2832–39. https://doi.org/10.1021/jp804867u.

61. Walkley B, San Nicolas R, Sani M-A, Rees GJ, Hanna JV, van Deventer JS, Provis JL. Phase evolution of C-(N)-A-S-H/NASH gel blends investigated via alkali-activation of synthetic calcium aluminosilicate precursors. Cem Concr Res. 2016;89:120–35. https://doi.org/10.1016/j.cemconres.2016.08.010.

62. Myers RJ, Bernal SA, San Nicolas R, Provis JL. Generalized structural description of calcium–sodium aluminosilicate hydrate gels: the cross-linked substituted tobermorite model. Langmuir. 2013;29(17):5294–306. https://doi.org/10.1021/la4000473.

63. Cai W, Xu Z, Zhang Z, Hu J, Huang H, Ma Y, Zhang Z, Wang H, Yin S, Wei J, Shi C, Yu Q. Chloride binding behavior of synthesized reaction products in alkali-activated slag. Compos Part B Eng. 2022;238:109919. https://doi.org/10.1016/j.compositesb.2022.109919.

64. Mohamed AK, Parker SC, Bowen P, Galmarini S. An atomistic building block description of C-S-H - Towards a realistic C-S-H model. Cem Concr Res. 2018;107:221–35. https://doi.org/10.1016/j.cemconres.2018.01.007.

65. Ihaddadene T, Claverie J, Bignonnet F, Amiri O. Diffusion and physicochemical behavior of chloride and sodium ions in CSH gel pores: A molecular dynamics investigation. Cem Concr Res. 2026;199:108053. https://doi.org/10.1016/j.cemconres.2025.108053.

66. Gong K, White CE. Predicting CaO-(MgO)-$Al_2O_3$-$SiO_2$ glass reactivity in alkaline environments from force field molecular dynamics simulations. Cem Concr Res. 2021;150:106588. https://doi.org/10.1016/j.cemconres.2021.106588.

67. Gong K, Olivetti EA. Development of structural descriptors to predict dissolution rate of volcanic glasses: Molecular dynamic simulations. J Am Ceram Soc. 2022;105(4):2575–94. https://doi.org/10.1111/jace.18265.

68. Du MR, Chen SJ, Duan WH, Chen WQ, Jing HW. Role of multiwalled carbon nanotubes as shear reinforcing nanopins in quasi-brittle matrices. ACS Appl Nano Mater. 2018;1(4):1731–40. https://doi.org/10.1021/acsanm.8b00162.

69. Pedone A, Malavasi G, Menziani MC, Cormack AN, Segre U. A new self-consistent empirical interatomic potential model for oxides, silicates, and silica-based glasses. J Phys Chem B. 2006;110(24):11780–95. https://doi.org/10.1021/jp0611018.

70. White CE, Page K, Henson NJ, Provis JL. In situ synchrotron X-ray pair distribution function analysis of the early stages of gel formation in metakaolin-based geopolymers. Appl Clay Sci. 2013;73:17–25. https://doi.org/10.1016/j.clay.2012.09.009.

71. Farrow C, Juhas P, Liu J, Bryndin D, Božin E, Bloch J, Proffen T, Billinge S. PDFfit2 and PDFgui: computer programs for studying nanostructure in crystals. J Phys Condens Matter. 2007;19(33):335219. https://doi.org/10.1088/0953-8984/19/33/335219.

72. Gong K, Özçelik VO, Yang K, White CE. Density functional modeling and total scattering analysis of the atomic structure of a quaternary CaO−MgO−$Al_2O_3$−$SiO_2$ (CMAS) glass: Uncovering the local environment of calcium and magnesium. Phys Rev Mater. 2021;5(1):015603. https://doi.org/10.1103/PhysRevMaterials.5.015603.

73. Lolli F, Manzano H, Provis JL, Bignozzi MC, Masoero E. Atomistic simulations of geopolymer models: the impact of disorder on structure and mechanics. ACS Appl Mater Interfaces. 2018;10(26):22809–20. https://doi.org/10.1021/acsami.8b03873.

74. Casar Z, López-Zorrilla J, Manzano H, Duque-Redondo E, Mohamed AK, Scrivener K, Bowen P. pyCSH: Automated atomic-level structure generation of bulk CSH and investigation of their intrinsic properties. Cem Concr Res. 2024;183:107593. https://doi.org/10.1016/j.cemconres.2024.107593.

75. Martínez L, Andrade R, Birgin EG, Martínez JM. PACKMOL: A package for building initial configurations for molecular dynamics simulations. J Comput Chem. 2009;30(13):2157–64. https://doi.org/10.1002/jcc.21224.

76. Liu L, Zhang C, Jiang W, Li X, Dai Y, Jia H. Understanding the sorption behaviors of heavy metal ions in the interlayer and nanopore of montmorillonite: A molecular dynamics study. J Hazard Mater. 2021;416:125976. https://doi.org/10.1016/j.jhazmat.2021.125976.

77. The Engineering ToolBox. Densities of Aqueous Solutions of Inorganic Chlorides. https://www.engineeringtoolbox.com/density-aqueous-solution-inorganic-chlorides-salt-concentration-d_1955.html. Accessed Aug 15, 2025.

78. Rowland D. Density of caesium chloride, CsCl(aq). https://advancedthermo.com/electrolytes/density_CsCl_Jun2021.html. Accessed Aug 15, 2025.

79. Thompson AP, Aktulga HM, Berger R, Bolintineanu DS, Brown WM, Crozier PS, In't Veld PJ, Kohlmeyer A, Moore SG, Nguyen TD. LAMMPS-a flexible simulation tool for particle-based materials modeling at the atomic, meso, and continuum scales. Comput Phys Commun. 2022;271:108171. https://doi.org/10.1016/j.cpc.2021.108171.

80. Stukowski A. Visualization and analysis of atomistic simulation data with OVITO–the Open Visualization Tool. Model Simul Mater Sci Eng. 2009;18(1):015012. https://doi.org/10.1088/0965-0393/18/1/015012.

81. Nosé S. A molecular dynamics method for simulations in the canonical ensemble. Mol Phys. 1984;52(2):255–68. https://doi.org/10.1080/00268970110089108.

82. Hoover WG. Canonical dynamics: Equilibrium phase-space distributions. Phys Rev A. 1985;31(3):1695. https://doi.org/10.1103/PhysRevA.31.1695.

83. Cygan RT, Liang J-J, Kalinichev AG. Molecular models of hydroxide, oxyhydroxide, and clay phases and the development of a general force field. J Phys Chem B. 2004;108(4):1255–66. https://doi.org/10.1021/jp0363287.

84. Koneshan S, Rasaiah JC, Lynden-Bell R, Lee S. Solvent structure, dynamics, and ion mobility in aqueous solutions at 25 °C. J Phys Chem B. 1998;102(21):4193–204. https://doi.org/10.1021/jp980642x.

85. Smith DE, Dang LX. Computer simulations of NaCl association in polarizable water. J Chem Phys. 1994;100(5):3757–66. https://doi.org/10.1063/1.466363.

86. Aqvist J. Ion-water interaction potentials derived from free energy perturbation simulations. J Phys Chem. 1990;94(21):8021–24. https://doi.org/10.1021/j100384a009.

87. Smith DE, Dang LX. Computer simulations of cesium–water clusters: Do ion–water clusters form gas-phase clathrates? J Chem Phys. 1994;101(9):7873–81. https://doi.org/10.1063/1.468213.

88. Lorentz HA. Ueber die Anwendung des Satzes vom Virial in der kinetischen Theorie der Gase. Ann Phys. 1881;248(1):127–36. https://doi.org/10.1002/andp.18812480110.

89. Berthelot D. Sur le mélange des gaz. C R Hebd Seances Acad Sci. 1898;126(3):1703–06. https://www.biodiversitylibrary.org/item/111847#page/757/mode/1up.

90. Hockney RW, Eastwood JW. Computer simulation using particles. 1st ed. Boca Raton, Florida, USA: CRC Press; 1988.

91. Van Duin AC, Dasgupta S, Lorant F, Goddard WA. ReaxFF: a reactive force field for hydrocarbons. J Phys Chem A. 2001;105(41):9396–409. https://doi.org/10.1021/jp004368u.

92. Kalinichev AG, Kirkpatrick RJ. Molecular dynamics modeling of chloride binding to the surfaces of calcium hydroxide, hydrated calcium aluminate, and calcium silicate phases. Chem Mater. 2002;14(8):3539–49. https://doi.org/10.1021/cm0107070.

93. Wang F, Zhang Y, Jiang J, Yin B, Li Z. Effect of temperature on the capillary transport of sodium sulfate solution in calcium silicate hydrate nanopore: A molecular dynamics study. Constr Build Mater. 2020;231:117111. https://doi.org/10.1016/j.conbuildmat.2019.117111.

94. Liu Z, Xu D, Gao S, Zhang Y, Jiang J. Assessing the adsorption and diffusion behavior of multicomponent ions in saturated calcium silicate hydrate gel pores using molecular dynamics. ACS Sustain Chem Eng. 2020;8(9):3718–27. https://doi.org/10.1021/acssuschemeng.9b06817.

95. Tu Y, Cao J, Wen R, Shi P, Yuan L, Ji Y, Das O, Försth M, Sas G, Elfgren L. Molecular dynamics simulation study of the transport of pairwise coupled ions confined in CSH gel nanopores. Constr Build Mater. 2022;318:126172. https://doi.org/10.1016/j.conbuildmat.2021.126172.

96. Hayat U, Kai M-F, Hu-Bao A, Liew J-X, Dai J-G. Atomic-level investigation into the transport of NaCl solution in porous cement paste: The effects of pore size and temperature. J Build Eng. 2024;86:108976. https://doi.org/10.1016/j.jobe.2024.108976.

97. Chowdhuri S, Chandra A. Molecular dynamics simulations of aqueous NaCl and KCl solutions: Effects of ion concentration on the single-particle, pair, and collective dynamical properties

of ions and water molecules. J Chem Phys. 2001;115(8):3732–41. https://doi.org/10.1063/1.1387447.

98. Chen WQ, Jivkov AP, Sedighi M. Thermo-osmosis in charged nanochannels: effects of surface charge and ionic strength. ACS Appl Mater Interfaces. 2023;15(28):34159–71. https://doi.org/10.1021/acsami.3c02559.

99. Honorio T, Carasek H, Cascudo O. Water self-diffusion in CSH: Effect of confinement and temperature studied by molecular dynamics. Cem Concr Res. 2022;155:106775. https://doi.org/10.1016/j.cemconres.2022.106775.

100. Collin M, Gin S, Dazas B, Mahadevan T, Du J, Bourg IC. Molecular dynamics simulations of water structure and diffusion in a 1 nm diameter silica nanopore as a function of surface charge and alkali metal counterion identity. J Phys Chem C. 2018;122(31):17764–76. https://doi.org/10.1021/acs.jpcc.8b03902.

101. Boon JP, Yip S. Molecular hydrodynamics. New York: Dover Publications, Inc.; 1991.

102. Pranami G, Lamm MH. Estimating error in diffusion coefficients derived from molecular dynamics simulations. J Chem Theory Comput. 2015;11(10):4586–92. https://doi.org/10.1021/acs.jctc.5b00574.

103. Duque-Redondo E, Yamada K, Manzano H. Cs retention and diffusion in CSH at different Ca/Si ratio. Cem Concr Res. 2021;140:106294. https://doi.org/10.1016/j.cemconres.2020.106294.

104. Chakrabarti H. Strong evidence of an isotope effect in the diffusion of a NaCl and CsCl solution. Phys Rev B. 1995;51(18):12809. https://doi.org/10.1103/PhysRevB.51.12809.

105. Sato H, Yui M, Yoshikawa H. Ionic diffusion coefficients of $Cs^+$, $Pb^{2+}$, $Sm^{3+}$, $Ni^{2+}$, $SeO^{2-}{}_4$ and $TcO^-{}_4$ in free water determined from conductivity measurements. J Nucl Sci Technol. 1996;33(12):950–55. https://doi.org/10.1080/18811248.1996.9732037.

106. Vitagliano V, Lyons PA. Diffusion coefficients for aqueous solutions of sodium chloride and barium chloride. J Am Chem Soc. 1956;78(8):1549–52. https://doi.org/10.1021/ja01589a011.

107. Li Y-H, Gregory S. Diffusion of ions in sea water and in deep-sea sediments. Geochim Cosmochim Acta. 1974;38(5):703–14. https://doi.org/10.1016/0016-7037(74)90145-8.

108. Marcus Y. Ion properties. In: Kreysa G, Ota K-i, Savinell RF, eds. Encyclopedia of Applied Electrochemistry. New York, NY, USA: Springer; 2014. https://doi.org/10.1007/978-1-4419-6996-5_15.

109. Mamatkulov S, Schwierz N. Force fields for monovalent and divalent metal cations in TIP3P water based on thermodynamic and kinetic properties. J Chem Phys. 2018;148(7):074504. https://doi.org/10.1063/1.5017694.

110. Marcus Y. Thermodynamics of solvation of ions. Part 5.—Gibbs free energy of hydration at 298.15 K. J Chem Soc, Faraday Trans. 1991;87(18):2995–99. https://doi.org/10.1039/FT9918702995.

111. Marcus Y. Ions in Solution and their Solvation. 1st ed. Hoboken, New Jersey, USA: John Wiley & Sons; 2015.

112. Bourg IC, Steefel CI. Molecular dynamics simulations of water structure and diffusion in silica nanopores. J Phys Chem C. 2012;116(21):11556–64. https://doi.org/10.1021/jp301299a.

113. Gin S, Collin M, Jollivet P, Fournier M, Minet Y, Dupuy L, Mahadevan T, Kerisit S, Du J. Dynamics of self-reorganization explains passivation of silicate glasses. Nat Commun. 2018;9(1):2169. https://doi.org/10.1038/s41467-018-04511-2.

114. Hou D, Li Z, Zhao T, Zhang P. Water transport in the nano-pore of the calcium silicate phase: reactivity, structure and dynamics. Phys Chem Chem Phys. 2015;17(2):1411–23. https://doi.org/10.1039/c4cp04137b.

115. Ma J, Li K, Li Z, Qiu Y, Si W, Ge Y, Sha J, Liu L, Xie X, Yi H. Drastically reduced ion mobility in a nanopore due to enhanced pairing and collisions between dehydrated ions. J Am Chem Soc. 2019;141(10):4264–72. https://doi.org/10.1021/jacs.8b08488.

116. Chen WQ, Sedighi M, Jivkov AP. Thermo-osmosis in hydrophilic nanochannels: Mechanism and size effect. Nanoscale. 2021;13(3):1696–716. https://doi.org/10.1039/d0nr06687g.

117. Han M. Thermophoresis in liquids: a molecular dynamics simulation study. J Colloid Interface Sci. 2005;284(1):339–48. https://doi.org/10.1016/j.jcis.2004.09.067.

118. Chen WQ, Sedighi M, Curvalle F, Jivkov AP. Elevated temperature effects (T> 100° C) on the interfacial water and microstructure swelling of Na-montmorillonite. Chem Eng J. 2024;481:148647. https://doi.org/10.1016/j.cej.2024.148647.

119. Gong K, White CE. Development of physics-based compositional parameters for predicting the reactivity of amorphous aluminosilicates in alkaline environments. Cem Concr Res. 2023;174:107296. https://doi.org/10.1016/j.cemconres.2023.107296.

120. Romero V, Vega V, García J, Prida V, Hernando B, Benavente J. Ionic transport across tailored nanoporous anodic alumina membranes. J Colloid Interface Sci. 2012;376(1):40–46. https://doi.org/10.1016/j.jcis.2012.02.066.

121. Chen WQ, Sedighi M, Jivkov AP. Thermal diffusion of ionic species in charged nanochannels. Nanoscale. 2023;15(1):215–29. https://doi.org/10.1039/d2nr05504j.

122. Holmboe M, Bourg IC. Molecular dynamics simulations of water and sodium diffusion in smectite interlayer nanopores as a function of pore size and temperature. J Phys Chem C. 2014;118(2):1001–13. https://doi.org/10.1021/jp408884g.

123. Zhang H, Hou J, Hu Y, Wang P, Ou R, Jiang L, Liu JZ, Freeman BD, Hill AJ, Wang H. Ultrafast selective transport of alkali metal ions in metal organic frameworks with subnanometer pores. Sci Adv. 2018;4(2):eaaq0066. https://doi.org/10.1126/sciadv.aaq0066.

124. Frankel GS, Vienna JD, Lian J, Scully JR, Gin S, Ryan JV, Wang J, Kim SH, Windl W, Du J. A comparative review of the aqueous corrosion of glasses, crystalline ceramics, and metals. npj Mater Degrad. 2018;2(1):15. https://doi.org/10.1038/s41529-018-0037-2.

125. Putnis A. Why mineral interfaces matter. Science. 2014;343(6178):1441–42. https://doi.org/10.1126/science.1250884.

126. You Y, Ismail A, Nam G-H, Goutham S, Keerthi A, Radha B. Angstrofluidics: Walking to the limit. Annu Rev Mater Res. 2022;52(1):189–218. https://doi.org/10.1146/annurev-matsci-081320-032747.

127. Heiranian M, DuChanois RM, Ritt CL, Violet C, Elimelech M. Molecular simulations to elucidate transport phenomena in polymeric membranes. Environ Sci Technol. 2022;56(6):3313–23. https://doi.org/10.1021/acs.est.2c00440.

128. Gao Y, Chen W, Liu Y, Wu J, Jing H. Graphene kirigami as an ultra-permeable water desalination membrane. Carbon. 2022;195:183–90. https://doi.org/10.1016/j.carbon.2022.04.031.

129. Feng Y, Zhu Y, Chen W, Huang X, Weng X, Meyer MD, Chen T-H, Liu Y, He Z, Hou C-H. A rationally designed scalable thin film nanocomposite cation exchange membrane for precise lithium extraction. Nat Commun. 2025;16(1):8618. https://doi.org/10.1038/s41467-025-63660-3.

130. Hao Z, Zhang Q, Xu X, Zhao Q, Wu C, Liu J, Wang H. Nanochannels regulating ionic transport for boosting electrochemical energy storage and conversion: A review. Nanoscale. 2020;12(30):15923–43. https://doi.org/10.1039/d0nr02464c.

131. Ma Z, Gamage RP, Rathnaweera T, Kong L. Review of application of molecular dynamic simulations in geological high-level radioactive waste disposal. Appl Clay Sci. 2019;168:436–49. https://doi.org/10.1016/j.clay.2018.11.018.

132. Lu X, Deng L, Gin S, Du J. Quantitative structure–property relationship (QSPR) analysis of $ZrO_2$-containing soda-lime borosilicate glasses. J Phys Chem B. 2019;123(6):1412–22. https://doi.org/10.1021/acs.jpcb.8b11108.

133. Lu X, Du J. Quantitative structure-property relationship (QSPR) analysis of calcium aluminosilicate glasses based on molecular dynamics simulations. J Non-Cryst Solids. 2020;530:119772. https://doi.org/10.1016/j.jnoncrysol.2019.119772.

134. Lu X, Vienna JD, Du J. Glass formulation and composition optimization with property models: A review. J Am Ceram Soc. 2024;107(3):1603–24. https://doi.org/10.1111/jace.19333.

135. Kalahe J, Lu X, Riley BJ, Vienna JD, Du J. Structure–property relations of sodium iron phosphate nuclear waste glasses: effects of iron redox ratio and glass composition. J Am Ceram Soc. 2025;108(2):e20174. https://doi.org/10.1111/jace.20174.

136. Scrivener K, Snellings R, Lothenbach B, eds. A practical guide to microstructural analysis of cementitious materials. 1st ed. Boca Raton, FL, USA: CRC Press 2016. https://doi.org/10.1201/b19074.

137. Metzler R, Jeon J-H, Cherstvy AG, Barkai E. Anomalous diffusion models and their properties: non-stationarity, non-ergodicity, and ageing at the centenary of single particle tracking. Phys Chem Chem Phys. 2014;16(44):24128–64. https://doi.org/10.1039/c4cp03465a.

# Supplementary Material for

# Gel-Chemistry-Controlled Heavy-Metal Transport and Retention in Cementitious Gel Nanopores: A Molecular Dynamics Study

Weiqiang Chen[a,b,c], Qiyao He[a,b,c], Kai Gong[a,b,c,*]

[a] Department of Civil and Environmental Engineering, Rice University, Houston, Texas 77005, United States

[b] Rice Advanced Materials Institute, Rice University, Houston, Texas 77005, United States

[c] Ken Kennedy Institute, Rice University, Houston, Texas 77005, United States

* Corresponding author. E-mail: kg51@rice.edu

## Section S1 The potential parameters for the sodium aluminosilicate (N–A–S) glass

Based on the empirical interatomic potential model of Pedone et al. [1], the total energy of the system is expressed as:

$$E_{\text{Total}} = E_{\text{Morse}} + E_{\text{Coul}} + E_{\text{Repulsive}}, \tag{S1}$$

where, $E_{\text{Morse}}$, $E_{\text{Coul}}$, and $E_{\text{Repulsive}}$ represent the short-range Morse potential, the long-range Coulombic interaction, and an additional repulsive term relevant under high-temperature and high-pressure conditions, respectively.

The bonded interactions in covalent systems are described by the Morse potential:

$$E_{\text{Morse}} = \sum M_{0,ij}\left[\left\{1 - e^{-m_{0,ij}(r_{ij}-r_{0,ij})}\right\}^2 - 1\right], \tag{S2}$$

where, $r_{ij}$ is the interatomic distance between species $i$ and species $j$, $r_{0,ij}$ the equilibrium bond distance/length, and $M_{0,ij}$ and $m_{0,ij}$ empirical parameters.

The electrostatic interaction is given by:

$$E_{\text{Coul}} = \frac{e^2}{4\pi\varepsilon_0}\sum\frac{q_i q_j}{r_{ij}}, \tag{S3}$$

where, *e* is the electronic charge, $\varepsilon_0$ the dielectric permititivity of vacuum ($8.85419 \times 10^{-12}\,\text{F/m}$), $q_i$ and $q_j$ the partial charges of species $i$ and $j$, respectively.

The short-range repulsion term is defined as:

$$E_{\text{Repulsive}} = \sum\frac{C_{0,ij}}{r^{12}}, \tag{S4}$$

where, $C_{0,ij}$ is the empirical potential parameter. The corresponding pairwise parameters adopted in this work are summarized in **Table S1**, and a cutoff radius of 1.5 nm was applied for the short-range interactions.

**Table S1.** Interatomic potential parameters for the sodium aluminosilicate (N–A–S) glass.

| Species $i$ ($q_i$) | Species $j$ ($q_j$) | $M_{0,ij}$ (eV) | $m_{0,ij}$ (Å$^{-2}$) | $r_{0,ij}$ (Å) | $C_{0,ij}$ (eVÅ$^{12}$) |
|---|---|---|---|---|---|
| Na (+0.6) | O (−1.2) | 0.023363 | 1.763867 | 3.006315 | 5.0 |
| Al (+1.8) | O (−1.2) | 0.361581 | 1.900442 | 2.164818 | 0.9 |
| Si (+2.4) | O (−1.2) | 0.340554 | 2.006700 | 2.100000 | 1.0 |
| O (−1.2) | O (−1.2) | 0.042395 | 1.379316 | 3.618701 | 22.0 |

## Section S2 The structural characteristics of the generated N−A−S glass

Following the "melt-and-quench" procedure described in the main text, the final atomic structure of the N–A–S glass is shown in **Figure S1**a.

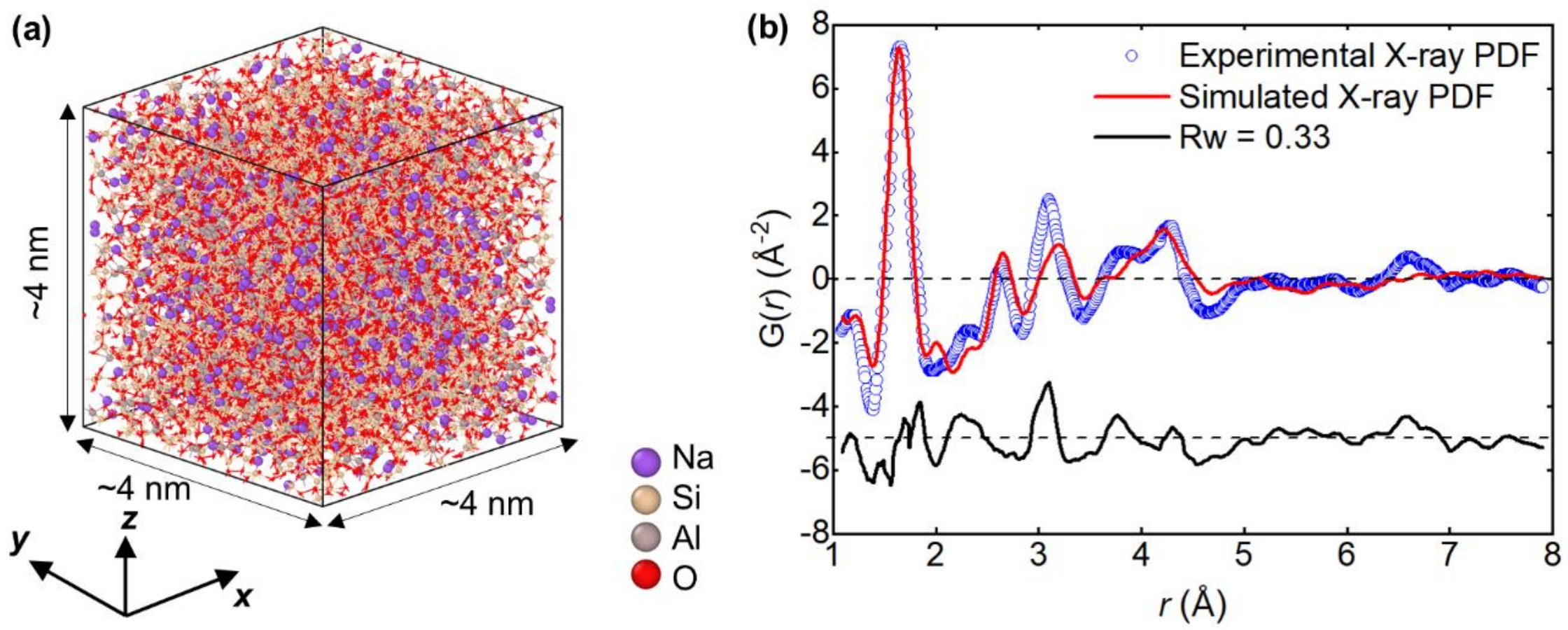

**Figure S1**. (a) The generated N–A–S glass model. (b) Comparison between the experimental X-ray pair distribution function (PDF) reported by White et al. [2] and the simulated X-ray PDF calculated from the model in (a).

The pair distribution function (PDF), $G(r)$, of the generated glass structure was calculated using **PDFgui** [3] and compared with the experimental X-ray PDF data reported by White et al. [2] for a high-alkali silicate-activated metakaolin-based geopolymer binder after 128 days of curing, with a gel composition of $(SiO_2)_3(Al_2O_3)(Na_2O)$. The parameters used for the data set configuration were $Q_{max} = 20Å^{-1}$ and an instrumental damping parameter $Q_{damp} = 0.05$. The refined parameters in the phase configuration included the scale factor, "delta2," and lattice length parameters. The isotropic atomic displacement parameters were fixed at $u_{ii} = 0.003Å^2$. The resulting simulated PDF, shown in **Figure S1**b, exhibits good agreement with the experimental data, confirming that the generated N–A–S glass structure captures the essential short- and medium-range order features of the real geopolymer gel. It should be noted that the experimental sample possesses a higher Al/Si ratio than the simulated glass, which accounts for the observed discrepancies.

The PDF comparison robustly validates the short-range coordination motifs (Si–O/Al–O) that govern the adsorption environments discussed in this study. Differences at larger $r$ likely arise from composition and heterogeneity differences between the simulated glass and the experimental geopolymer, and complementary metrics (e.g., $Q^n$ and ring statistics) will be considered in future work to further assess medium-range connectivity.

**Section S3 $Q^n$ species and mean silicate chain length (MCL) of different gel models**

The $Q^n$ notation characterizes the connectivity of silicate tetrahedra, where (n) denotes the number of bridging oxygen atoms connecting a given $SiO_4$ tetrahedron to neighboring silicate or aluminate tetrahedra. Accordingly, $Q^0$ represents an isolated silicate monomer, $Q^1$ represents a chain-end silicate tetrahedron, $Q^2$ represents a silicate tetrahedron within a linear chain, $Q^3$ represents a branched silicate tetrahedron connected to three neighboring tetrahedra, and $Q^4$ represents a fully polymerized silicate tetrahedron connected to four neighboring silicate or aluminate tetrahedra. For Al-containing structures, the notation $Q^n$((m)Al) may be used when it is necessary to specify that (m) of the neighboring tetrahedra are aluminate units.

For the predominantly chain-like C–S–H and C–(N)–A–S–H models, the mean silicate chain length (MCL) was estimated from the relative populations of $Q^1$ and $Q^2$ silicate species using $\mathrm{MCL} \approx 2 \times (1 + Q^2/Q^1)$ [4]. The calculated $Q^n$ distributions and MCL values are summarized in **Table S2**. The MCL of the C–S–H model with Ca/Si = 1.67 is approximately 3.31, consistent with the range of 3.14–3.47 reported by Ihaddadene et al. [4] for C–S–H models with Ca/Si ratios of 1.6–1.8. The calculated $Q^n$ distribution is also consistent with the $^{29}$Si magic-angle-spinning (MAS) NMR observations reported by Cong and Kirkpatrick [5], which showed that C–S–H gels are dominated by $Q^1$ and $Q^2$ species, with $Q^1$ being more abundant than $Q^2$ at a Ca/Si ratio of approximately 1.7.

For comparison, a nominal MCL was also calculated for the N–A–S–H model using the same expression and is reported in **Table S2**. This value is higher than that of C–S–H. However, because N–A–S–H contains a highly branched and cross-linked aluminosilicate network with substantial $Q^3$ and $Q^4$ populations, the conventional $Q^1$–$Q^2$-based MCL expression does not provide a rigorous measure of its network size. The reported value should therefore be interpreted only as an approximate comparative descriptor. Among the investigated models, N–A–S–H exhibits the largest combined fraction of $Q^3$ and $Q^4$ species, indicating the highest degree of network polymerization.

**Table S2.** The $Q^n$ species and mean chain length (MCL) of different gel models.

| Gel type | $Q^0$ (%) | $Q^1$ (%) | $Q^2$ (%) | $Q^3$ (%) | $Q^4$ (%) | MCL |
|---|---|---|---|---|---|---|
| C–(N)–A–S–H (Al/Si = 0) | 0 | 0 | 100 | 0 | 0 | Infinity |
| C–(N)–A–S–H (Al/Si = 0.059) | 0 | 0 | 100 | 0 | 0 | Infinity |
| C–(N)–A–S–H (Al/Si = 0.125) | 0 | 0 | 100 | 0 | 0 | Infinity |
| C–S–H (Ca/Si = 1.67) | 0 | 60.38 | 39.62 | 0 | 0 | 3.31 |
| N–A–S–H (Al/Si = 0.33) | 0 | 0.76 | 6.20 | 26.14 | 66.90 | 18.32 |

## Section S4 The temporal evolution of the system's potential energy during the relaxation stage

**Figure S2** shows the evolution of potential energy during the relaxation stage for all simulated systems. The energy profiles indicate that equilibrium was reached within 8 ns.

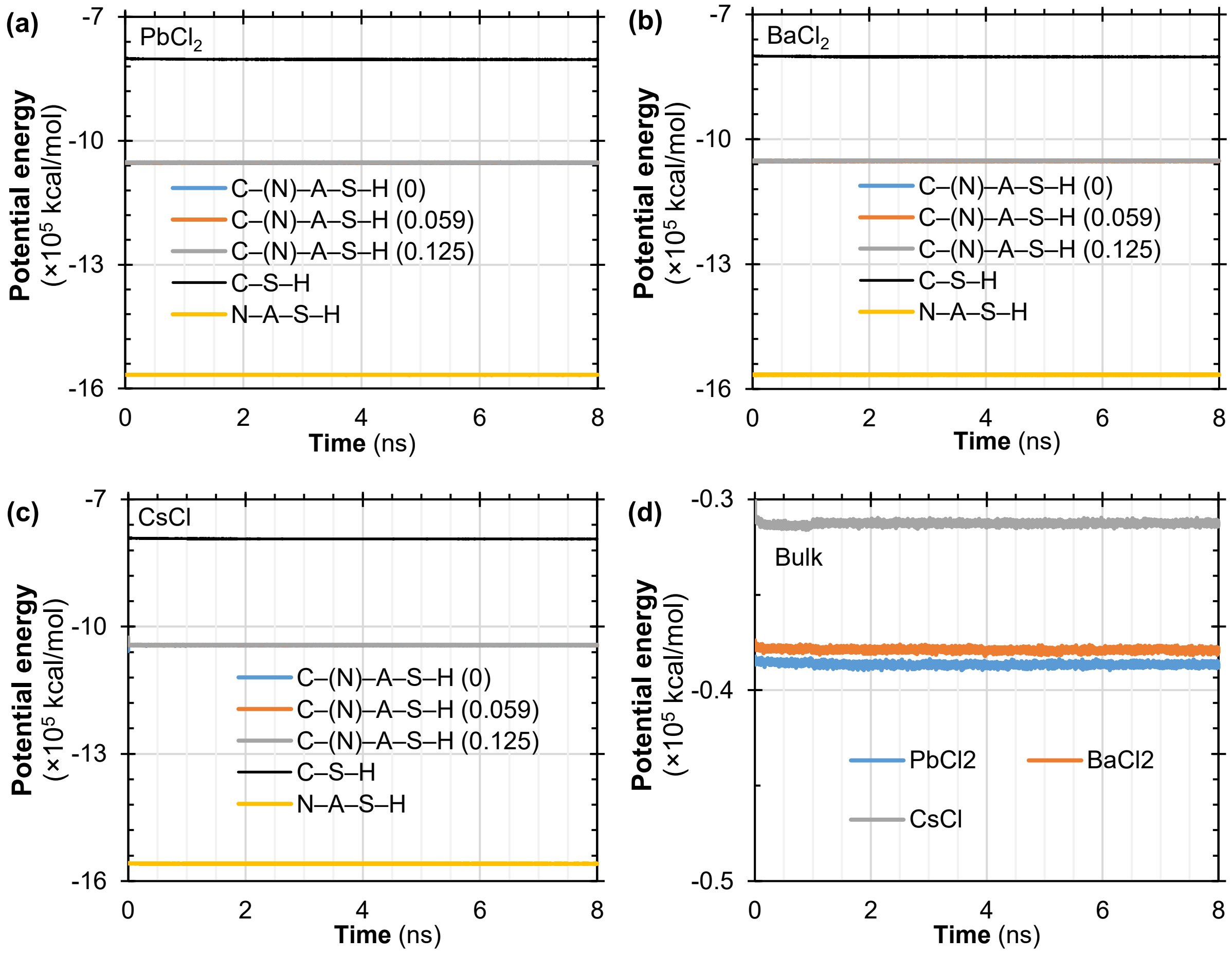

**Figure S2.** The temporal evolution of the system's potential energy during the relaxation stage for (a) nanoconfined $PbCl_2$ solutions, (b) nanoconfined $BaCl_2$ solutions, (c) nanoconfined CsCl solutions, and (d) the corresponding bulk $PbCl_2$, $BaCl_2$, and CsCl solutions.

## Section S5 The potential parameters of the ClayFF force field

The total potential energy of the simulated system, as described by the ClayFF force field [6], is expressed as,

$$E_{\text{Total}} = E_{\text{vdW}} + E_{\text{Coul}} + E_{\text{Bond}} + E_{\text{Angle}}, \quad \text{(S5)}$$

where, $E_{\text{vdW}}$, $E_{\text{Coul}}$, $E_{\text{Bond}}$, and $E_{\text{Angle}}$ represent the contributions from van der Waals, electrostatic, bond-stretching, and angle-bending interactions, respectively.

The short-range dispersion and repulsion energies are evaluated using the conventional Lennard–Jones 12-6 potential:

$$E_{\text{vdW}} = \sum 4D_{ij}\left[\left(\frac{R_{0,ij}}{r_{ij}}\right)^{12} - \left(\frac{R_{0,ij}}{r_{ij}}\right)^{6}\right], \quad \text{(S6)}$$

where, $r_{ij}$ is the interatomic distance between atoms $i$ and $j$, while $D_{ij}$ and $R_{0,ij}$ are the well-depth and equilibrium separation parameters, respectively. For identical atom types, these parameters are listed in **Table S3**. For dissimilar pairs, Lorentz–Berthelot mixing rules are applied:

$$D_{ij} = \sqrt{D_{ii}D_{jj}}, \qquad R_{0,ij} = \frac{1}{2}(R_{0,ii} + R_{0,jj}), \quad \text{(S7)}$$

Long-range Coulombic interactions are computed as,

$$E_{\text{Coul}} = \frac{e^2}{4\pi\varepsilon_0}\sum\frac{q_i q_j}{r_{ij}}, \quad \text{(S8)}$$

where, $e$ is the elementary charge, $\varepsilon_0 = 8.85419 \times 10^{-12}\text{F/m}$ is the vacuum permittivity, and $q_i$, $q_j$ are the partial atomic charges derived from quantum-mechanical calculations.

Bond stretching and angular distortions within water molecules and hydroxyl groups are modelled harmonically as,

$$E_{\text{Bond}} = \sum B_{ij}(r_{ij} - r_{0,ij})^2, \quad \text{(S9)}$$

$$E_{\text{Angle}} = \sum A_{ijk}(\theta_{ijk} - \theta_{0,ijk})^2, \quad \text{(S10)}$$

where, $B_{ij}$ and $A_{ijk}$ are the corresponding force constants, and $r_{0,ij}$, $\theta_{0,ijk}$ denote the equilibrium bond length and bond angle, respectively. The complete parameter sets employed in this study are summarized from **Table S3** to **Table S5**.

Interatomic interaction/bond/coordination strengths between different atomic species (e.g., Pb–Ow) can be quantified using the total energy expression (**Eq. (S5)**) or its first derivative (force-based approach). The equilibrium bond distances (i.e., the nearest interatomic distances) used in these calculations were taken from the position of the first maximum of the corresponding radial distribution functions (RDFs) as $r_{ij}$.

For reference and the convenience of readers, additional ClayFF parameters applicable to other heavy-metal species not explicitly investigated in this work are provided in **Table S3**.

**Table S3.** Nonbonded parameters for van der Waals and electrostatic interactions.

| **Species** $i$ | **Relative Atomic Mass** | $q_i$ $(e)$ | $D_{ii}$ **(kcal/mol)** | $R_{0,ii}$ **(Å)** |
|---|---|---|---|---|
| Intralayer calcium (Cai) | 40.08 | +1.05 | $5.0219 \times 10^{-6}$ | 5.5624 |
| Aqueous interlayer calcium ion (Ca) [7] | 40.08 | +2 | 0.1000 | 2.8720 |
| Tetrahedral silicon (St) | 28.0855 | +2.1000 | $1.8405 \times 10^{-6}$ | 3.3020 |
| Tetrahedral aluminum (At) | 26.98154 | +1.5750 | $1.8405 \times 10^{-6}$ | 3.7064 |
| Bridging oxygen (Ob) | 15.9994 | −1.0500 | 0.1554 | 3.1655 |
| Bridging oxygen with tetrahedral substitution (Obts) | 15.9994 | −1.1688 | 0.1554 | 3.1655 |
| Bridging oxygen with octahedral substitution (Obos) | 15.9994 | −1.1808 | 0.1554 | 3.1655 |
| Water oxygen (Ow) | 15.9994 | −0.8200 | 0.1554 | 3.1655 |
| Water hydrogen (Hw) | 1.00797 | +0.4100 | 0.0000 | 0.0000 |
| Hydroxyl oxygen (Oh) | 15.9994 | −0.9500 | 0.1554 | 3.1655 |
| Hydroxyl oxygen with substitution (Ohs) | 15.9994 | −1.0808 | 0.1554 | 3.1655 |
| Hydroxyl hydrogen (Hh) | 1.00797 | +0.4250 | 0.0000 | 0.0000 |
| Aqueous sodium ion (Na) [8] | 22.99 | +1 | 0.1301 | 2.3500 |
| Aqueous potassium ion (K) | 39.1 | +1 | 0.1000 | 3.3340 |
| Aqueous magnesium ion (Mg) | 24.305 | +2 | 0.8750 | 1.6445 |
| Aqueous strontium ion (Sr) | 87.62 | +2 | 0.1000 | 3.4620 |
| Aqueous zinc ion (Zn) | 65.38 | +2 | 0.2500 | 1.9422 |
| Aqueous cadmium ion (Cd) | 112.414 | +2 | 0.0545 | 2.5370 |
| Aqueous lead ion (Pb) | 207.2 | +2 | 0.1182 | 3.3243 |
| Aqueous barium ion (Ba) [9] | 137.33 | +2 | 0.0471 | 3.8166 |
| Aqueous cesium ion (Cs) [10] | 132.91 | +1 | 0.1000 | 3.8310 |
| Aqueous chloride ion (Cl) [8] | 35.453 | −1 | 0.1001 | 4.4000 |

*For the calcium silicate hydrate (C–S–H) gel, minor adjustments to the partial atomic charges were made following Ref. [11] to ensure overall charge neutrality of the simulation system. The assigned partial charges were +0.4238$e$ for hydroxyl hydrogen atoms, −1.4238$e$ for hydroxyl oxygen atoms, and −1.164$e$ for the remaining oxygen atoms.

**Table S4.** Bond stretching parameters.

| **Species $i$** | **Species $j$** | **$B_{ij}$ (kcal/mol/Å$^2$)** | **$r_{0,ij}$ (Å)** |
|---|---|---|---|
| Ow | Hw | 553.9350 | 1.0000 |
| Oh | Hh | 553.9350 | 1.0000 |
| Ohs | Hh | 553.9350 | 1.0000 |

**Table S5.** Angle bending parameters.

| **Species $i$** | **Species $j$** | **Species $k$** | **$A_{ijk}$ (kcal/mol)** | **$\theta_{0,ijk}$ (Å)*** |
|---|---|---|---|---|
| Hw | Ow | Hw | 45.7530 | 109.4700 |

*In LAMMPS, angle values specified in degrees are automatically converted to radians for internal calculations.

## Section S6 Average diffusion coefficient of interlayer solution species

**Figure S3** shows the nanoconfinement effects of various gel surfaces on the diffusion coefficients of interlayer solution species, including heavy metal ions, chloride ions, and water molecules. **Figure S3**a and b show that the pore-averaged diffusivity of $Pb^{2+}$ and $Ba^{2+}$ ions follows the sequence C–(N)–A–S–H (Al/Si = 0) > N–A–S–H > C–S–H, whereas for $Cs^{+}$ ions it follows C–(N)–A–S–H (Al/Si = 0) > C–S–H > N–A–S–H. Increasing the Al/Si ratio to 0.059 and 0.125 in the C–(N)–A–S–H gels markedly reduces the diffusivity of heavy metal ions. Among all gel systems studied, the C–(N)–A–S–H gel with an Al/Si ratio of 0.125 exhibits the strongest immobilization capacity for all three heavy metal ions.

The pore-averaged diffusivities of $Cl^{-}$ and water decrease in the order C–(N)–A–S–H (Al/Si = 0) > N–A–S–H > C–S–H (**Figure S3**c–f). Increasing the Al/Si ratio produces only a marginal additional reduction in diffusivity.

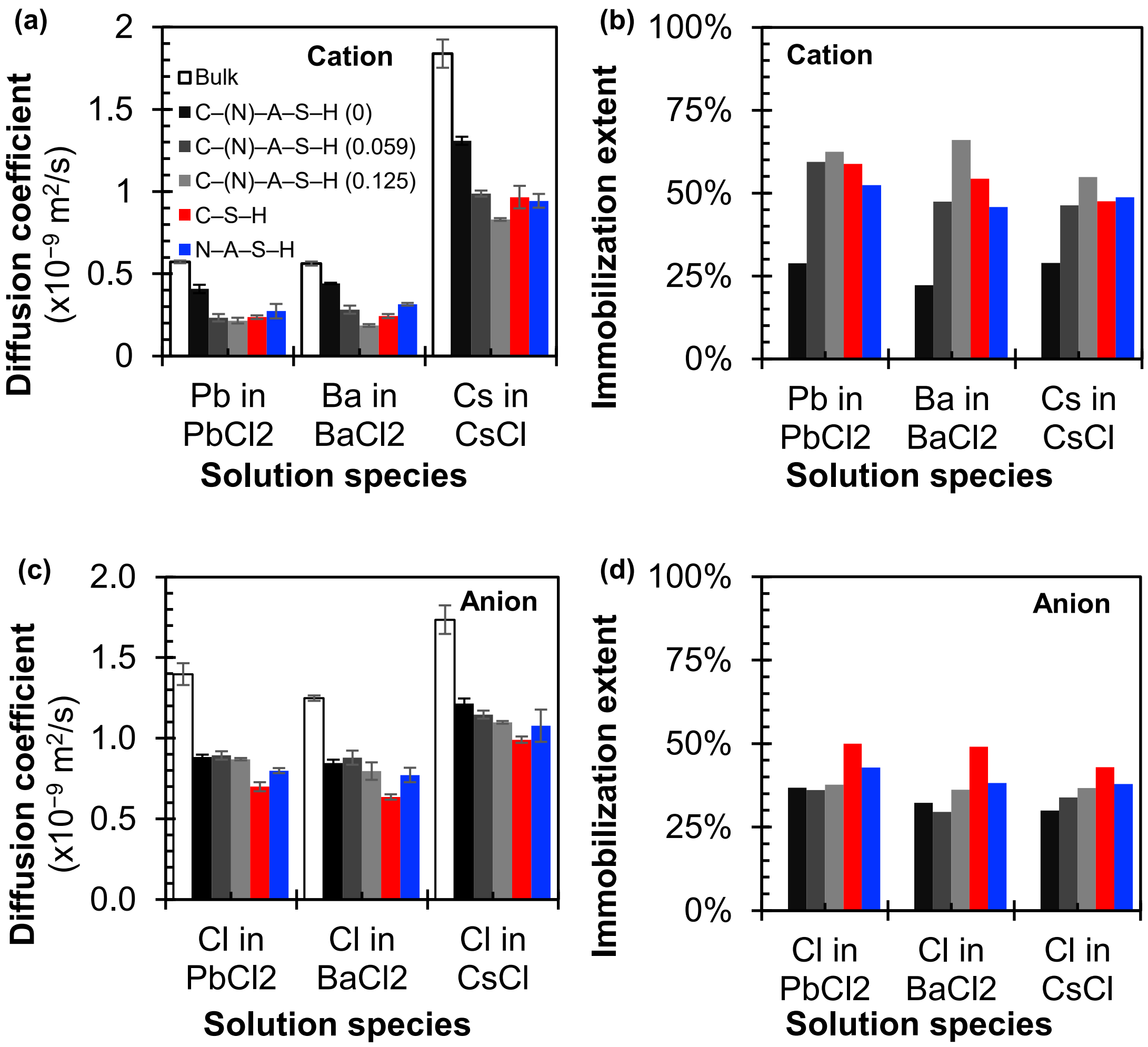

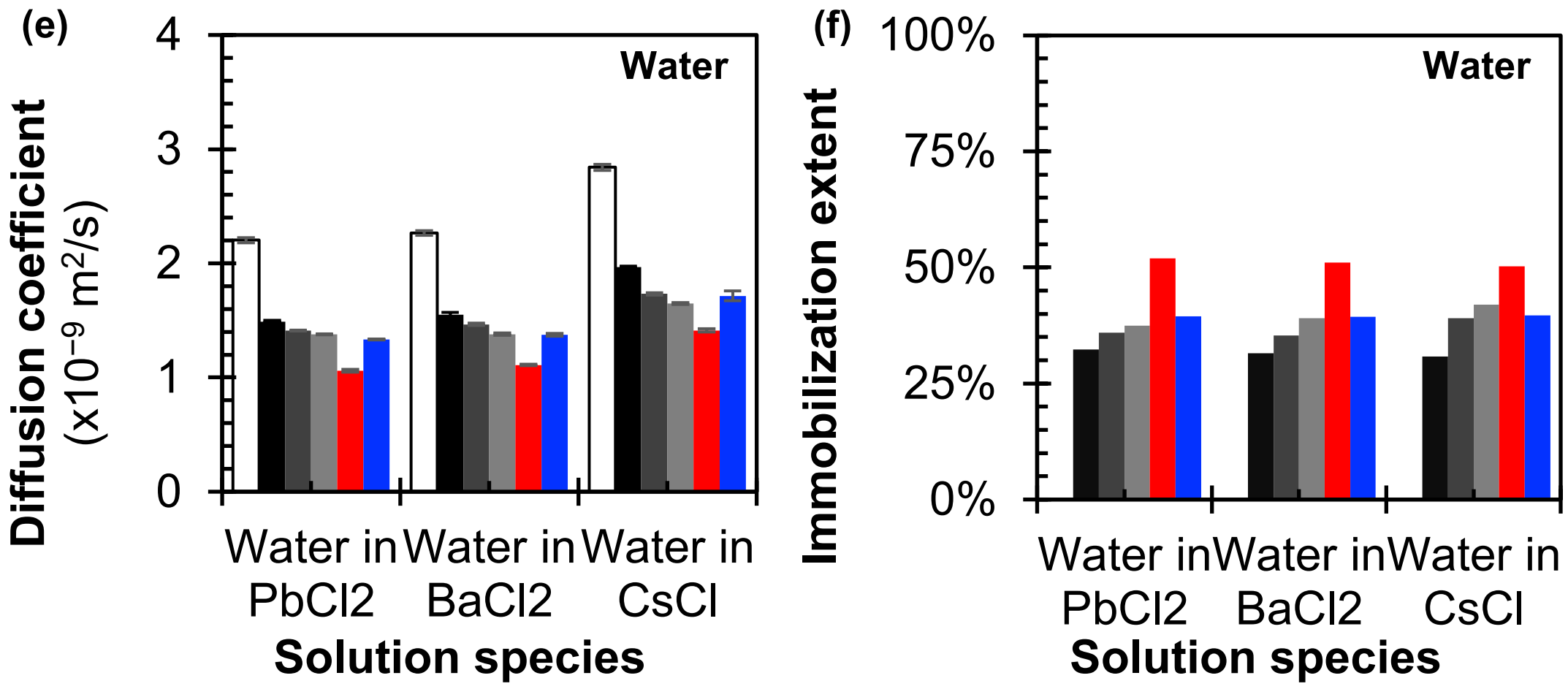

**Figure S3.** Pore-averaged diffusion coefficients of interlayer solution species confined within various gel nanopores, where (a, c, e) show the pore-scale averaged diffusion coefficients, and (b, d, f) depict the corresponding extent of immobilization. Error bars represent standard deviations from three independent simulations.

**Section S7 The spatial distribution of water and ion diffusivity inside different gel nanopores**

The spatial heterogeneity of water and ion mobility within the gel nanopore was evaluated using a binning procedure along the *z*-axis, following the methodology established in our earlier work [12]. The simulation box was partitioned into bins of uniform width $b$, each centered at position $z$ and spanning the range from $z - b/2$ to $z + b/2$. During molecular dynamics trajectories, water molecules and ions may transiently move in and out of each bin. For a particle $i$ that remains continuously within a bin for a period $\tau$ after entering at time $t_0$, the local diffusivity at position $z$ was computed as,

$$D(z) = \langle \frac{(\boldsymbol{r}_i(\tau+t_0)-\boldsymbol{r}_i(t_0))^2}{2d\tau} \rangle_{i,t_0,\tau;z_i(\tau+t_0)\in[z-b/2,z+b/2]}, \quad \text{(S11)}$$

where, $\langle\cdot\rangle_{i,t_0,\tau}$ denotes averaging over all relevant particles, time origins, and residence durations. Here $\boldsymbol{r}_i(t_0) = (x_i(t_0), y_i(t_0), z_i(t_0))$ and $\boldsymbol{r}_i(\tau + t_0) = (x_i(\tau + t_0), y_i(\tau + t_0), z_i(\tau + t_0))$ are the coordinates of particle $i$ at the beginning and end of the interval, respectively, and $d$ represents the dimensionality of diffusion (1, 2, or 3). In this work, $d = 3$ was adopted to represent three-dimensional mobility. The functional form of **Eq. (S11)** follows that of **Eq. (4)** in the main text, which ensures methodological consistency and allows direct demonstration of how local diffusivities integrate to yield pore-averaged values.

The analysis was implemented in MATLAB R2023a (The MathWorks, Inc.) [13] by following **Eq. (S11)** and post-processing unwrapped atomic trajectories exported from LAMMPS. Trajectory snapshots were saved every 8 ps for a total of 8 ns, yielding 1001 frames. The MATLAB routine iterates through all particles, time origins, and residence intervals in each spatial bin to accumulate mean-squared displacements that meet the residence criterion. Parallel processing was enabled using the parfor (parallel for loop) command in MATLAB's Parallel Computing Toolbox to accelerate computation.

It should be noted that this binning scheme primarily probes the in-plane (*x*–*y*) mobility parallel to the pore surface and does not explicitly capture motion perpendicular to the surface. The resulting spatially resolved diffusivity profiles of water and ions for each gel nanopore type are presented from **Figure S4** to **Figure S6**, which reflect the per-ion diffusivity inside different gel nanochannels.

A bin width of 1 Å was selected to balance spatial resolution and statistical reliability: smaller bins yield insufficient sampling for stable averaging and lead to pronounced fluctuations in the profiles, whereas larger bins smooth near-surface gradients, obscure interfacial features, and increase the computational overhead of trajectory analysis, particularly for rougher N–A–S–H surfaces. A representative example is shown in **Figure S5**e, demonstrating that although minor statistical variations occur with different bin sizes, the resulting profiles remain consistent in their overall distributions and qualitative trends. **Figure S5**f further demonstrates the trajectory-length convergence of the spatially resolved $Ba^{2+}$ diffusivity profiles in C–(N)–A–S–H (Al/Si = 0.125), calculated using cumulative trajectory intervals from 0–1 ns to 0–8 ns.

The analysis assumes normal Fickian diffusion. Because strongly adsorbed species near the gel surfaces may exhibit subdiffusive behavior [4, 12, 14], the spatially resolved values are interpreted as apparent local diffusion coefficients used primarily for relative comparison among pore regions and gel chemistries.

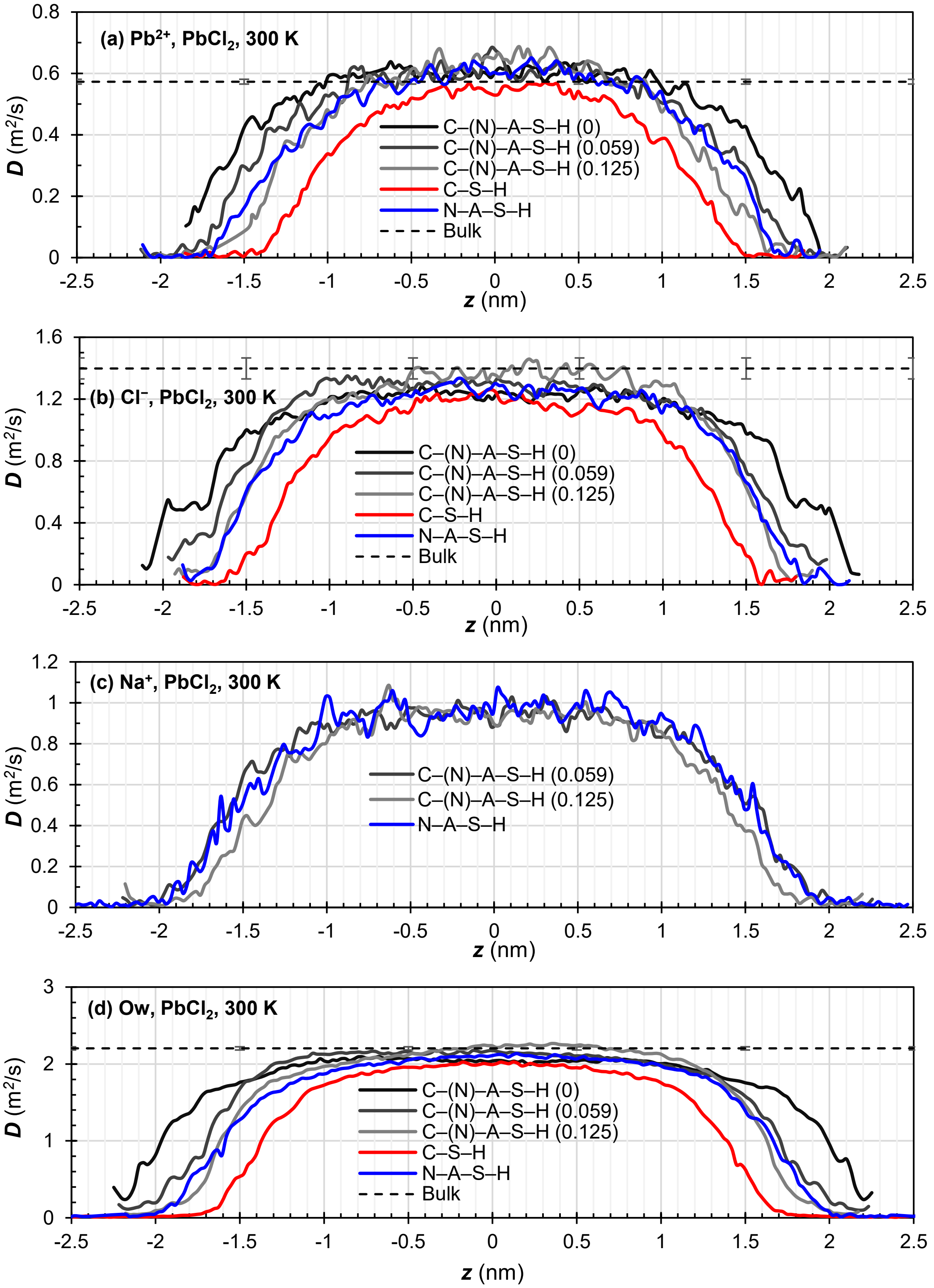

**Figure S4**. The spatial distribution of (a) heavy-metal cation diffusivity, (b) anion diffusivity, (c) possible counter-ion (Na) diffusivity, and (d) water (Ow) diffusivity of $PbCl_2$ solutions inside different gel nanopores, where a bin size of 1 Å is used in the *z* direction to produce the spatial distribution profiles. The dashed lines with error bars indicate the corresponding bulk values.

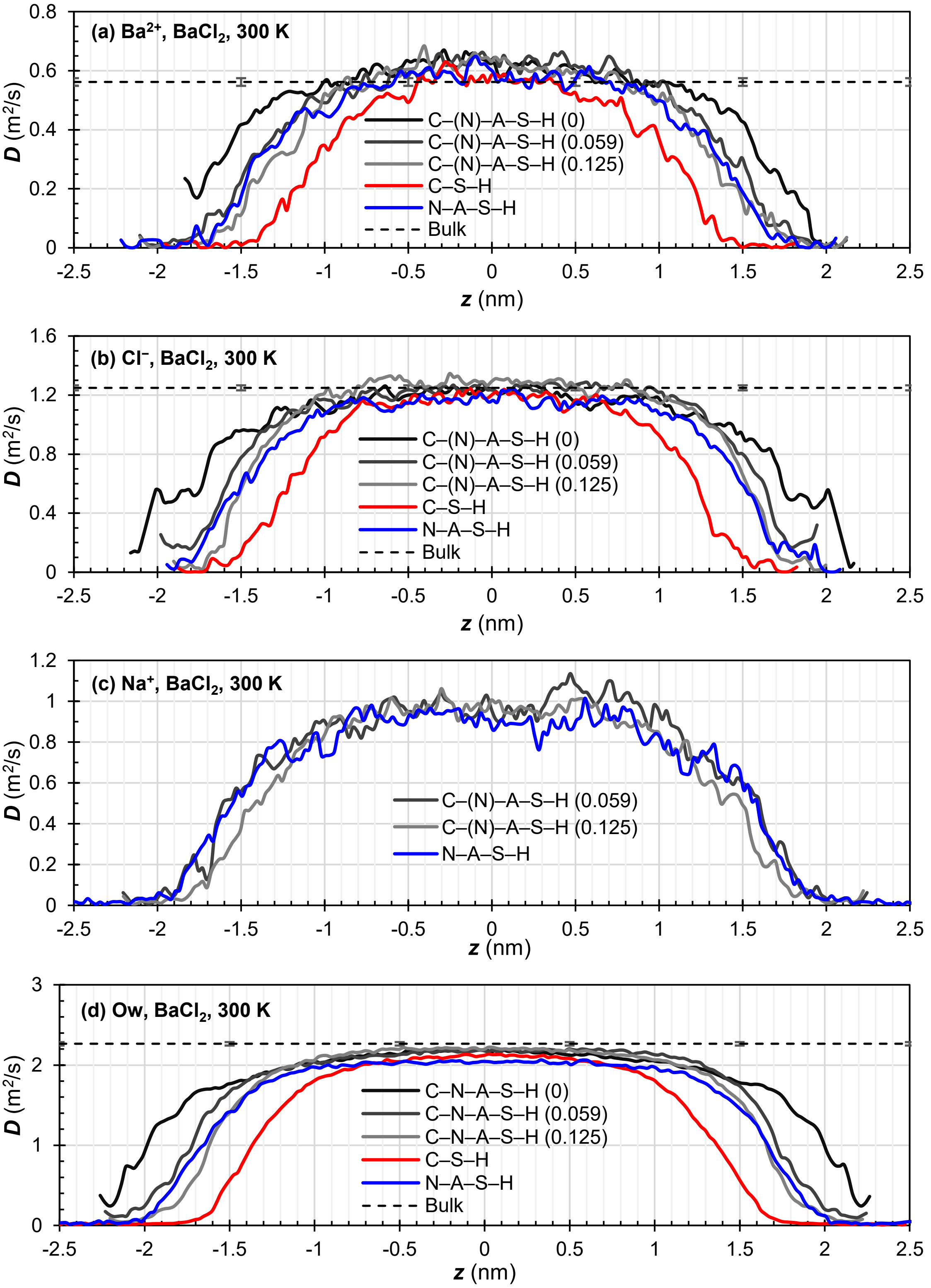
(a) Ba$^{2+}$, BaCl$_2$, 300 K
D (m$^2$/s)
z (nm)
C–(N)–A–S–H (0)
C–(N)–A–S–H (0.059)
C–(N)–A–S–H (0.125)
C–S–H
N–A–S–H
Bulk
(b) Cl$^-$, BaCl$_2$, 300 K
D (m$^2$/s)
z (nm)
C–(N)–A–S–H (0)
C–(N)–A–S–H (0.059)
C–(N)–A–S–H (0.125)
C–S–H
N–A–S–H
Bulk
(c) Na$^+$, BaCl$_2$, 300 K
D (m$^2$/s)
z (nm)
C–(N)–A–S–H (0.059)
C–(N)–A–S–H (0.125)
N–A–S–H
(d) Ow, BaCl$_2$, 300 K
D (m$^2$/s)
z (nm)
C–N–A–S–H (0)
C–N–A–S–H (0.059)
C–N–A–S–H (0.125)
C–S–H
N–A–S–H
Bulk

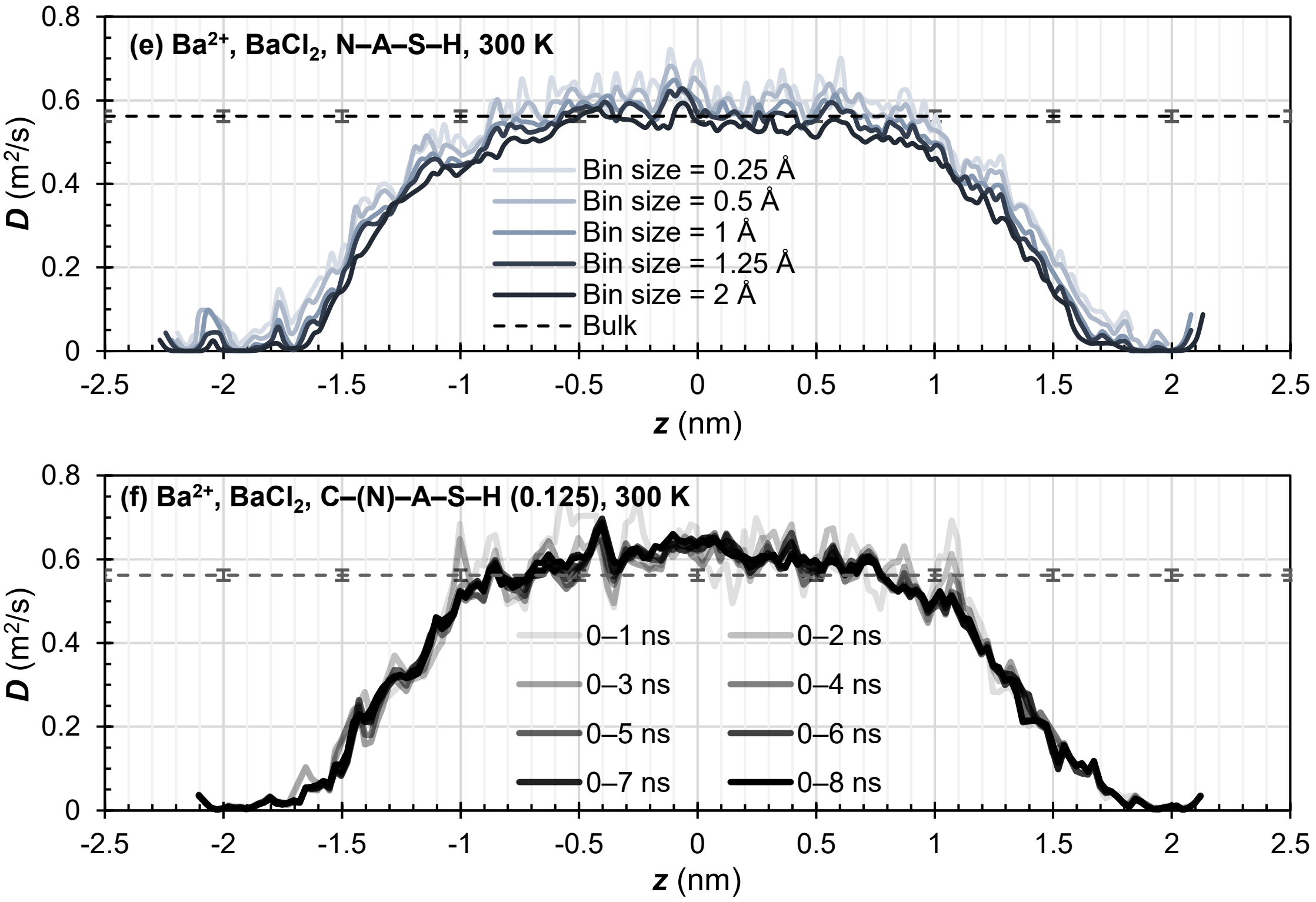

**Figure S5**. The spatial distribution of (a) heavy-metal cation diffusivity, (b) anion diffusivity, (c) possible counter-ion (Na) diffusivity, and (d) water (Ow) diffusivity of $BaCl_2$ solutions inside different gel nanopores, where a bin size of 1 Å is used in the *z* direction to produce the spatial distribution profiles. (e) Effect of bin size on the spatial distribution profiles of heavy-metal cation diffusivity in the N–A–S–H gel. (f) Convergence of the spatially resolved $Ba^{2+}$ diffusivity profiles in C–(N)–A–S–H (Al/Si = 0.125), calculated using cumulative trajectory intervals from 0–1 ns to 0–8 ns. The dashed lines with error bars indicate the corresponding bulk values.

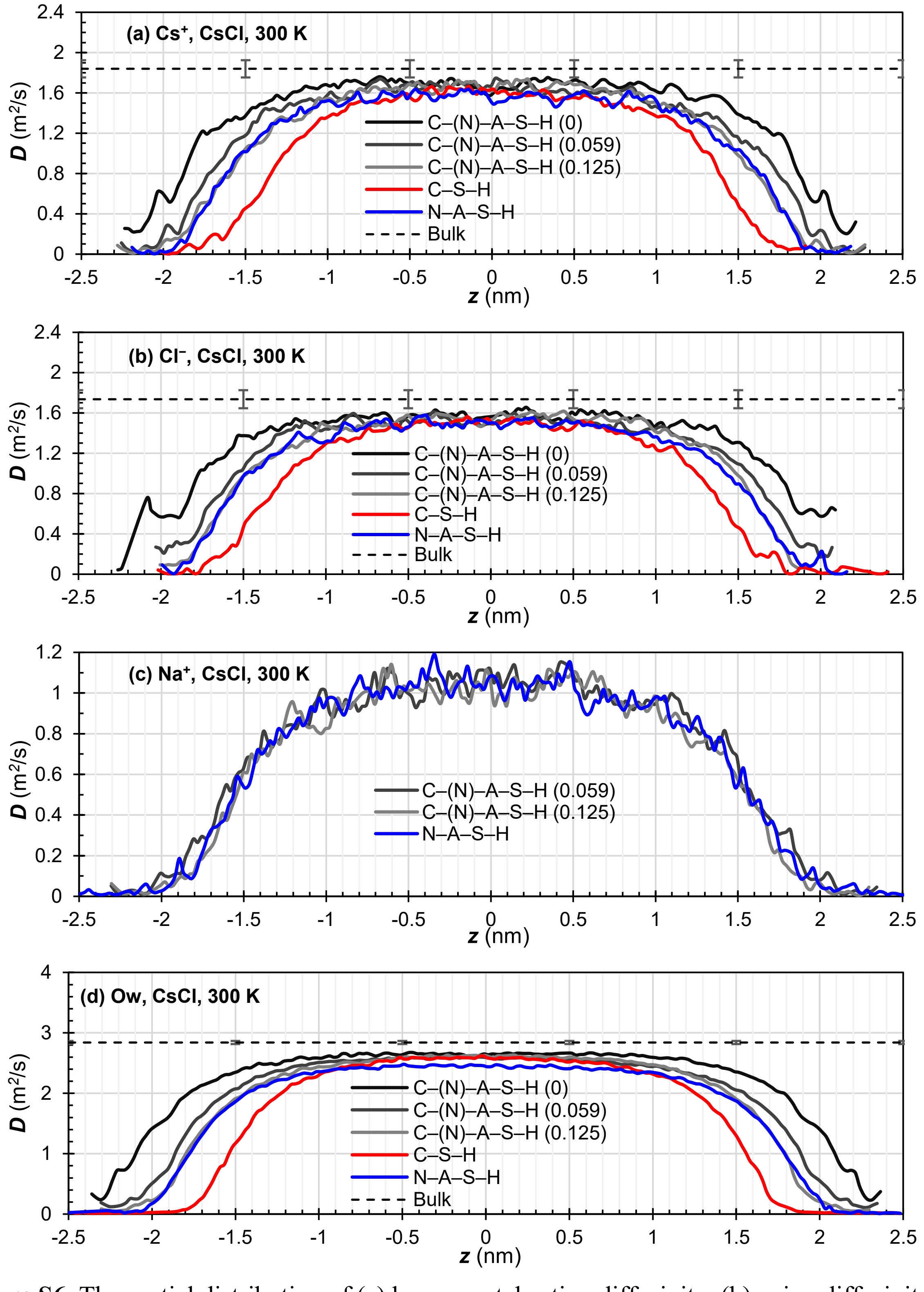

**Figure S6**. The spatial distribution of (a) heavy-metal cation diffusivity, (b) anion diffusivity, (c) possible counter-ion (Na) diffusivity, and (d) water (Ow) diffusivity of CsCl solutions inside different gel nanopores, where a bin size of 1 Å is used in the *z* direction to produce the spatial distribution profiles. The dashed lines with error bars indicate the corresponding bulk values.

**Determination of the nanochannel middle plane for aligning different nanochannels.** Due to the different gel surfaces used in this work, the nanochannels are aligned by setting their middle plane at $z = 0$ nm. The middle plane is determined by integrating the number density profile of the interlayer water oxygen (Ow) and identifying the plane that separates half of the total number of water oxygens. The corresponding positions of the middle planes for each system are listed in **Table S6** and are used to align various distribution profiles.

**Table S6.** Position of the nanochannel middle plane (nm) for each case.

| Simulation case | Gel type | Solution type | Number of atoms | Position of the nanochannel middle plane (nm) |
|---|---|---|---|---|
| Case 1 | C–(N)–A–S–H (Al/Si = 0) | 1 M $PbCl_2$ | 13200 | 3.623425 |
| Case 2 | C–(N)–A–S–H (Al/Si = 0.59) | 1 M $PbCl_2$ | 13251 | 3.61777 |
| Case 3 | C–(N)–A–S–H (Al/Si = 0.125) | 1 M $PbCl_2$ | 13263 | 3.65352 |
| Case 4 | C–S–H (Ca/Si = 1.67) | 1 M $PbCl_2$ | 12363 | 3.527355 |
| Case 5 | N–A–S–H (Al/Si = 0.33) | 1 M $PbCl_2$ | 11995 | 4.031875 |
| Case 6 | C–(N)–A–S–H (Al/Si = 0) | 1 M $BaCl_2$ | 13200 | 3.63615 |
| Case 7 | C–(N)–A–S–H (Al/Si = 0.59) | 1 M $BaCl_2$ | 13251 | 3.655915 |
| Case 8 | C–(N)–A–S–H (Al/Si = 0.125) | 1 M $BaCl_2$ | 13263 | 3.628515 |
| Case 9 | C–S–H (Ca/Si = 1.67) | 1 M $BaCl_2$ | 12363 | 3.527375 |
| Case 10 | N–A–S–H (Al/Si = 0.33) | 1 M $BaCl_2$ | 11995 | 4.01849 |
| Case 11 | C–(N)–A–S–H (Al/Si = 0) | 2 M CsCl | 13247 | 3.63583 |
| Case 12 | C–(N)–A–S–H (Al/Si = 0.59) | 2 M CsCl | 13298 | 3.630745 |
| Case 13 | C–(N)–A–S–H (Al/Si = 0.125) | 2 M CsCl | 13310 | 3.654675 |
| Case 14 | C–S–H (Ca/Si = 1.67) | 2 M CsCl | 12411 | 3.516865 |
| Case 15 | N–A–S–H (Al/Si = 0.33) | 2 M CsCl | 12033 | 3.991405 |
| Case 16 | Bulk | 1 M $PbCl_2$ | 6195 | – |
| Case 17 | Bulk | 1 M $BaCl_2$ | 6195 | – |
| Case 18 | Bulk | 2 M CsCl | 6232 | – |

**Section S8 The spatial evolution of the diffusion coefficient ratio for water/cation, water/anion, and anion/cation**

Using the datasets presented in **Section S7**, the spatial distributions of diffusion-coefficient ratios for different solution-species pairs in nanoconfined $PbCl_2$, $BaCl_2$, and CsCl solutions were evaluated across various cementitious gels, as shown in **Figure S7**, **Figure S8**, and **Figure S9**, respectively. The corresponding bulk ratios are indicated by horizontal dashed lines for reference. Only the half-pore region ($-2.5$ nm < $z$ < 0 nm) is shown. The profiles were obtained by averaging the values from the two symmetric halves of the nanopore, with the 0 nm < $z$ < 2.5 nm half reflected about $z$ = 0 nm, and the error bars represent the standard deviation between the two symmetry-related values. When the diffusivity of the species in the denominator approaches zero, the calculated ratio can become spuriously large and highly sensitive to small fluctuations in diffusivity. This effect is reflected by the large error bars observed in the diffusion-coefficient ratio profiles near the pore surface.

**Figure S7** shows that near the pore surfaces, the $H_2O/Pb^{2+}$ and $H_2O/Cl^-$ ratios are substantially higher than in the bulk, indicating that ionic motion is more strongly hindered than that of water due to pronounced ion–surface interactions. Toward the channel center, the $H_2O/Pb^{2+}$ ratio drops slightly below the bulk value, suggesting that $Pb^{2+}$ mobility becomes relatively enhanced once interfacial effects weaken. Comparison between the two ionic species shows a $Cl^-/Pb^{2+}$ ratio above its bulk value near the surface, signifying that $Pb^{2+}$ experiences stronger adsorption than $Cl^-$.

Comparable spatial trends are observed for $BaCl_2$ and CsCl systems (**Figure S8** and **Figure S9**), independent of gel chemistry. A notable exception occurs for the $Cl^-/Cs^+$ ratio in CsCl, which drops below its bulk value near the surface, reflecting stronger immobilization of $Cl^-$ relative to $Cs^+$.

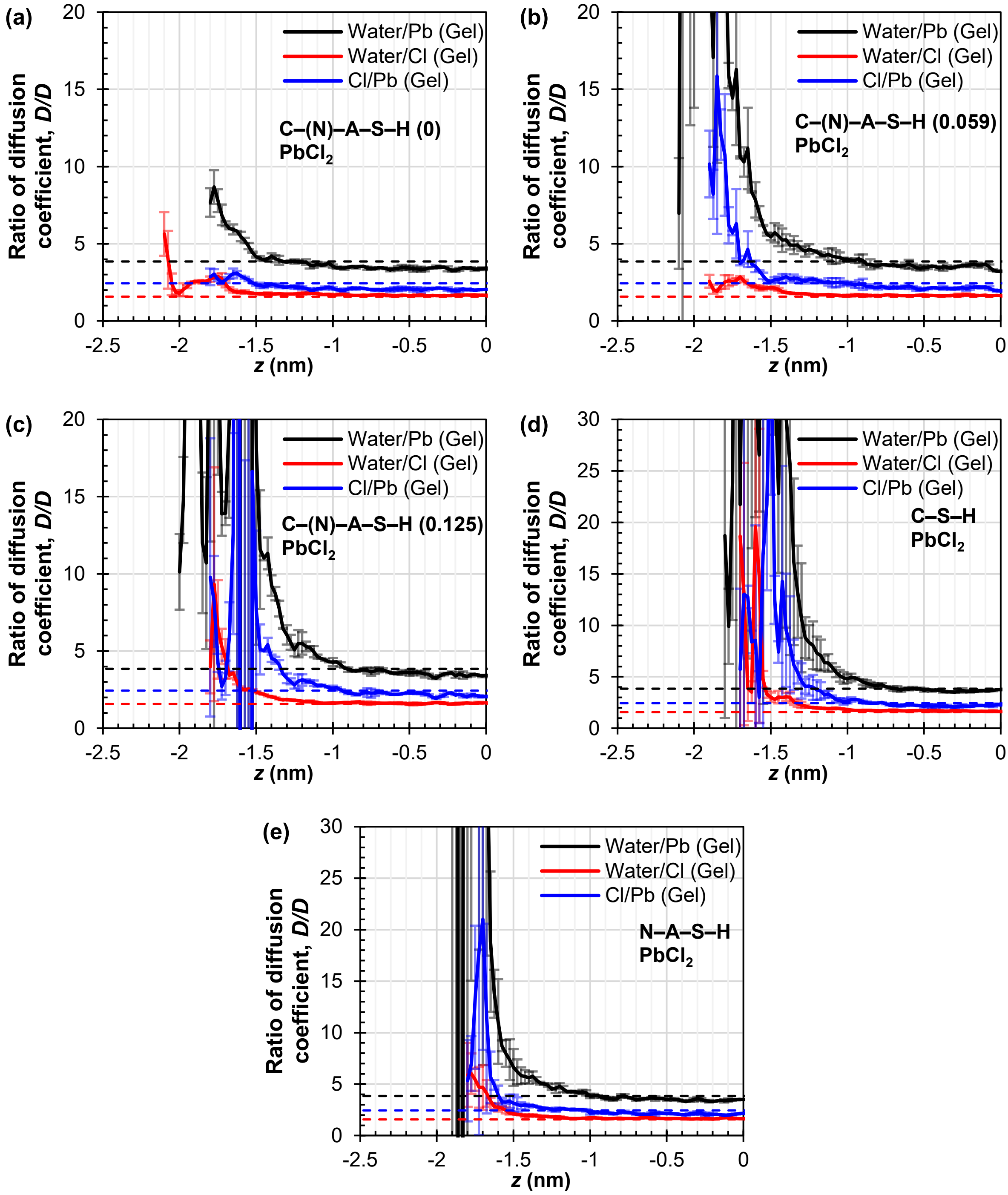

**Figure S7.** Evolution of the diffusion coefficient ratio for water/Pb, water/Cl, and Cl/Pb across the nanopore for different gel nanopores: (a) C–(N)–A–S–H (0), (b) C–(N)–A–S–H (0.059), (c) C–(N)–A–S–H (0.125), (d) C–S–H, and (e) N–A–S–H. Only the half-pore region (−2.5 nm < $z$ < 0 nm) is shown. The profiles were obtained by averaging the values from the two symmetric halves of the nanopore, with the 0 nm < $z$ < 2.5 nm half reflected about $z$ = 0 nm, and the error bars represent the standard deviation between the two symmetry-related values. Dashed lines in the figures represent the corresponding values for the bulk solution.

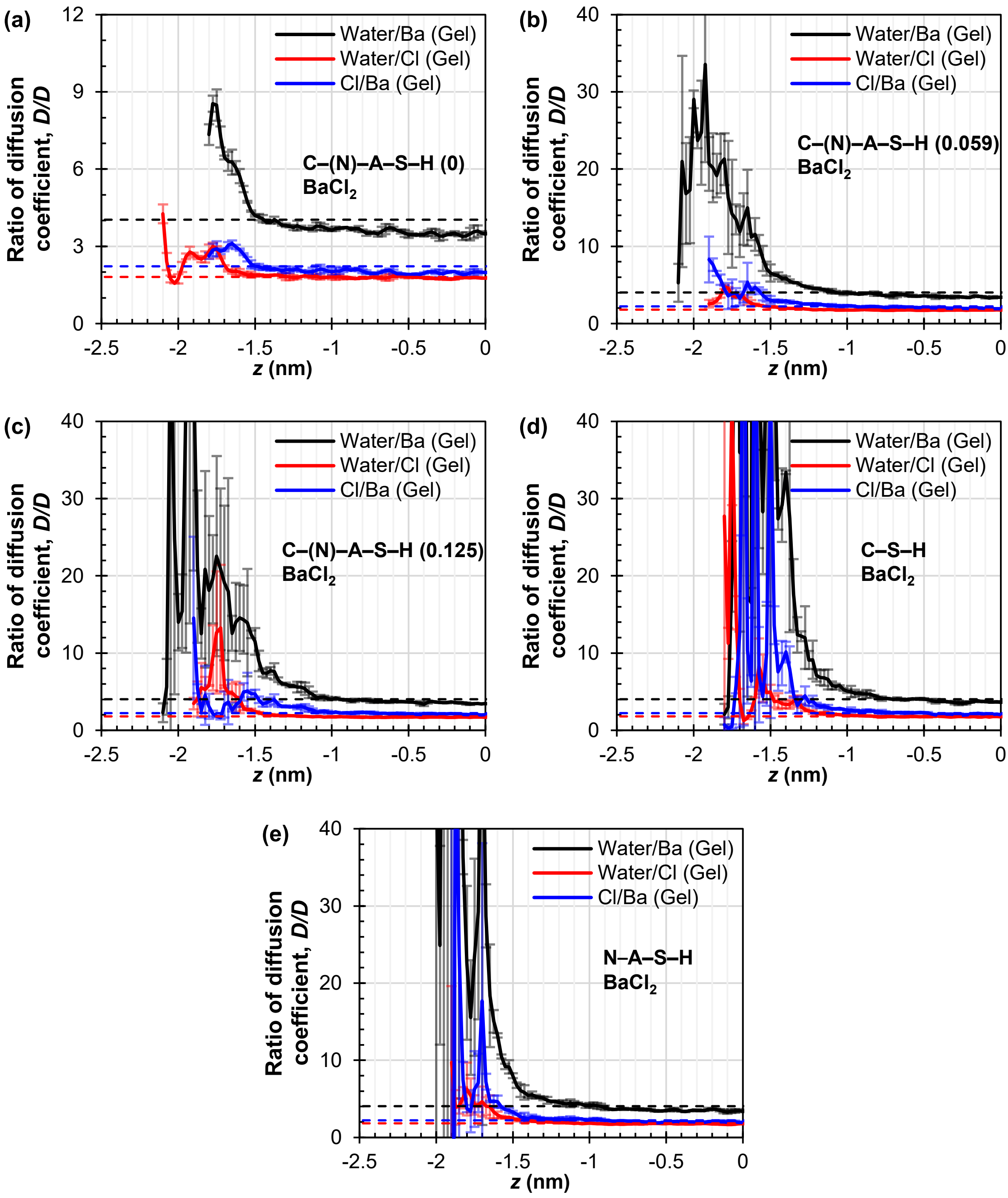

**Figure S8.** Evolution of the diffusion coefficient ratio for water/Ba, water/Cl, and Cl/Ba across the nanopore for different gel nanopores: (a) C–(N)–A–S–H (0), (b) C–(N)–A–S–H (0.059), (c) C–(N)–A–S–H (0.125), (d) C–S–H, and (e) N–A–S–H. Only the half-pore region ($-2.5$ nm < $z$ < 0 nm) is shown. The profiles were obtained by averaging the values from the two symmetric halves of the nanopore, with the 0 nm < $z$ < 2.5 nm half reflected about $z$ = 0 nm, and the error bars represent the standard deviation between the two symmetry-related values. Dashed lines in the figures represent the corresponding values for the bulk solution.

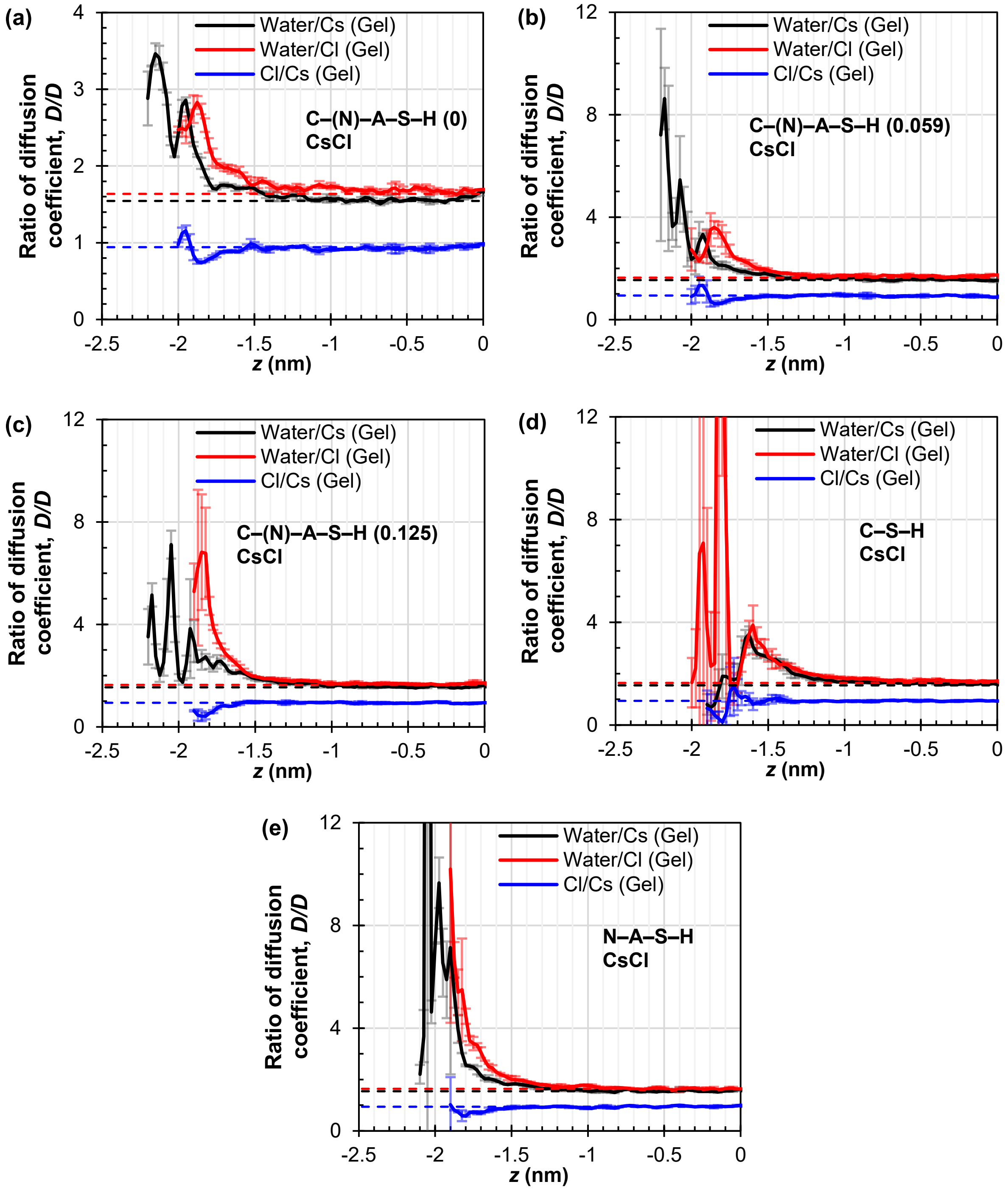

**Figure S9.** Evolution of the diffusion coefficient ratio for water/Cs, water/Cl, and Cl/Cs across the nanopore for different gel nanopores: (a) C–(N)–A–S–H (0), (b) C–(N)–A–S–H (0.059), (c) C–(N)–A–S–H (0.125), (d) C–S–H, and (e) N–A–S–H. Only the half-pore region (−2.5 nm < $z$ < 0 nm) is shown. The profiles were obtained by averaging the values from the two symmetric halves of the nanopore, with the 0 nm < $z$ < 2.5 nm half reflected about $z$ = 0 nm, and the error bars represent the standard deviation between the two symmetry-related values. Dashed lines in the figures represent the corresponding values for the bulk solution.

**Section S9 The spatial evolution of the residence time (RT) of different interlayer solution species inside various gel nanopores**

The residence time (RT) characterizes the duration that solution species remain at specific positions within the nanochannel, serving as a measure of their adsorption tendency and mobility.

To evaluate the spatial variation in RT, the simulation box was divided into bins of width ($b = 1$) Å along the $z$-direction. The bin centered at $z$ was defined as ($[z - \frac{b}{2}, z + \frac{b}{2}]$). Each continuous period during which a particle remained within a bin was treated as an individual residence event. The mean residence time at position ($z$) was calculated as

$$RT(z) = \frac{\sum_{i=1}^{N} \sum_{k=1}^{n_i(z)} [t_{i,k}^{\mathrm{out}}(z) - t_{i,k}^{\mathrm{in}}(z)]}{\sum_{i=1}^{N} n_i(z)}, \tag{S12}$$

where $t_{i,k}^{\mathrm{in}}(z)$ and $t_{i,k}^{\mathrm{out}}(z)$ are the entry and exit times, respectively, of particle $i$ during its ($k$)th residence event in the bin centered at ($z$), and $n_i(z)$ is the number of completed residence events for particle $i$.

The analysis was performed in MATLAB R2023a (The MathWorks, Inc.) [13] using unwrapped atomic trajectories generated by LAMMPS. Trajectories were recorded every 8 ps over 8 ns, yielding 1001 frames. Parallel processing was implemented using the parfor command in the MATLAB Parallel Computing Toolbox. A bin width of 1 Å was selected to balance spatial resolution and statistical reliability. The resulting RT distributions are presented in **Figure S10** for $PbCl_2$, **Figure S11** for $BaCl_2$, and **Figure S12** for CsCl solutions confined within different gel nanopores.

The results show that all investigated heavy-metal ions exhibit markedly longer residence times near the pore surface, which gradually approach bulk-like values toward the channel center. This near-wall RT enhancement mirrors the parabolic diffusivity profiles shown in **Figures S4–S6**, confirming that immobilization primarily arises from strong interfacial interactions.

The magnitude of RT enhancement near the surface varies systematically with gel chemistry. For all heavy-metal ions, the near-surface RT follows the trend C–S–H > N–A–S–H > C–(N)–A–S–H (Al/Si = 0). Increasing the Al/Si ratio in C–(N)–A–S–H further amplifies the RT near the pore surface. The corresponding RT profiles for $Cl^-$, $Na^+$, and water molecules, exhibit broadly similar spatial behaviors and compositional dependencies.

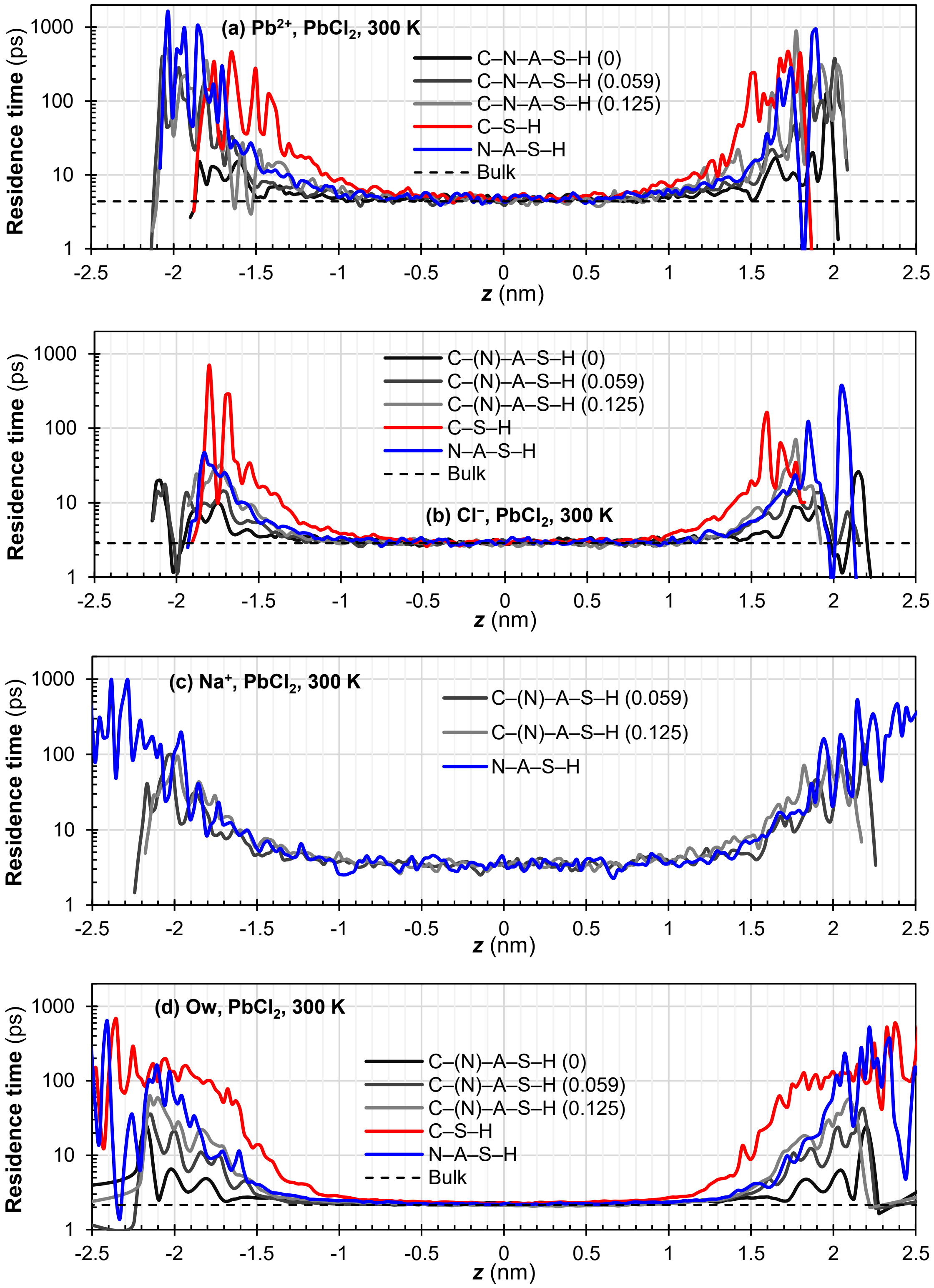

**Figure S10**. The residence time distribution of different solution species in $PbCl_2$ solution across various gel nanopores. Dashed lines in the figures represent the corresponding values for the bulk solution.

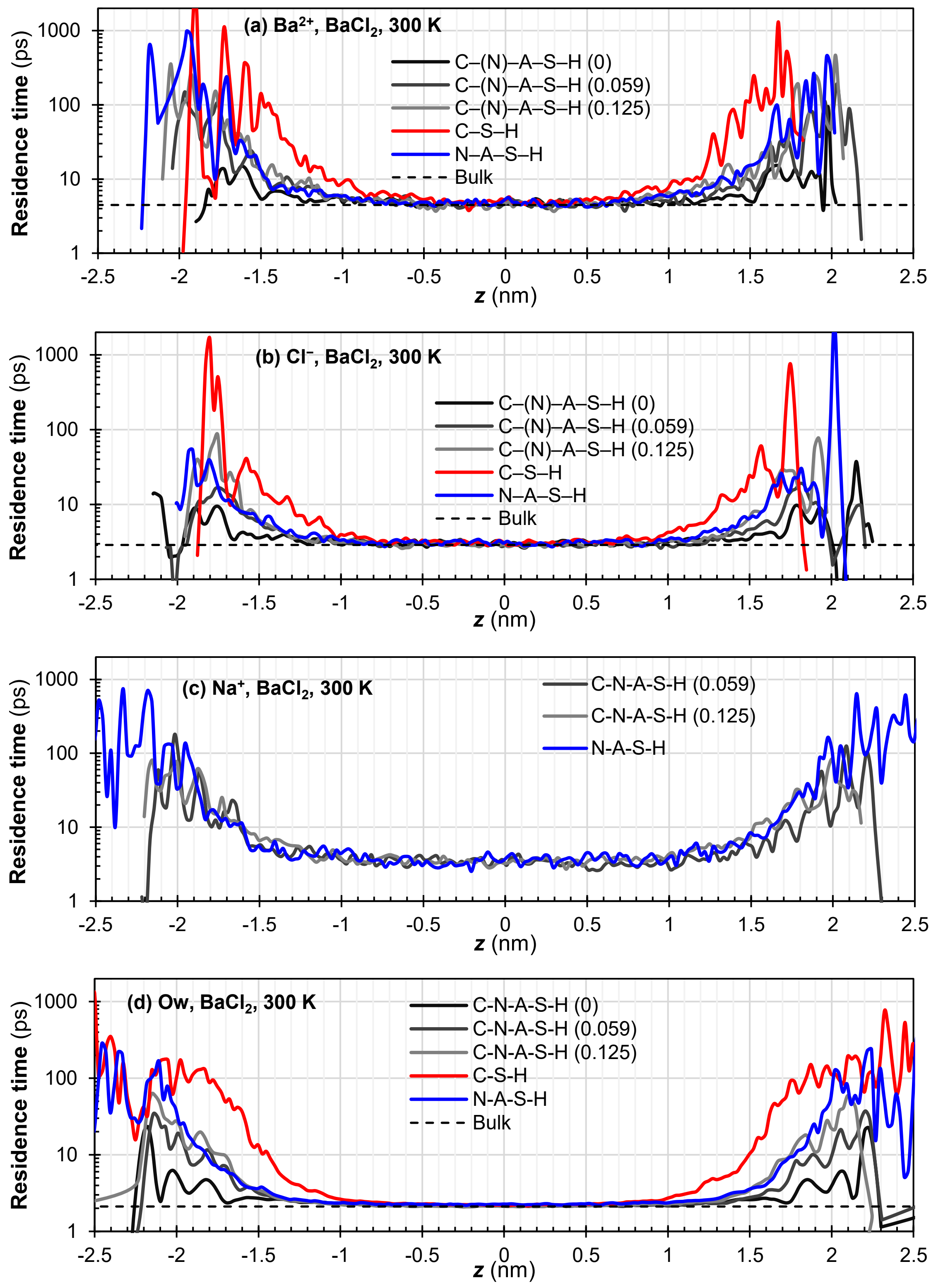

**Figure S11**. The residence time distribution of different solution species in $BaCl_2$ solution across various gel nanopores. Dashed lines in the figures represent the corresponding values for the bulk solution.

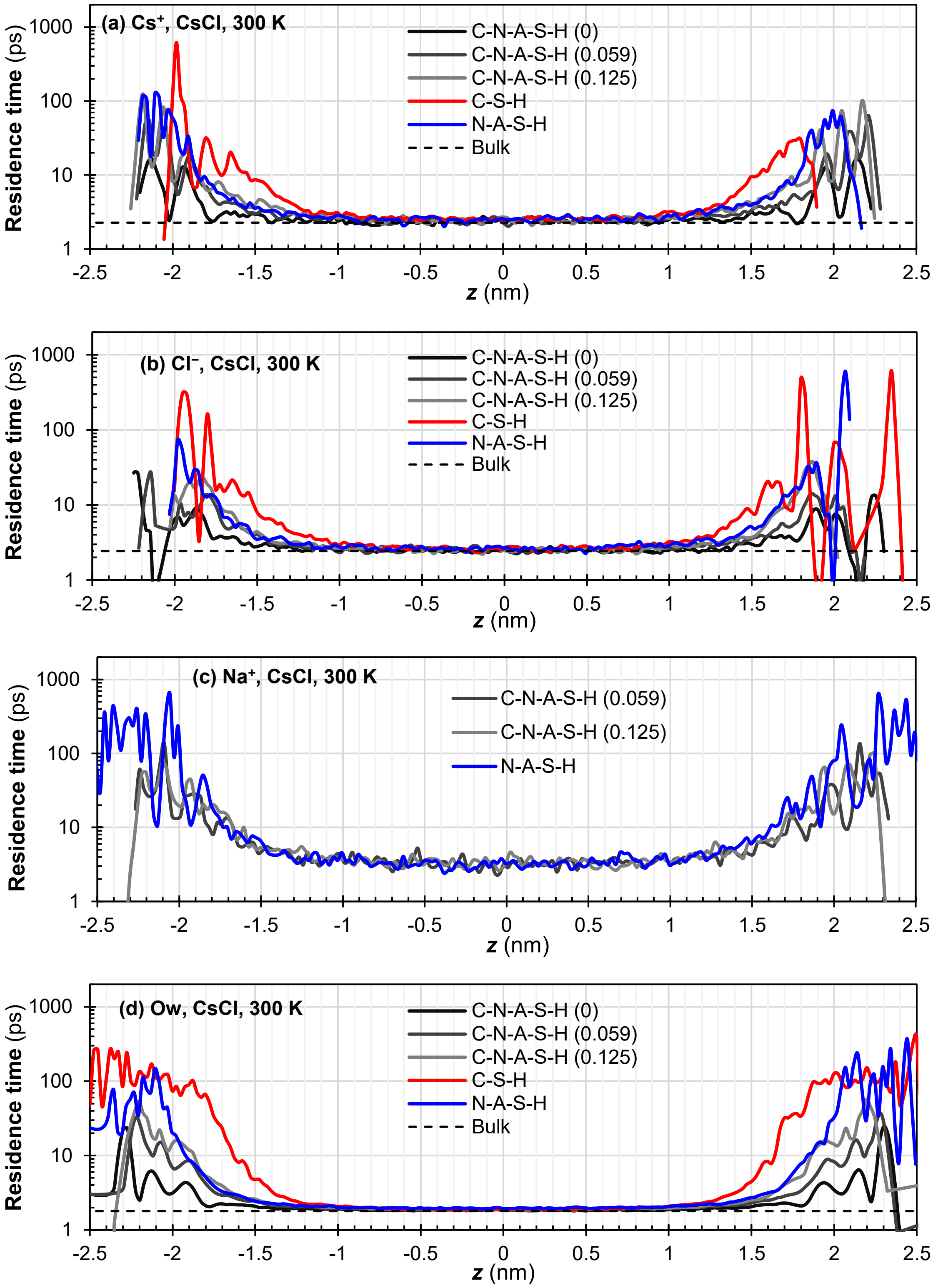

**Figure S12**. The residence time distribution of different solution species in CsCl solution across various gel nanopores. Dashed lines in the figures represent the corresponding values for the bulk solution.

**Section S10 Theoretical calculations of correlation between diffusivity (*D*) and residence time (RT)**

We evaluate the residence time (RT) using a binwise analysis along the $z$ direction (normal to the gel surface). In this framework, the following relation holds:

$$RT(z) \times D(z) = (\Delta l)^2. \quad \text{(S13)}$$

Here, $RT(z)$ and $D(z)$ are the residence time and diffusion coefficient of a given solution species at position $z$, and $\Delta l$ is the characteristic distance the species must travel to exit its original bin. Accordingly, the theoretical prediction of $RT(z)$ is:

$$RT(z) = \frac{(\Delta l)^2}{D(z)}, \quad \text{(S14)}$$

The shortest exit distance is defined by the bin half-width, $\Delta l = \Delta z/2$. Although the effective exit distance may increase under isotropic diffusion of water and ions, we adopt $\Delta l = \Delta z/2$ in this work. This choice yields the theoretical $RT(z)$ curve shown in **Figure 6**b of the main text, representing the lower-limit bound for all simulation data.

**Section S11 One-dimensional number density distribution of different solution species across the gel nanopore**

The one-dimensional number-density distributions of solution species in $PbCl_2$, $BaCl_2$, and CsCl solutions within various gel nanopores are shown in **Figure S13**, **Figure S14**, and **Figure S15**, respectively, with bulk densities indicated by horizontal dashed lines. Density oscillations are pronounced near the gel surface for all systems but gradually converge to bulk-like values at the pore center. In C–(N)–A–S–H and C–S–H gels, the oscillatory profiles of water oxygens and hydrogens (Ow, Hw) indicate distinct interfacial layering, whereas the less ordered N–A–S–H surface produces weaker structuring and small density fluctuations around −2.5 to −1.5 nm. Beyond −1.0 nm, both water and ions exhibit uniform, bulk-like densities and diffusivities.

The $Pb^{2+}$/$Ba^{2+}$/$Cs^{+}$ and $Cl^{-}$ profiles exhibit similarly layered structures near the surfaces of all gel pores, with densities significantly higher than in the pore center and bulk solution, confirming strong adsorption. The precise peak positions and intensities, however, vary with surface composition. In C–(N)–A–S–H pores, increasing the Al/Si ratio from 0 to 0.059 and 0.125 progressively enhances the $Pb^{2+}$/$Ba^{2+}$/$Cs^{+}$ adsorption peak at the surface, reflecting greater cation accumulation. Both $Pb^{2+}$/$Ba^{2+}$/$Cs^{+}$ and $Cl^{-}$ peaks also shift closer to the gel surface, with $Pb^{2+}$/$Ba^{2+}$/$Cs^{+}$ preceding $Cl^{-}$, consistent with preferential cation adsorption onto negatively charged Al-tetrahedral sites. Correspondingly, the $Na^{+}$ profiles reveal substantial leaching of surface $Na^{+}$ that originally balanced the negatively charged Al tetrahedra. This behavior suggests a competitive interaction in which $Pb^{2+}$/$Ba^{2+}$/$Cs^{+}$ adsorption at these sites is accompanied by partial $Na^{+}$ release into the solution, indicative of cation-exchange-like processes that maintain surface charge balance. In the C–S–H and N–A–S–H pores, $Pb^{2+}$/$Ba^{2+}$/$Cs^{+}$ and $Cl^{-}$ densities are also elevated near the surface relative to the bulk, but remain lower than in C–(N)–A–S–H with Al/Si = 0.059–0.125.

The yellow shaded regions denote the immobile layers of the corresponding heavy-metal ions. Their approximate ranges were initially determined from the z-directional diffusivity profiles in **Figure S4**a, **Figure S5**a, and **Figure S6**a, where ion mobility begins to increase from the near-zero interfacial region. Because these near-surface regions correspond to strongly adsorbed ions, the one-dimensional number-density profiles were further used to refine the layer boundaries. Specifically, the peaks and valleys in the number-density profiles indicate preferred adsorption positions of solution species; therefore, the final immobile-layer boundaries were assigned to the corresponding peak or valley positions, as shown in **Figure S13**, **Figure S14**, and **Figure S15**.

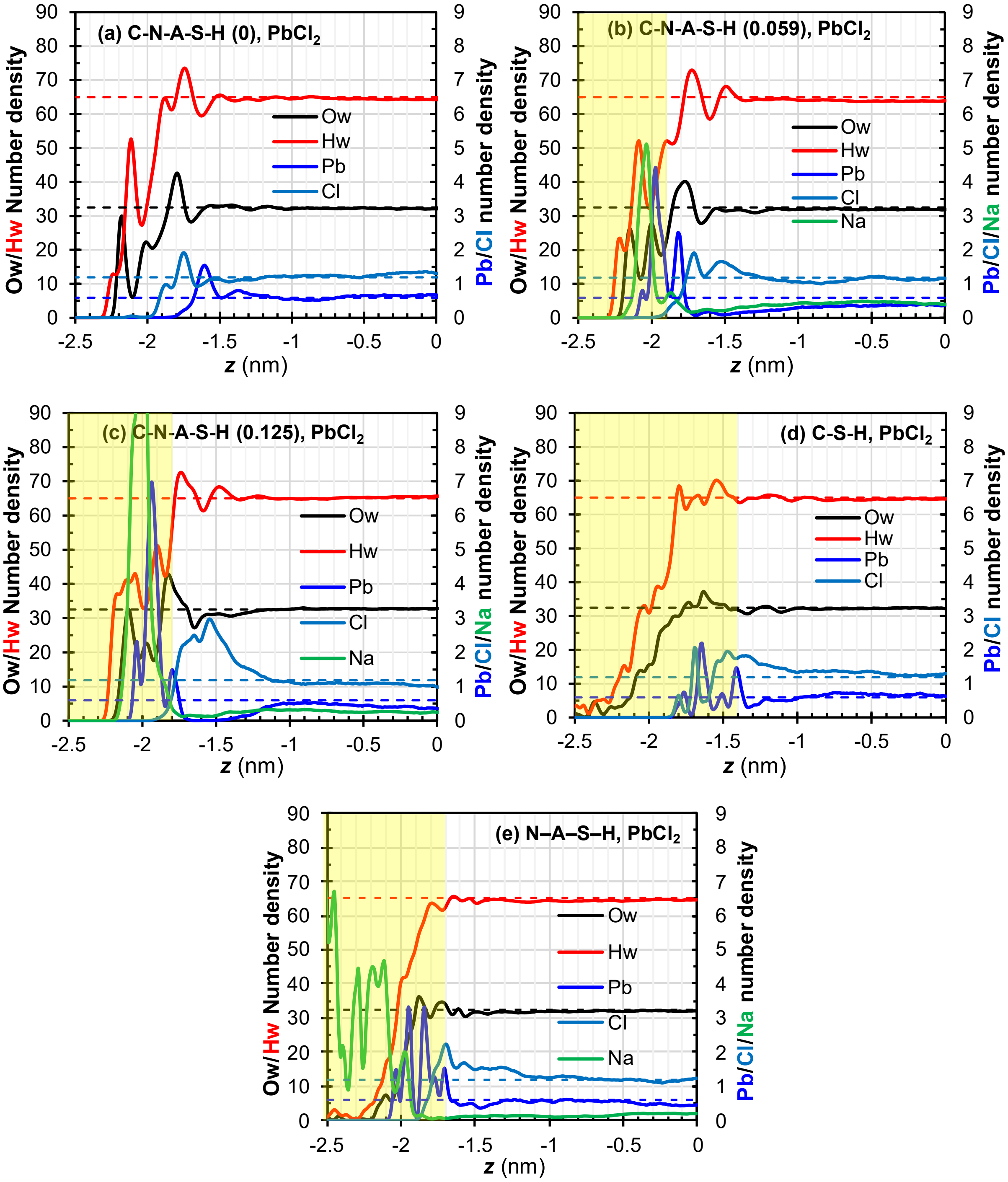

**Figure S13**. Number density distribution of different solution species for 1M $PbCl_2$ solutions inside various gel nanochannels: (a) C–(N)–A–S–H with an Al/Si ratio of 0, (b) 0.059, and (c) 0.125; (d) C–S–H; and (e) N–A–S–H, where the pore center is located at $z \approx 0$ nm and the pore surface at $z \approx -2$ nm. The yellow-shaded regions denote the "immobile layer" for heavy-metal ions.

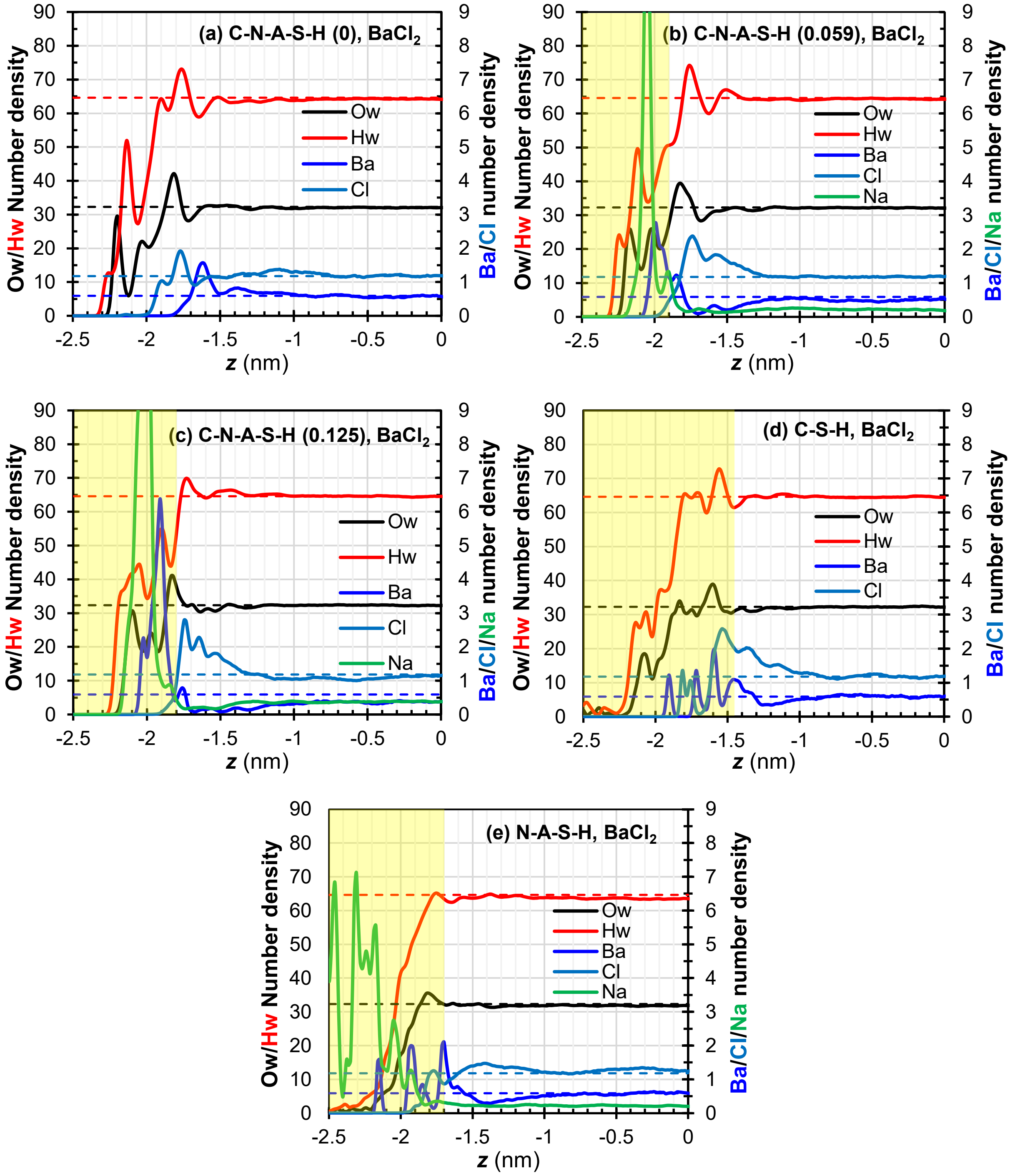

**Figure S14**. Number density distribution of different solution species for 1M $BaCl_2$ solutions inside various gel nanochannels: (a) C–(N)–A–S–H with an Al/Si ratio of 0, (b) 0.059, and (c) 0.125; (d) C–S–H; and (e) N–A–S–H, where the pore center is located at $z \approx 0$ nm and the pore surface at $z \approx -2$ nm. The yellow-shaded regions denote the "immobile layer" for heavy-metal ions.

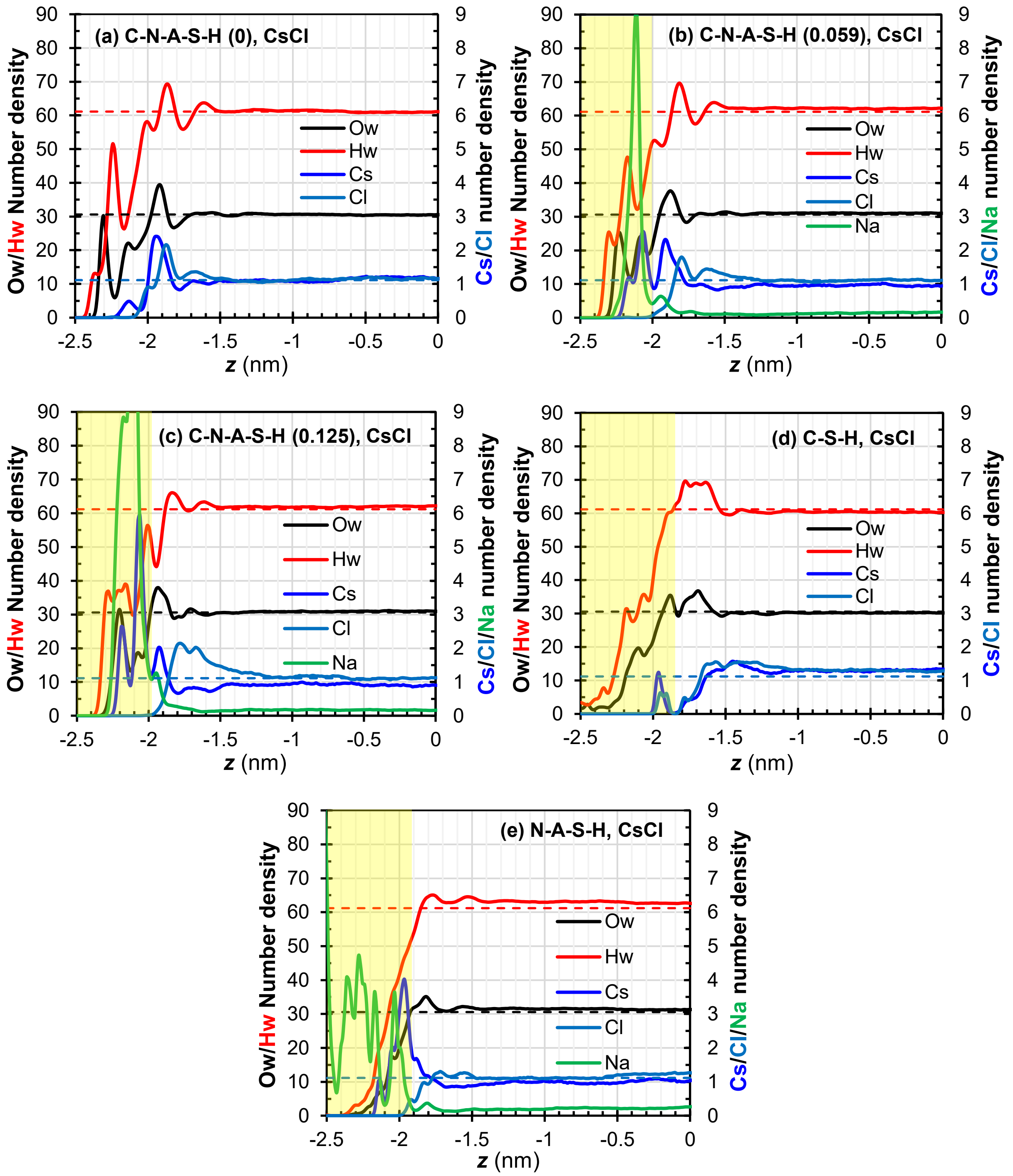

**Figure S15**. Number density distribution of different solution species for 2M CsCl solutions inside various gel nanochannels: (a) C–(N)–A–S–H with an Al/Si ratio of 0, (b) 0.059, and (c) 0.125; (d) C–S–H; and (e) N–A–S–H, where the pore center is located at $z \approx 0$ nm and the pore surface at $z \approx -2$ nm. The yellow-shaded regions denote the "immobile layer" for heavy-metal ions.

## Section S12 Two-dimensional adsorption pattern of ions on the gel surface

The two-dimensional adsorption of ions in nanoconfined $PbCl_2$, $BaCl_2$, and CsCl solutions was analyzed by calculating their in-plane number density distributions (bin size = 0.5 Å) along the $x$ and $y$ directions within the range $-2.5 < z < -1.5$ nm (approximately yellow-shaded zones in **Figure S13**, **Figure S14**, and **Figure S15**), covering the primary condensed adsorption layers near the gel surface. The resulting 2D density maps (**Figure S16**, **Figure S17**, and **Figure S18**) reveal heterogeneous ion distributions that vary with gel chemistry. In C–(N)–A–S–H with Al/Si = 0, ion adsorption is diffuse and weakly correlated with surface silicate trenches, indicating limited affinity for hydroxyl sites. Increasing Al/Si produces distinct adsorption hotspots where $Pb^{2+}$/$Ba^{2+}$/$Cs^{+}$ and $Na^{+}$ ions concentrate near Al/Si substitution sites, reflecting localized ion–surface interactions that underlie immobilization. These cation-rich sites coincide with reduced $Na^{+}$ densities, suggesting partial ion exchange. The spatial distribution of $Cl^{-}$ largely mirrors that of pre-adsorbed cations. For C–S–H and N–A–S–H gels, elevated but disordered ion densities occur without clear periodic motifs, consistent with their less ordered surfaces.

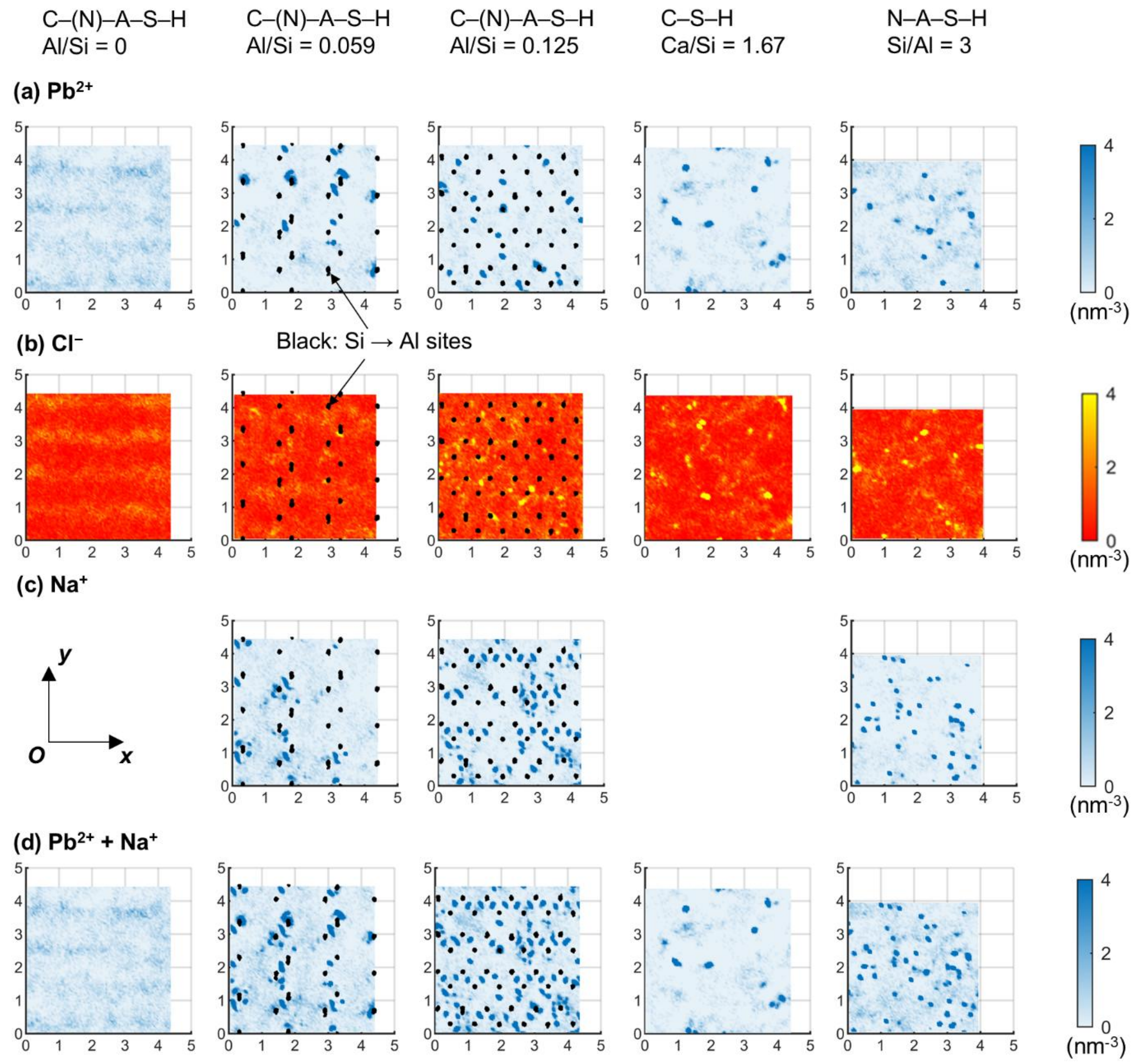

**Figure S16.** Two-dimensional spatial distributions of ions in a 1 M $PbCl_2$ aqueous solution on various gel surfaces, including (a) $Pb^{2+}$, (b) $Cl^-$, (c) $Na^+$, and (d) $Pb^{2+}$ + $Na^+$ ions.

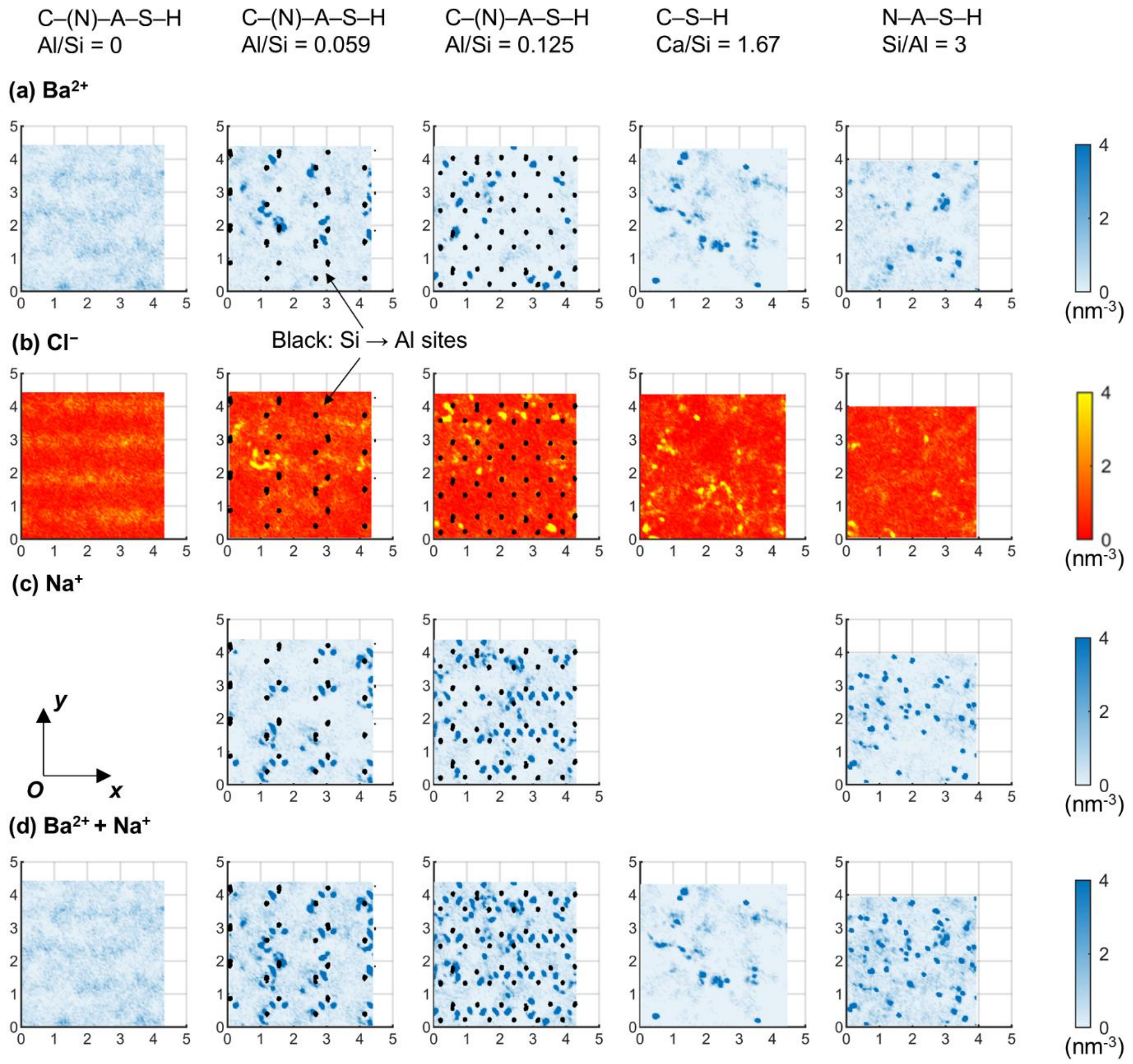

**Figure S17.** Two-dimensional spatial distributions of ions in a 1 M $BaCl_2$ aqueous solution on various gel surfaces, including (a) $Ba^{2+}$, (b) $Cl^-$, (c) $Na^+$, and (d) $Ba^{2+}$ + $Na^+$ ions.

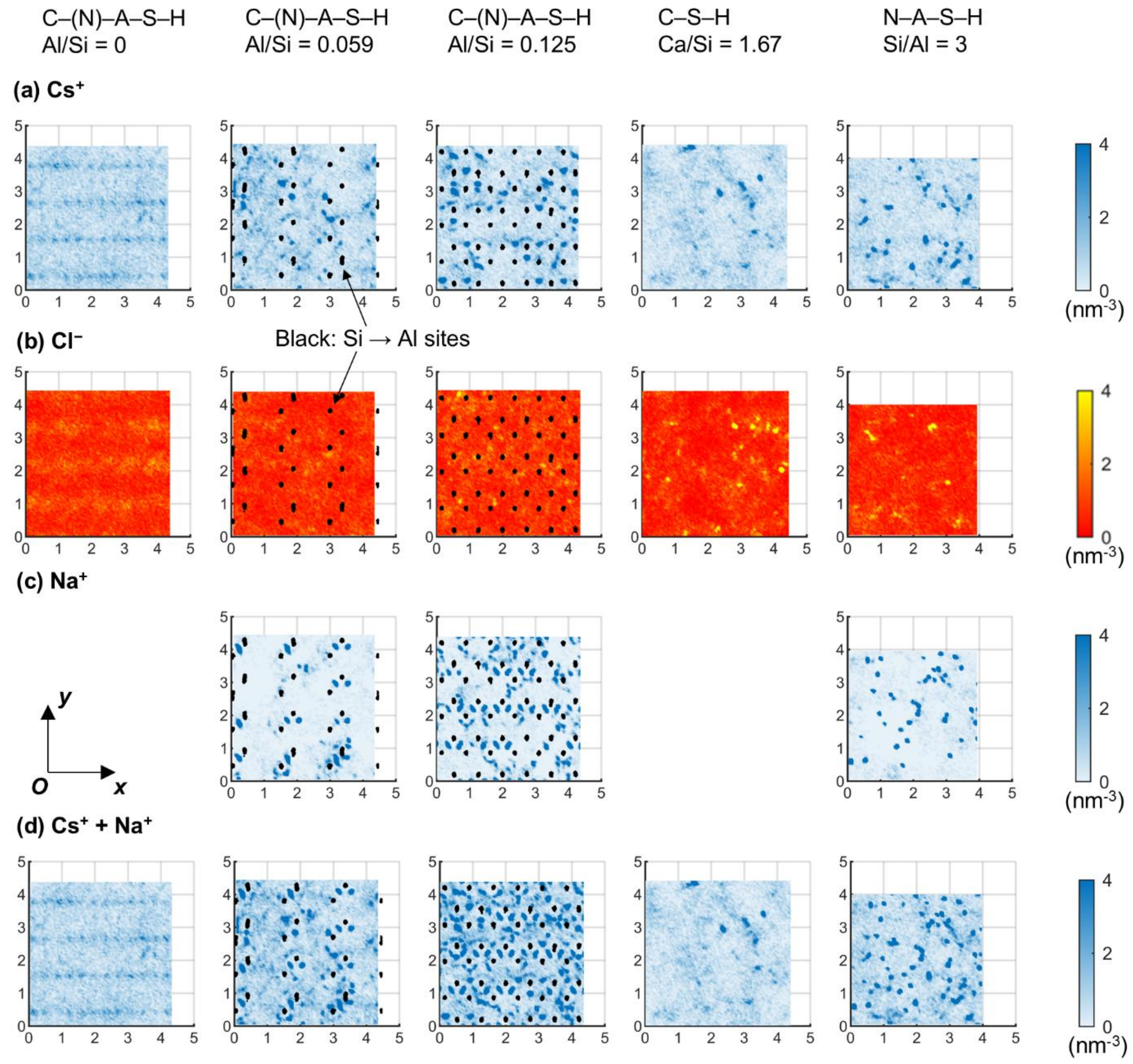

**Figure S18.** Two-dimensional spatial distributions of ions in a 2 M CsCl aqueous solution on various gel surfaces, including (a) $Cs^+$, (b) $Cl^-$, (c) $Na^+$, and (d) $Cs^+$ + $Na^+$ ions.

**Section S13 The partial radial distribution function (RDF) for different ion species**

To further elucidate the adsorption and transport behaviors discussed above, we conducted a detailed local structural analysis of both bulk and nanoconfined ion species across different gel nanopores. The partial radial distribution functions (RDFs) for various Pb/Ba/Cs–X and Cl–X atom pairs in nanoconfined and bulk $PbCl_2$, $BaCl_2$, and CsCl solutions at 300 K are presented in **Figure S19**, **Figure S20**, and **Figure S21**, respectively. The results show that: (1) for Pb/Ba/Cs–X pairs, the primary neighboring atoms involved in attractive interactions are O and Cl within the C–(N)–A–S–H, C–S–H, and N–A–S–H gels; and (2) for Cl–X pairs, attractive neighbors include H, Pb/Ba/Cs, and Na in the C–(N)–A–S–H and N–A–S–H gels, and additionally Ca in the C–S–H gels. Accordingly, **Figure 10** in the main text displays only the nearest neighbors exhibiting attractive interactions for the $BaCl_2$ systems. The equilibrium bond lengths (i.e., nearest interatomic distances) for each Pb/Ba/Cs–X and Cl–X pair were determined from the position of the first maximum in the corresponding RDFs.

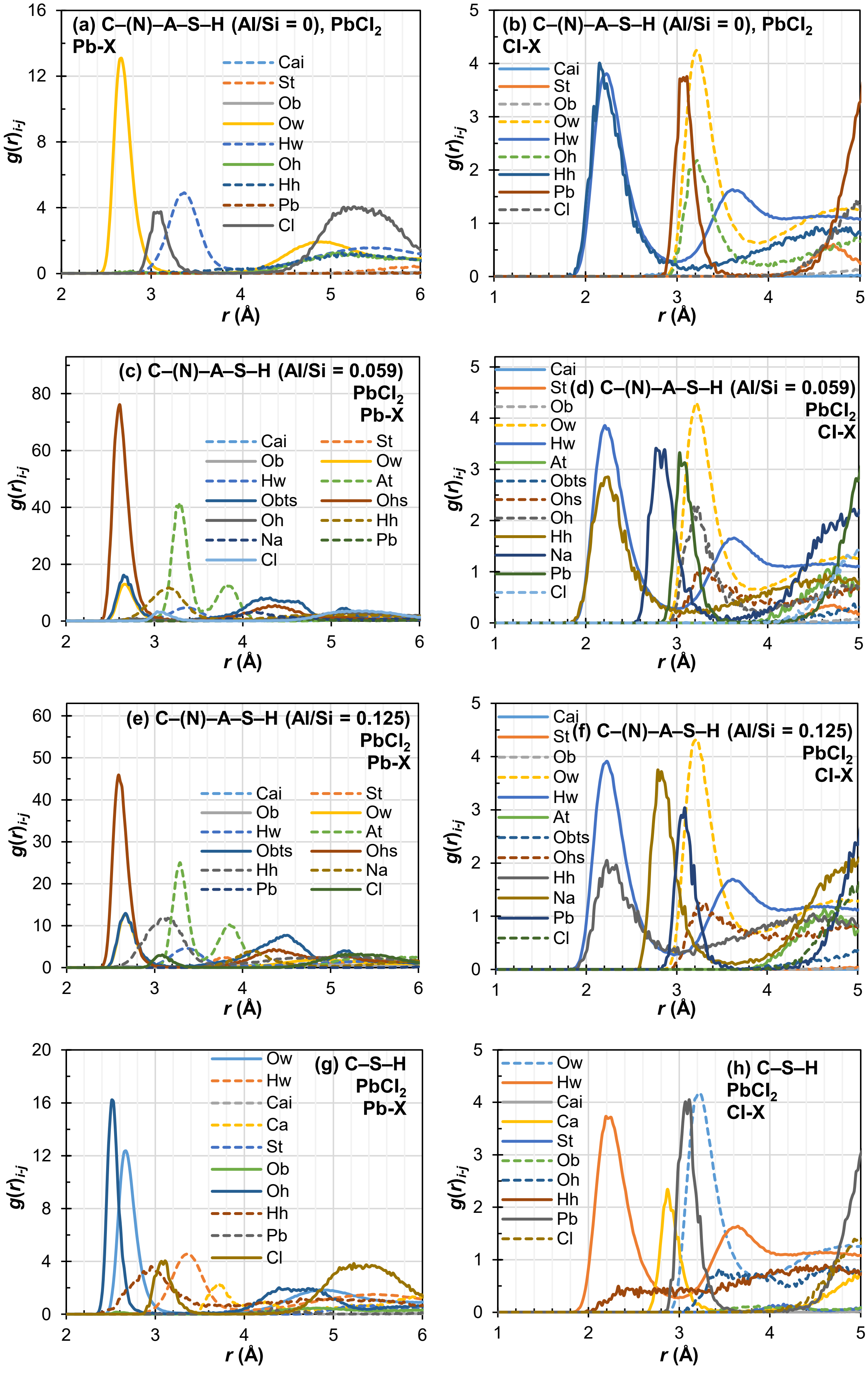
(a) C–(N)–A–S–H (Al/Si = 0), PbCl2
Pb-X
Cai
St
Ob
Ow
Hw
Oh
Hh
Pb
Cl
g(r)i-j
r (Å)
(b) C–(N)–A–S–H (Al/Si = 0), PbCl2
Cl-X
Cai
St
Ob
Ow
Hw
Oh
Hh
Pb
Cl
(c) C–(N)–A–S–H (Al/Si = 0.059)
PbCl2
Pb-X
Cai
St
Ob
Ow
Hw
At
Obts
Ohs
Oh
Hh
Na
Pb
Cl
(d) C–(N)–A–S–H (Al/Si = 0.059)
PbCl2
Cl-X
Cai
St
Ob
Ow
Hw
At
Obts
Ohs
Oh
Hh
Na
Pb
Cl
(e) C–(N)–A–S–H (Al/Si = 0.125)
PbCl2
Pb-X
Cai
St
Ob
Ow
Hw
At
Obts
Ohs
Hh
Na
Pb
Cl
(f) C–(N)–A–S–H (Al/Si = 0.125)
PbCl2
Cl-X
Cai
St
Ob
Ow
Hw
At
Obts
Ohs
Hh
Na
Pb
Cl
(g) C–S–H
PbCl2
Pb-X
Ow
Hw
Cai
Ca
St
Ob
Oh
Hh
Pb
Cl
(h) C–S–H
PbCl2
Cl-X
Ow
Hw
Cai
Ca
St
Ob
Oh
Hh
Pb
Cl

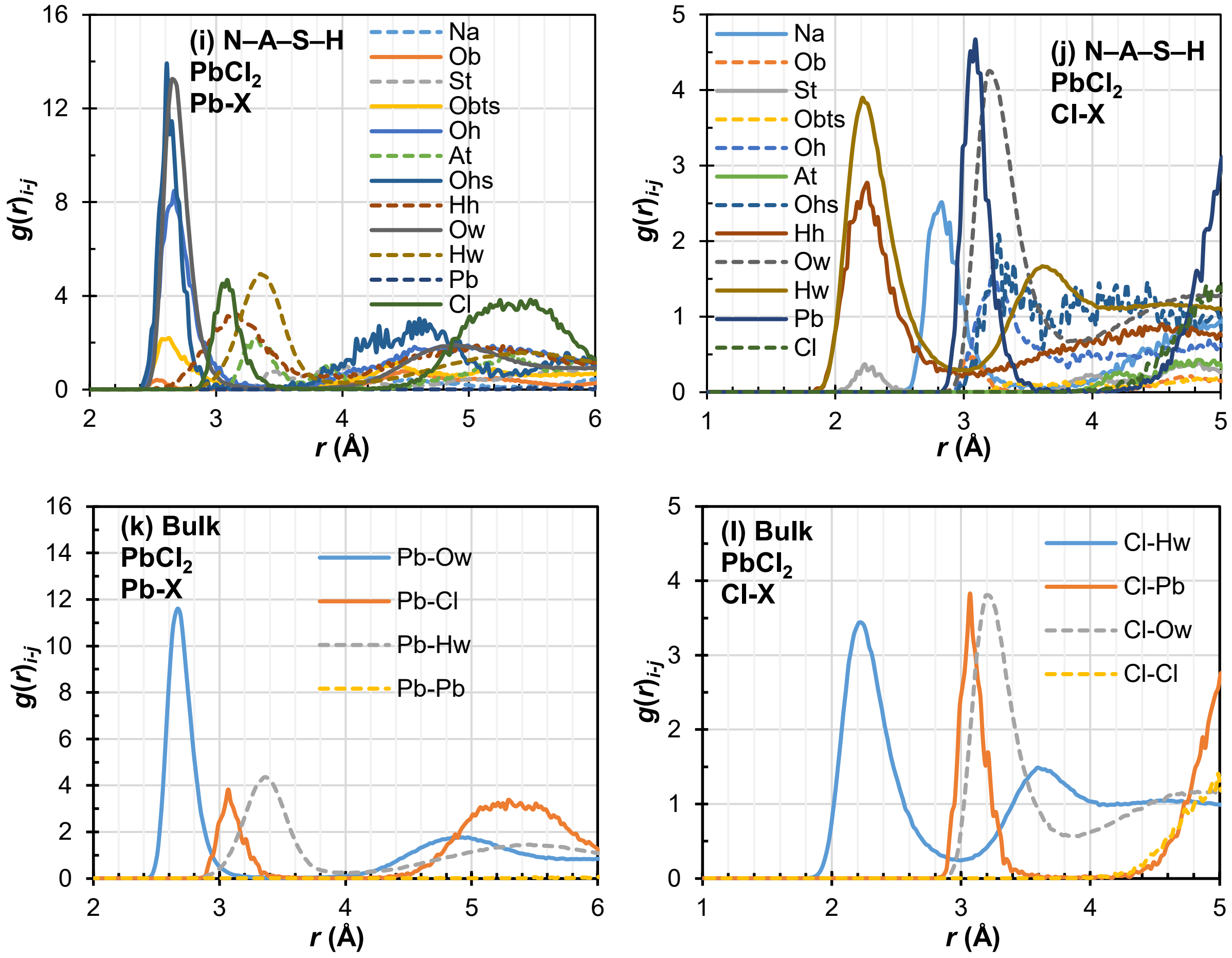

**Figure S19**. (a, c, e, g, i, k) Partial RDFs in the nanoconfined 1 M $PaCl_2$ aqueous solution for Pb–X pairs inside (a) the C–(N)–A–S–H gel with an Al/Si ratio of 0, (c) 0.059, (e) 0.125, (g) C–S–H gel, (i) N–A–S–H gel, (k) bulk solution, with their respective nearest neighbors. (b, d, f, h, j, l) The corresponding partial RDFs for Cl–X pairs. Atom labels are as follows: St = tetrahedral silicon atoms; At = tetrahedral aluminum atoms; Ca = aqueous interlayer calcium ion; Cai = intralayer calcium; Ow = oxygen atoms in water molecules; Hw = hydrogen atoms in water molecules; Ob = bridging oxygen atoms connected to tetrahedral Si atoms; Obts = bridging oxygen atoms connecting tetrahedral Si and Al; Oh = oxygen atoms in surface hydroxyl groups; Ohs = hydroxyl oxygen atoms bonded to Al tetrahedral sites; Hh = hydrogen atoms in surface hydroxyl groups.

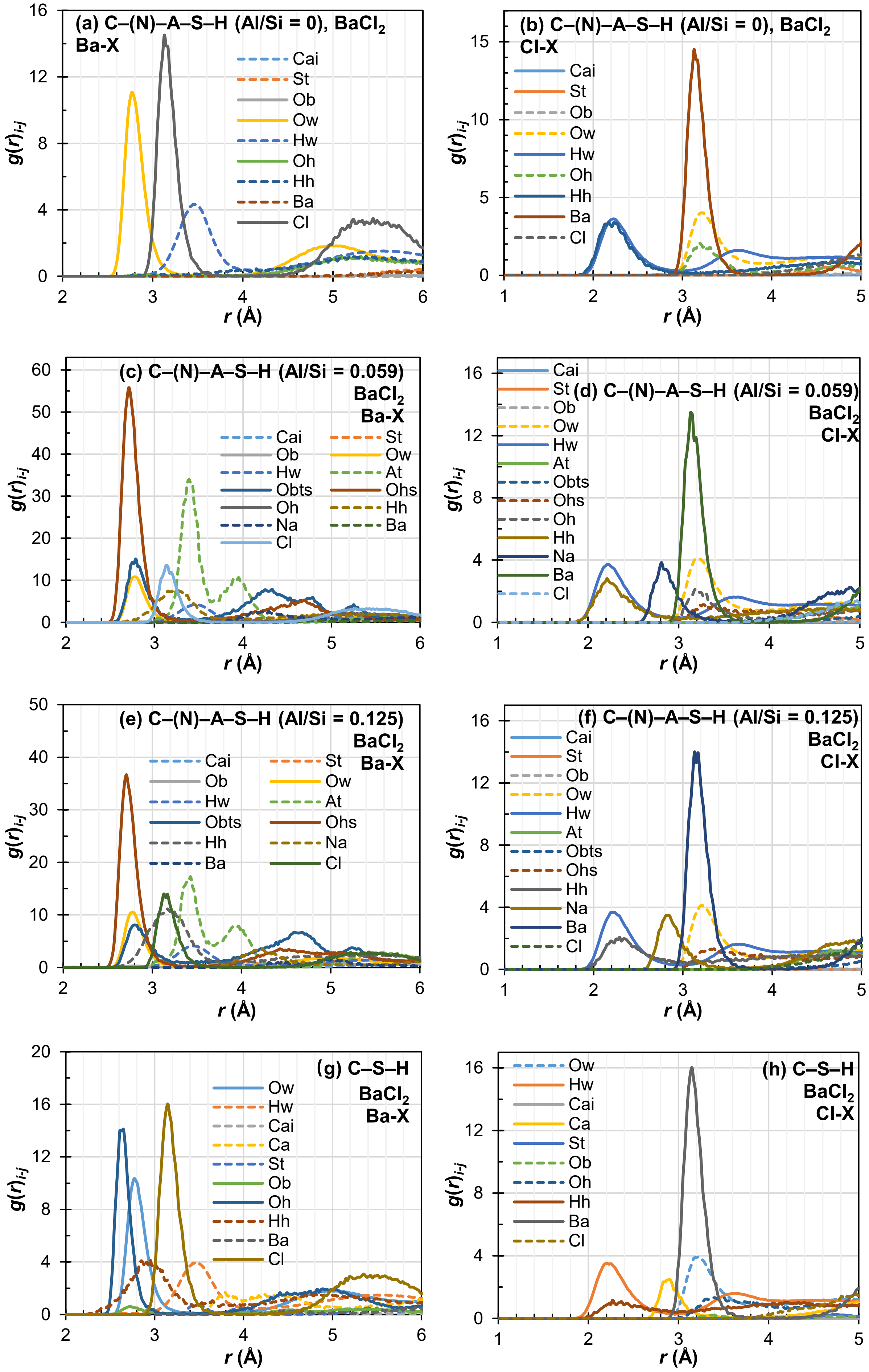
(a) C–(N)–A–S–H (Al/Si = 0), BaCl2
Ba-X
Cai
St
Ob
Ow
Hw
Oh
Hh
Ba
Cl
g(r)i-j
r (Å)
(b) C–(N)–A–S–H (Al/Si = 0), BaCl2
Cl-X
Cai
St
Ob
Ow
Hw
Oh
Hh
Ba
Cl
g(r)i-j
r (Å)
(c) C–(N)–A–S–H (Al/Si = 0.059)
BaCl2
Ba-X
Cai
St
Ob
Ow
Hw
At
Obts
Ohs
Oh
Hh
Na
Ba
Cl
g(r)i-j
r (Å)
(d) C–(N)–A–S–H (Al/Si = 0.059)
BaCl2
Cl-X
Cai
St
Ob
Ow
Hw
At
Obts
Ohs
Oh
Hh
Na
Ba
Cl
g(r)i-j
r (Å)
(e) C–(N)–A–S–H (Al/Si = 0.125)
BaCl2
Ba-X
Cai
St
Ob
Ow
Hw
At
Obts
Ohs
Hh
Na
Ba
Cl
g(r)i-j
r (Å)
(f) C–(N)–A–S–H (Al/Si = 0.125)
BaCl2
Cl-X
Cai
St
Ob
Ow
Hw
At
Obts
Ohs
Hh
Na
Ba
Cl
g(r)i-j
r (Å)
(g) C–S–H
BaCl2
Ba-X
Ow
Hw
Cai
Ca
St
Ob
Oh
Hh
Ba
Cl
g(r)i-j
r (Å)
(h) C–S–H
BaCl2
Cl-X
Ow
Hw
Cai
Ca
St
Ob
Oh
Hh
Ba
Cl
g(r)i-j
r (Å)

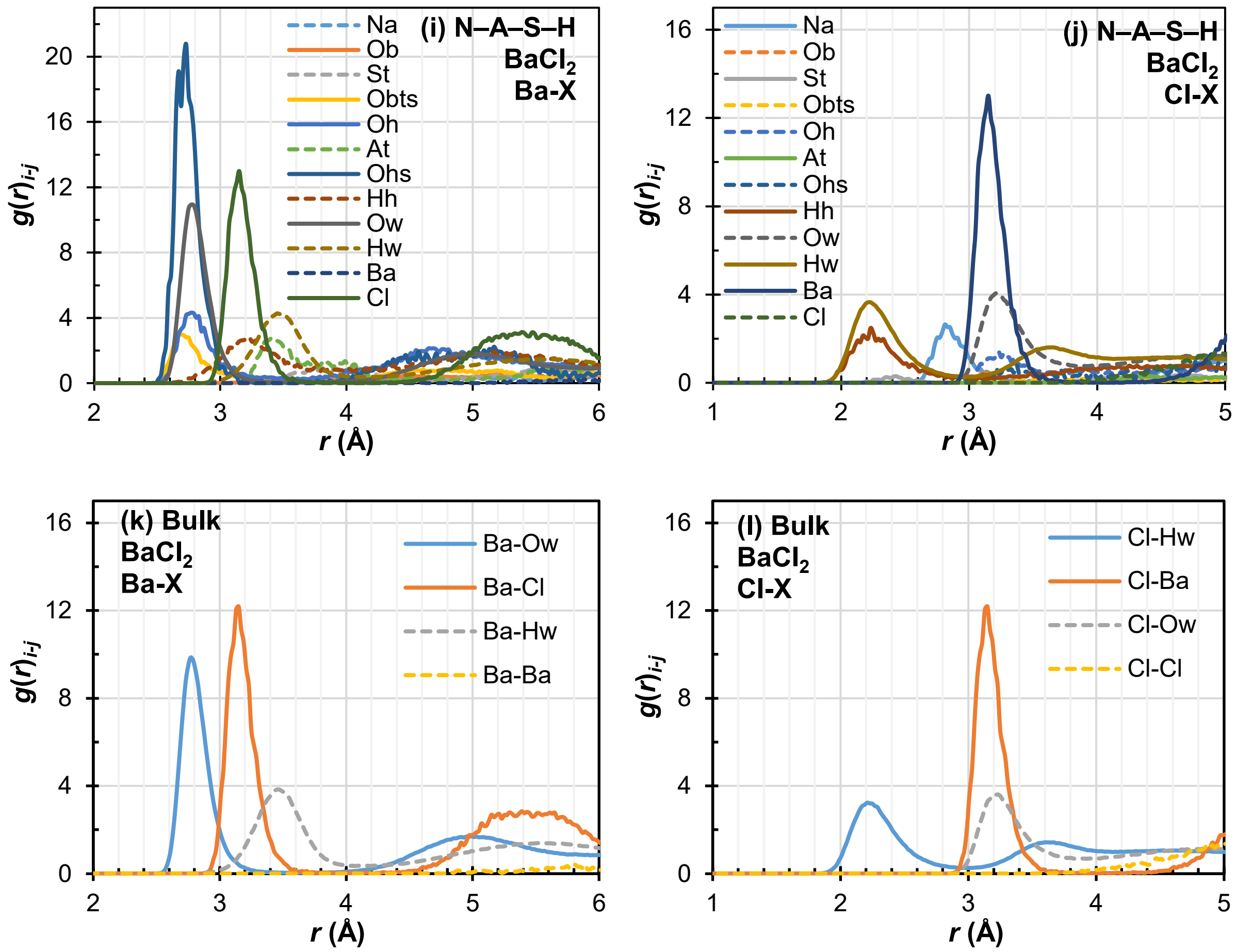

**Figure S20**. (a, c, e, g, i, k) Partial RDFs in the nanoconfined 1 M $BaCl_2$ aqueous solution for Ba–X pairs inside (a) the C–(N)–A–S–H gel with an Al/Si ratio of 0, (c) 0.059, (e) 0.125, (g) C–S–H gel, (i) N–A–S–H gel, (k) bulk solution, with their respective nearest neighbors. (b, d, f, h, j, l) The corresponding partial RDFs for Cl–X pairs. Atom labels are as follows: St = tetrahedral silicon atoms; At = tetrahedral aluminum atoms; Ca = aqueous interlayer calcium ion; Cai = intralayer calcium; Ow = oxygen atoms in water molecules; Hw = hydrogen atoms in water molecules; Ob = bridging oxygen atoms connected to tetrahedral Si atoms; Obts = bridging oxygen atoms connecting tetrahedral Si and Al; Oh = oxygen atoms in surface hydroxyl groups; Ohs = hydroxyl oxygen atoms bonded to Al tetrahedral sites; Hh = hydrogen atoms in surface hydroxyl groups.

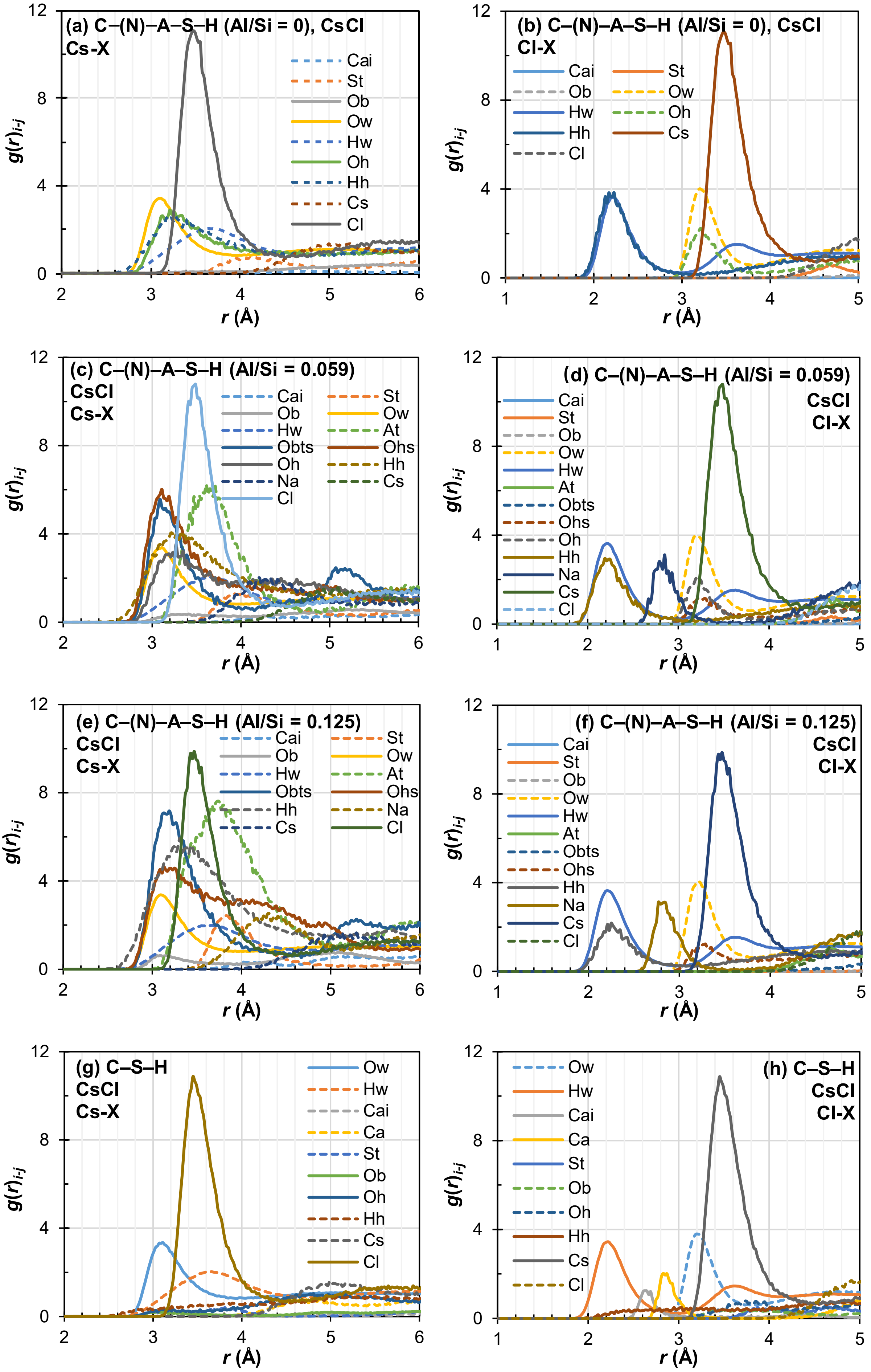
(a) C–(N)–A–S–H (Al/Si = 0), CsCl
Cs-X
Cai St Ob Ow Hw Oh Hh Cs Cl
(b) C–(N)–A–S–H (Al/Si = 0), CsCl
Cl-X
Cai St Ob Ow Hw Oh Hh Cs Cl
(c) C–(N)–A–S–H (Al/Si = 0.059)
CsCl
Cs-X
Cai St Ob Ow Hw At Obts Ohs Oh Hh Na Cs Cl
(d) C–(N)–A–S–H (Al/Si = 0.059)
CsCl
Cl-X
Cai St Ob Ow Hw At Obts Ohs Oh Hh Na Cs Cl
(e) C–(N)–A–S–H (Al/Si = 0.125)
CsCl
Cs-X
Cai St Ob Ow Hw At Obts Ohs Hh Na Cs Cl
(f) C–(N)–A–S–H (Al/Si = 0.125)
CsCl
Cl-X
Cai St Ob Ow Hw At Obts Ohs Hh Na Cs Cl
(g) C–S–H
CsCl
Cs-X
Ow Hw Cai Ca St Ob Oh Hh Cs Cl
(h) C–S–H
CsCl
Cl-X
Ow Hw Cai Ca St Ob Oh Hh Cs Cl
g(r)i-j
r (Å)

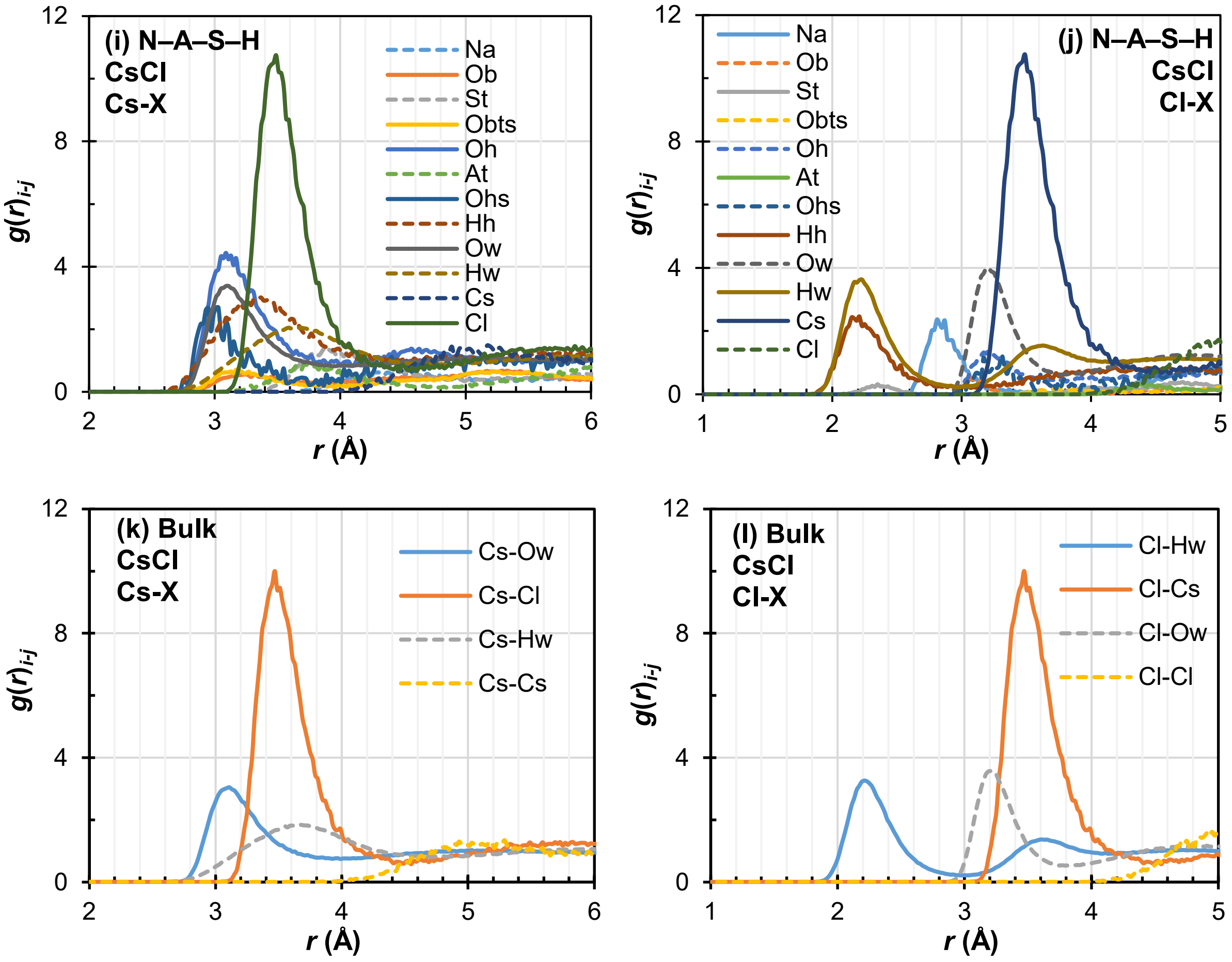

**Figure S21.** (a, c, e, g, i, k) Partial RDFs in the nanoconfined 2 M CsCl aqueous solution for Cs–X pairs inside (a) the C–(N)–A–S–H gel with a Al/Si ratio of 0, (c) 0.059, (e) 0.125, (g) C–S–H gel, (i) N–A–S–H gel, (k) bulk solution, with their respective nearest neighbors. (b, d, f, h, j, l) The corresponding partial RDFs for Cl–X pairs. Atom labels are as follows: St = tetrahedral silicon atoms; At = tetrahedral aluminum atoms; Ca = aqueous interlayer calcium ion; Cai = intralayer calcium; Ow = oxygen atoms in water molecules; Hw = hydrogen atoms in water molecules; Ob = bridging oxygen atoms connected to tetrahedral Si atoms; Obts = bridging oxygen atoms connecting tetrahedral Si and Al; Oh = oxygen atoms in surface hydroxyl groups; Ohs = hydroxyl oxygen atoms bonded to Al tetrahedral sites; Hh = hydrogen atoms in surface hydroxyl groups.

**Section S14 Summary of nearest interatomic distances, interaction strengths, coordination numbers, and total binding strengths within the first coordination shell of heavy metal ions**

For the primary $Pb^{2+}$/$Ba^{2+}$/$Cs^{+}$–X and $Cl^{-}$–X atom pairs involved in attractive interactions identified in the previous section, **Table S7**, **Table S12** and **Table S17** summarize the nearest interatomic distances in the first coordination shell of $Pb^{2+}$, $Ba^{2+}$ and $Cs^{+}$ ions for bulk and nanoconfined $PbCl_2$/$BaCl_2$/CsCl solutions in different gel environments, derived from the peak positions of the partial radial distribution functions (RDFs) shown in **Figure S19**, **Figure S20**, and **Figure S21**, respectively.

The corresponding bond strengths/energies, expressed as the pair energies between the studied atomic species, are calculated and presented in **Table S8**, **Table S13**, and **Table S18**, respectively. A more negative pair energy corresponds to a stronger attractive interaction between the two species. Across all heavy-metal–oxygen/chloride interactions, the trend $Pb^{2+}$ > $Ba^{2+}$ > $Cs^{+}$ holds, indicating that $Pb^{2+}$ exhibits the strongest hydration structure, pairing capability to chloride ions, and binding affinity to the gel surface, whereas $Cs^{+}$ shows the weakest. This observation is consistent with the findings presented throughout this work. For both divalent heavy metal ions (M = $Pb^{2+}$ or $Ba^{2+}$) and univalent $Cs^{+}$ ions, the bond strength ordering varies with the gel type:

- **C–(N)–A–S–H gel:** M–Obts > M–Ohs > M–Ob > (M–Oh) > M–Cl > M–Ow;
- **C–S–H gel:** M–Oh > M–Ob > M–Cl > M–Ow;
- **N–A–S–H gel:** M–Obts > M–Ohs > M–Ob > M–Oh > M–Cl > M–Ow.

The corresponding coordination numbers (CNs), obtained by integrating the RDFs up to their first minima, are presented in **Table S9**, **Table S14**, and **Table S19**, respectively. The total binding strengths (TBS)—obtained by summing the bond strengths and coordination numbers between the heavy-metal ions and surface oxygen atoms—are reported in **Table S10**, **Table S15**, and **Table S20**, respectively.

Furthermore, after accounting for the effect of multiple ion–surface atom interactions acting on the same ion, the refined TBS values are presented in **Table S11**, **Table S16**, and **Table S21**, respectively. This analysis was performed using an in-house MATLAB script (R2023a, The MathWorks, Inc.) [13], which automatically identifies ions coordinated to at least one surface oxygen site and obtains their count. The resulting number is then divided by the total number of ions in the nanopore to calculate the surface-bound ion fraction, $f_{\mathrm{Y-j}}$, where Y = $Pb^{2+}$, $Ba^{2+}$, and $Cs^{+}$. Comparisons across heavy-metal ions indicate that gels exhibit weak immobilization of $Cs^{+}$ because the bond strengths of its individual coordination pairs are both low and closely comparable—specifically, $Cs^{+}$–water-oxygen and $Cs^{+}$–surface-oxygen interactions are of similar magnitude, suggesting that $Cs^{+}$ ions are both easy to dehydrate and to detach from the gel surface. In contrast, $Pb^{2+}$ and $Ba^{2+}$ develop stronger hydration structures and stronger binding to surface oxygen atoms. Therefore, the immobilization of $Cs^{+}$ by gel surfaces is much weaker than for $Pb^{2+}$ and $Ba^{2+}$.

The correlation between the immobilization extent of heavy-metal ions and the calculated total binding strength (TBS) in various gel nanopores is presented in **Figure 12** of the main text.

**Table S7.** Summary of the nearest interatomic distances (bond lengths, $r_0$) in the first coordination shell of $Pb^{2+}$ ions for bulk and nanoconfined $PbCl_2$ solutions within various gel environments, derived from the peak positions of the partial radial distribution functions (RDF) shown in **Figure S19**. The bond lengths obtained are consistent with literature values for Pb–Ow pairs, reported as 2.60 Å [15] and 2.70 Å [16]. Atom labels: Ow = oxygen atoms in water molecules; Ob = bridging oxygen atoms connected to tetrahedral Si atoms; Obts = bridging oxygen atoms connecting tetrahedral Si and Al; Oh = oxygen atoms in surface hydroxyl groups; Ohs = hydroxyl oxygen atoms bonded to Al tetrahedral sites. The standard deviations reported in the table were calculated based on three independent simulations.

| **Gel type** | **Bond length: $r_{0,\mathrm{Pb-j}}$ (Å)** | | | | | |
|---|---|---|---|---|---|---|
| | **Ow** | **Obts** | **Ob** | **Oh** | **Ohs** | **Cl** |
| Bulk | 2.67±0.00 | – | – | – | – | 3.06±0.01 |
| C–(N)–A–S–H (Al/Si = 0) | 2.67±0.00 | – | – | 2.77±0.00 | – | 3.09±0.03 |
| C–(N)–A–S–H (Al/Si = 0.59) | 2.67±0.00 | 2.66±0.01 | 2.67±0.02 | 2.76±0.04 | 2.60±0.01 | 3.05±0.02 |
| C–(N)–A–S–H (Al/Si = 0.125) | 2.67±0.00 | 2.66±0.01 | 2.68±0.02 | – | 2.59±0.00 | 3.09±0.00 |
| C–S–H (Ca/Si = 1.67) | 2.67±0.00 | – | 2.57±0.00 | 2.51±0.00 | – | 3.08±0.02 |
| N–A–S–H (Al/Si = 0.33) | 2.66±0.01 | 2.60±0.02 | 2.54±0.01 | 2.64±0.02 | 2.62±0.01 | 3.09±0.00 |

**Table S8.** Summary of nearest-neighbor interatomic interacting strength (bond energies) for Pb–j pairs ($E_{\mathrm{Pb-j}}$) in the first coordination shell of $Pb^{2+}$ ions for bulk and nanoconfined $PbCl_2$ solutions within various gel environments. Bond energies were calculated from the total potential energy function using the corresponding Pb–j bond lengths listed in **Table S7**, following the method described in **Section S5**. More negative bond energy values indicate stronger attractive interactions. The standard deviations reported in the table were calculated based on three independent simulations.

| **Gel type** | **Bond energy: $E_{\mathrm{Pb-j}}$ ($10^{-19}$ J)** | | | | | |
|---|---|---|---|---|---|---|
| | **Ow** | **Obts** | **Ob** | **Oh** | **Ohs** | **Cl** |
| Bulk | −13.90±0.00 | – | – | – | – | −14.70±0.03 |
| C–(N)–A–S–H (Al/Si = 0) | −13.90±0.00 | – | – | −15.67±0.00 | – | −14.61±0.09 |
| C–(N)–A–S–H (Al/Si = 0.59) | −13.90±0.00 | −19.97±0.06 | −17.88±0.09 | −15.70±0.20 | −18.80±0.05 | −14.74±0.05 |
| C–(N)–A–S–H (Al/Si = 0.125) | −13.90±0.00 | −19.97±0.06 | −17.81±0.10 | – | −18.84±0.00 | −14.61±0.00 |
| C–S–H (Ca/Si = 1.67) | −13.90±0.00 | – | −20.43 | −25.53±0.00 | – | −14.65±0.08 |
| N–A–S–H (Al/Si = 0.33) | −13.95±0.04 | −20.33±0.11 | −18.54±0.04 | −16.27±0.09 | −18.70±0.05 | −14.61±0.00 |

**Table S9.** Summary of the coordination numbers ($CN_{Pb-j}$) for Pb–j pairs in the first coordination shell of $Pb^{2+}$ ions for bulk and nanoconfined $PbCl_2$ solutions within various gel environments. Coordination numbers were obtained by integrating the corresponding partial radial distribution functions (RDFs) shown in **Figure S19** up to the cutoff radius ($r_c$) listed in the table. The standard deviations reported in the table were calculated based on three independent simulations.

| Gel type | Coordination number: $CN_{Pb-j}$ | | | | | |
|---|---|---|---|---|---|---|
| | Ow | Obts | Ob | Oh | Ohs | Cl |
| | ($r_c$ = 3.19Å) | | | | | ($r_c$ = 3.99Å) |
| Bulk | 8.30±0.01 | – | – | – | – | 0.12±0.01 |
| C–(N)–A–S–H (Al/Si = 0) | 8.28±0.01 | – | – | 0.01±0.01 | – | 0.11±0.00 |
| C–(N)–A–S–H (Al/Si = 0.59) | 7.09±0.04 | 0.26±0.02 | 0.02±0.01 | 0.02±0.00 | 0.89±0.05 | 0.09±0.00 |
| C–(N)–A–S–H (Al/Si = 0.125) | 6.79±0.02 | 0.34±0.00 | 0.07±0.00 | – | 1.08±0.01 | 0.09±0.01 |
| C–S–H (Ca/Si = 1.67) | 7.92±0.02 | – | 0.00±0.00 | 0.28±0.01 | – | 0.15±0.02 |
| N–A–S–H (Al/Si = 0.33) | 7.61±0.03 | 0.23±0.01 | 0.03±0.01 | 0.36±0.02 | 0.06±0.02 | 0.12±0.01 |

**Table S10.** Total binding strength (TBS) and its individual Pb–j components in the first coordination shell of $Pb^{2+}$ ions for bulk and nanoconfined $PbCl_2$ solutions within various gel environments. Each component of the total binding strength ($-CN_{Pb-j}E_{Pb-j}$) was obtained by multiplying the individual bond energies (**Table S8**) with corresponding coordination numbers (**Table S9**). Since attractive Pb–j interactions have negative energies, the leading negative sign converts them into positive binding-strength magnitudes for easier comparison. The total TBS values ($-\sum(CN_{Pb-j}E_{Pb-j})$, where *j* = Obts, Ob, Oh, Ohs) represent the collective contribution of all Pb–surface interactions. The standard deviations reported in the table were calculated based on three independent simulations.

| Gel type | Individual components of TBS: $-CN_{Pb-j}E_{Pb-j}$ ($10^{-19}$ J) | | | | | | Total binding strength (TBS): $-\sum CN_{Pb-j}E_{Pb-j}$ | Diffusion coefficient: $D$ ($10^{-9} m^2/s$) | Immobilization extent (%) |
|---|---|---|---|---|---|---|---|---|---|
| | Ow | Obts | Ob | Oh | Ohs | Cl | | | |
| Bulk | 115.35 ±0.08 | 0.00 ±0.00 | 0.00 ±0.00 | 0.00 ±0.00 | 0.00 ±0.00 | 1.75 ±0.07 | 0.00±0.00 | 0.57±0.01 | 0.00±0.00 |
| C–(N)–A–S–H (Al/Si = 0) | 115.16 ±0.09 | 0.00 ±0.00 | 0.00 ±0.00 | 0.11 ±0.12 | 0.00 ±0.00 | 1.64 ±0.03 | 0.00±0.00 | 0.41±0.03 | 28.83±4.40 |
| C–(N)–A–S–H (Al/Si = 0.59) | 98.55 ±0.49 | 5.19 ±0.41 | 0.33 ±0.15 | 0.24 ±0.08 | 16.78 ±0.85 | 1.32 ±0.03 | 22.53±0.65 | 0.23±0.02 | 59.39±3.99 |
| C–(N)–A–S–H (Al/Si = 0.125) | 94.32 ±0.29 | 6.72 ±0.10 | 1.28 ±0.05 | 0.00 ±0.00 | 20.27 ±0.25 | 1.31 ±0.11 | 28.27±0.26 | 0.22±0.02 | 62.44±3.07 |
| C–S–H (Ca/Si = 1.67) | 110.09 ±0.25 | 0.00 ±0.00 | 0.07 ±0.10 | 7.16 ±0.35 | 0.00 ±0.00 | 2.14 ±0.26 | 10.88±0.47* | 0.24±0.01 | 58.79±1.88 |
| N–A–S–H (Al/Si = 0.33) | 106.24 ±0.42 | 4.62 ±0.29 | 0.59 ±0.10 | 5.89 ±0.29 | 1.20 ±0.32 | 1.69 ±0.17 | 12.29±0.58 | 0.27±0.04 | 52.40±7.71 |

*The secondary Ca–Ow–Pb interaction for the C–S–H gel was also included in the calculation of the TBS due to its relatively high strength. The calculated amount of such interactions per $Pb^{2+}$ ion within the gel system is provided in **Table S11**. According to our previous study [12] and the present results, the Ca–Ow bond ($-15.14 \times 10^{-19}$J) is stronger than other interactions, such as Pb–Ow ($-13.90 \times 10^{-19}$J) and Na–Ow ($-7.84 \times 10^{-19}$J). In addition, interlayer $Ca^{2+}$ ions remain strongly bound to the gel surface throughout the simulation process. Therefore, the secondary immobilization pathway of $Pb^{2+}$ ions via the Pb–Ow–Ca linkage should be taken into consideration for the C–S–H gel, whereas other gels do not exhibit comparably strong secondary interactions.

**Table S11.** The total binding strength (TBS) of $Pb^{2+}$ ions within different gel nanopores, refined by correcting for multiple ion–surface atom interactions per ion, with cutoff radii of 3.4 Å for Ca/Cai–Ow and 3.19 Å for Pb–Obts/Oh/Ohs/Ob/Ow interactions. The standard deviations reported in the table were calculated based on three independent simulations.

| **Gel type** | **Surface-bound ion fraction: $f_{Pb-j}$** | | **Total binding strength (TBS): $-\sum f_{Pb-j}E_{Pb-j}$ (j=Obts/Ob/Oh/Ohs,** | **Diffusion coefficient: $D$ ($10^{-9}m^2/s$)** | **Immobilization extent (%)** |
|---|---|---|---|---|---|
| | **Ow−Ca/Cai*** | **Obts/Oh/Ohs/Ob** | | | |

| | | | **and Ow−Ca/Cai)**** | | |
|---|---|---|---|---|---|
| C–(N)–A–S–H (Al/Si = 0) | – | (0.50±0.28)/47=0.01±0.01 | 0.17±0.09 | 0.41±0.03 | 28.83±4.40 |
| C–(N)–A–S–H (Al/Si = 0.59) | – | (21.92±0.47)/47=0.47±0.01 | 9.31±0.20 | 0.23±0.02 | 59.39±3.99 |
| C–(N)–A–S–H (Al/Si = 0.125) | – | (27.32±0.06)/47=0.58±0.00 | 11.61±0.02 | 0.22±0.02 | 62.44±3.07 |
| C–S–H (Ca/Si = 1.67) | (12.60±0.78)/48=0.26±0.02 | (8.26±0.74)/48=0.17±0.02 | 8.04±0.62 | 0.24±0.01 | 58.79±1.88 |
| N–A–S–H (Al/Si = 0.33) | – | (11.05±0.61)/38=0.29±0.02 | 5.91±0.33 | 0.27±0.04 | 52.40±7.71 |

*The ion binds indirectly to the gel surface by coordinating with water oxygens that are themselves bonded to surface calcium atoms.

**$E_{\mathrm{Pb-Obts/Oh/Ohs/Ob}} = \min(E_{\mathrm{Pb-Obts}}, E_{\mathrm{Pb-Oh}}, E_{\mathrm{Pb-Ohs}}, E_{\mathrm{Pb-Ob}})$ and $E_{\mathrm{Pb-Ow-Ca/Cai}} = E_{\mathrm{Pb-Ow}}$

**Table S12.** Summary of the nearest interatomic distances (bond lengths, $r_0$) in the first coordination shell of $Ba^{2+}$ ions for bulk and nanoconfined $BaCl_2$ solutions within various gel environments, derived from the peak positions of the partial radial distribution functions (RDF) shown in **Figure S20**. The bond lengths obtained are consistent with literature values for Ba–Ow pairs, reported as 2.86 Å [15]. Atom labels: Ow = oxygen atoms in water molecules; Ob = bridging oxygen atoms connected to tetrahedral Si atoms; Obts = bridging oxygen atoms connecting tetrahedral Si and Al; Oh = oxygen atoms in surface hydroxyl groups; Ohs = hydroxyl oxygen atoms bonded to Al tetrahedral sites. The standard deviations reported in the table were calculated based on three independent simulations.

| **Gel type** | **Bond length: $r_{0,\mathrm{Ba-j}}$ (Å)** | | | | | |
|---|---|---|---|---|---|---|
| | **Ow** | **Obts** | **Ob** | **Oh** | **Ohs** | **Cl** |
| Bulk | 2.77±0.00 | − | − | − | − | 3.14±0.01 |
| C–(N)–A–S–H (Al/Si = 0) | 2.77±0.00 | − | − | − | − | 3.13±0.00 |
| C–(N)–A–S–H (Al/Si = 0.59) | 2.77±0.00 | 2.78±0.01 | 2.83±0.02 | − | 2.72±0.01 | 3.14±0.01 |
| C–(N)–A–S–H (Al/Si = 0.125) | 2.77±0.00 | 2.80±0.01 | 2.78±0.02 | − | 2.71±0.00 | 3.14±0.02 |
| C–S–H (Ca/Si = 1.67) | 2.77±0.00 | − | 2.72±0.01 | 2.62±0.02 | − | 3.14±0.01 |
| N–A–S–H (Al/Si = 0.33) | 2.77±0.00 | 2.68±0.01 | (2.77±0.04) | 2.78±0.01 | 2.70±0.02 | 3.14±0.01 |

**Table S13.** Summary of nearest-neighbor interatomic interacting strength (bond energies) for Ba–j pairs ($E_{\mathrm{Ba-j}}$) in the first coordination shell of $Ba^{2+}$ ions for bulk and nanoconfined $BaCl_2$ solutions within various gel environments. Bond energies were calculated from the total potential energy function using the corresponding Ba–j bond lengths listed in **Table S12**, following the method described in **Section S5**. More negative bond energy values indicate stronger attractive interactions. The standard deviations reported in the table were calculated based on three independent simulations.

| **Gel type** | **Bond energy: $E_{\mathrm{Ba-j}}$ ($10^{-19}$ J)** | | | | | |
|---|---|---|---|---|---|---|
| | **Ow** | **Obts** | **Ob** | **Oh** | **Ohs** | **Cl** |
| Bulk | −13.37±0.00 | − | − | − | − | −14.32±0.03 |
| C–(N)–A–S–H (Al/Si = 0) | −13.37±0.00 | − | − | − | − | −14.34±0.00 |
| C–(N)–A–S–H (Al/Si = 0.59) | −13.37±0.00 | −19.15±0.05 | −16.91±0.08 | − | −17.98±0.05 | −14.32±0.03 |
| C–(N)–A–S–H (Al/Si = 0.125) | −13.37±0.00 | −19.00±0.05 | −17.17±0.12 | − | −18.01±0.00 | −14.30±0.06 |
| C–S–H (Ca/Si = 1.67) | −13.37±0.00 | − | −19.36±0.05 | −24.44±0.12 | − | −14.32±0.03 |
| N–A–S–H (Al/Si = 0.33) | −13.37±0.00 | −19.65±0.05 | (−17.20±0.21) | −15.48±0.04 | −18.08±0.12 | −14.32±0.03 |

**Table S14.** Summary of the coordination numbers ($\mathrm{CN_{Ba-j}}$) for Ba–j pairs in the first coordination shell of $Ba^{2+}$ ions for bulk and nanoconfined $BaCl_2$ solutions within various gel environments. Coordination numbers were obtained by integrating the corresponding partial radial distribution functions (RDFs) shown in **Figure S20** up to the cutoff radius ($r_c$) listed in the table. The standard deviations reported in the table were calculated based on three independent simulations.

| Gel type | Coordination number: $\mathbf{CN_{Ba-j}}$ | | | | | |
|---|---|---|---|---|---|---|
| | **Ow** | **Obts** | **Ob** | **Oh** | **Ohs** | **Cl** |
| | ($r_c$ = 3.19Å) | | | | | ($r_c$ = 3.99Å) |
| Bulk | 8.08±0.02 | – | – | – | – | 0.45±0.02 |
| C–(N)–A–S–H (Al/Si = 0) | 8.04±0.01 | – | – | – | – | 0.46±0.01 |
| C–(N)–A–S–H (Al/Si = 0.59) | 7.20±0.03 | 0.19±0.01 | 0.01±0.01 | – | 0.66±0.01 | 0.42±0.03 |
| C–(N)–A–S–H (Al/Si = 0.125) | 6.71±0.08 | 0.29±0.01 | 0.07±0.01 | – | 1.06±0.04 | 0.37±0.03 |
| C–S–H (Ca/Si = 1.67) | 7.53±0.09 | – | 0.09±0.00 | 0.34±0.01 | – | 0.50±0.06 |
| N–A–S–H (Al/Si = 0.33) | 7.55±0.11 | 0.25±0.02 | 0.01±0.00 | 0.22±0.05 | 0.09±0.01 | 0.40±0.03 |

**Table S15.** Total binding strength (TBS) and its individual Ba–j components in the first coordination shell of $Ba^{2+}$ ions for bulk and nanoconfined $BaCl_2$ solutions within various gel environments. Each component of the total binding strength ($-\mathrm{CN_{Ba-j}}E_{\mathrm{Ba-j}}$) was obtained by multiplying the individual bond energies (**Table S13**) with corresponding coordination numbers (**Table S14**). Since attractive Ba–j interactions have negative energies, the leading negative sign converts them into positive binding-strength magnitudes for easier comparison. The total TBS values ($-\sum(\mathrm{CN_{Ba-j}}E_{\mathrm{Ba-j}})$, where $j$ = Obts, Ob, Oh, Ohs) represent the collective contribution of all Ba–surface interactions. The standard deviations reported in the table were calculated based on three independent simulations.

| Gel type | Individual components of TBS: $-\mathbf{CN_{Ba-j}}\boldsymbol{E}_{\mathbf{Ba-j}}$ ($10^{-19}$ J) | | | | | | Total binding strength (TBS): $-\sum \mathbf{CN_{Ba-j}}\boldsymbol{E}_{\mathbf{Ba-j}}$ | Diffusion coefficient: $D$ ($10^{-9}\mathrm{m^2/s}$) | Immobilization extent (%) |
|---|---|---|---|---|---|---|---|---|---|
| | **Ow** | **Obts** | **Ob** | **Oh** | **Ohs** | **Cl** | | | |
| Bulk | 108.03 ±0.29 | 0.00 ±0.00 | 0.00 ±0.00 | 0.00 ±0.00 | 0.00 ±0.00 | 6.39 ±0.24 | 0.00±0.00 | 0.56±0.01 | 0.00±0.00 |
| C–(N)–A–S–H (Al/Si = 0) | 107.53 ±0.18 | 0.00 ±0.00 | 0.00 ±0.00 | 0.00 ±0.00 | 0.00 ±0.00 | 6.58 ±0.15 | 0.00±0.00 | 0.44±0.00 | 21.52±0.75 |
| C–(N)–A–S–H (Al/Si = 0.59) | 96.34 ±0.34 | 3.65 ±0.19 | 0.13 ±0.14 | 0.00 ±0.00 | 11.82 ±0.11 | 6.03 ±0.47 | 15.60±0.34 | 0.28±0.02 | 49.92±4.36 |
| C–(N)–A–S–H (Al/Si = 0.125) | 89.71 ±1.07 | 5.52 ±0.18 | 1.13 ±0.13 | 0.00 ±0.00 | 19.12 ±0.74 | 5.30 ±0.44 | 25.77±0.81 | 0.19±0.01 | 67.01±1.51 |
| C–S–H (Ca/Si = 1.67) | 100.67 ±1.14 | 0.00 ±0.00 | 1.74 ±0.02 | 8.37 ±0.32 | 0.00 ±0.00 | 7.09 ±0.78 | 14.21±0.13 | 0.24±0.01 | 56.82±2.31 |
| N–A–S–H (Al/Si = 0.33) | 100.97 ±1.50 | 4.84 ±0.39 | 0.10 ±0.05 | 3.40 ±0.78 | 1.57 ±0.16 | 5.66 ±0.44 | 9.91±1.31 | 0.31±0.01 | 44.05±1.31 |

*The secondary Ca–Ow–Ba interaction for the C–S–H gel was also included in the calculation of the TBS due to its relatively high strength. The calculated amount of such interactions per $Ba^{2+}$ ion within the gel system is provided in **Table S16**. According to our previous study [12] and the present results, the Ca–Ow bond ($-15.14 \times 10^{-19}$J) is stronger than other interactions, such as Ba–Ow ($-13.37 \times 10^{-19}$J) and Na–Ow ($-7.84 \times 10^{-19}$J). In addition, interlayer $Ca^{2+}$ ions remain strongly bound to the gel surface throughout the simulation process. Therefore, the secondary immobilization pathway of $Ba^{2+}$ ions via the Ba–Ow–Ca linkage should be taken into consideration for the C–S–H gel, whereas other gels do not exhibit comparably strong secondary interactions.

**Table S16**. The total binding strength (TBS) of $Ba^{2+}$ ions within different gel nanopores, refined by correcting for multiple ion–surface atom interactions per ion, with cutoff radii of 3.4 Å for Ca/Cai–Ow and 3.19 Å for Ba–Obts/Oh/Ohs/Ob/Ow interactions. The standard deviations reported in the table were calculated based on three independent simulations.

| **Gel type** | **Surface-bound ion fraction:** $\boldsymbol{f_{\mathrm{Ba-j}}}$ | | **Total binding strength (TBS):** $-\sum f_{\mathrm{Pb-j}}E_{\mathrm{Pb-j}}$ **(j=Obts/Ob/Oh/Ohs,** | **Diffusion coefficient:** $D$ **(**$10^{-9}\mathrm{m}^2/\mathrm{s}$**)** | **Immobilization extent (%)** |
|---|---|---|---|---|---|
| | **Ow−Ca/Cai*** | **Obts/Oh/Ohs/Ob** | | | |

| | | | **and Ow−Ca/Cai)**** | | |
|---|---|---|---|---|---|
| C–(N)–A–S–H (Al/Si = 0) | – | (0.20±0.05)/47=0.00±0.00 | 0.00±0.00 | 0.44±0.00 | 21.52±0.75 |
| C–(N)–A–S–H (Al/Si = 0.59) | – | (18.49±0.66)/47=0.39±0.01 | 7.53±0.27 | 0.28±0.02 | 49.92±4.36 |
| C–(N)–A–S–H (Al/Si = 0.125) | – | (26.25±0.92)/47=0.56±0.02 | 10.61±0.37 | 0.19±0.01 | 67.01±1.51 |
| C–S–H (Ca/Si = 1.67) | (14.73±1.01)/48=0.31±0.02 | (11.29±0.29)/48=0.24±0.01 | 9.85±0.43 | 0.24±0.01 | 56.82±2.31 |
| N–A–S–H (Al/Si = 0.33) | – | (8.66±1.16)/38=0.23±0.03 | 4.48±0.60 | 0.31±0.01 | 44.05±1.31 |

*The ion binds indirectly to the gel surface by coordinating with water oxygens that are themselves bonded to surface calcium atoms.

**$E_{\mathrm{Ba-Obts/Oh/Ohs/Ob}} = \min(E_{\mathrm{Ba-Obts}}, E_{\mathrm{Ba-Oh}}, E_{\mathrm{Ba-Ohs}}, E_{\mathrm{Ba-Ob}})$ and $E_{\mathrm{Ba-Ow-Ca/Cai}} = E_{\mathrm{Ba-Ow}}$

**Table S17.** Summary of the nearest interatomic distances (bond lengths, $r_0$) in the first coordination shell of $Cs^+$ ions for bulk and nanoconfined CsCl solutions within various gel environments, derived from the peak positions of the partial radial distribution functions (RDF) shown in **Figure S21**. The bond lengths obtained are consistent with literature values for Cs–Ow pairs, reported as 3.25 Å [15]. Atom labels: Ow = oxygen atoms in water molecules; Ob = bridging oxygen atoms connected to tetrahedral Si atoms; Obts = bridging oxygen atoms connecting tetrahedral Si and Al; Oh = oxygen atoms in surface hydroxyl groups; Ohs = hydroxyl oxygen atoms bonded to Al tetrahedral sites. The standard deviations reported in the table were calculated based on three independent simulations.

| Gel type | Bond length: $r_{0,\mathrm{Cs-j}}$ (Å) | | | | | |
|---|---|---|---|---|---|---|
| | **Ow** | **Obts** | **Ob** | **Oh** | **Ohs** | **Cl** |
| Bulk | 3.09±0.02 | − | − | − | − | 3.46±0.01 |
| C–(N)–A–S–H (Al/Si = 0) | 3.09±0.02 | − | 3.56±0.01 | 3.20±0.02 | − | 3.47±0.02 |
| C–(N)–A–S–H (Al/Si = 0.59) | 3.10±0.01 | 3.10±0.01 | 3.40±0.01 | 3.26±0.05 | 3.10±0.02 | 3.48±0.02 |
| C–(N)–A–S–H (Al/Si = 0.125) | 3.10±0.01 | 3.18±0.02 | 3.12±0.02 | − | 3.16±0.03 | 3.47±0.02 |
| C–S–H (Ca/Si = 1.67) | 3.09±0.02 | − | 3.06±0.02 | 3.09±0.00 | − | 3.48±0.02 |
| N–A–S–H (Al/Si = 0.33) | 3.10±0.02 | 3.14±0.03 | 3.20±0.04 | 3.11±0.03 | 3.06±0.03 | 3.48±0.01 |
| C–(N)–A–S–H (Al/Si = 0.125, Pore size ~1 nm) | 3.10±0.02 | 3.16±0.01 | 3.16±0.02 | − | 3.09±0.00 | 3.48±0.02 |
| C–(N)–A–S–H (Al/Si = 0.125, Pore size ~2 nm) | 3.11±0.00 | 3.20±0.03 | 3.20±0.05 | − | 3.10±0.02 | 3.48±0.02 |
| C–(N)–A–S–H (Al/Si = 0.125, Pore size ~8 nm) | 3.10±0.01 | 3.17±0.03 | 3.12±0.01 | − | 3.16±0.01 | 3.46±0.01 |

**Table S18.** Summary of nearest-neighbor interatomic interacting strength (bond energies) for Cs–j pairs ($E_{\mathrm{Cs-j}}$) in the first coordination shell of $Cs^+$ ions for bulk and nanoconfined CsCl solutions within various gel environments. Bond energies were calculated from the total potential energy function using the corresponding Cs–j bond lengths listed in **Table S17**, following the method described in **Section S5**. More negative bond energy values indicate stronger attractive interactions. The standard deviations reported in the table were calculated based on three independent simulations.

| Gel type | Bond energy: $E_{\mathrm{Cs-j}}$ ($10^{-19}$ J) | | | | | |
|---|---|---|---|---|---|---|
| | **Ow** | **Obts** | **Ob** | **Oh** | **Ohs** | **Cl** |
| Bulk | −6.04±0.03 | − | − | − | − | −6.52±0.01 |
| C–(N)–A–S–H (Al/Si = 0) | −6.04±0.03 | − | −6.81±0.02 | −6.81±0.03 | − | −6.51±0.02 |
| C–(N)–A–S–H (Al/Si = 0.59) | −6.03±0.01 | −8.63±0.02 | −7.12±0.02 | −6.70±0.09 | −7.98±0.04 | −6.50±0.03 |
| C–(N)–A–S–H (Al/Si = 0.125) | −6.03±0.01 | −8.44±0.04 | −7.70±0.04 | – | −7.83±0.07 | −6.51±0.02 |
| C–S–H (Ca/Si = 1.67) | −6.04±0.03 | – | −8.67±0.06 | −10.55±0.00 | – | −6.50±0.03 |
| N–A–S–H (Al/Si = 0.33) | −6.02±0.04 | −8.54±0.08 | −7.53±0.09 | −6.97±0.05 | −8.05±0.07 | −6.49±0.01 |
| C–(N)–A–S–H (Al/Si = 0.125, Pore size ~1 nm) | −6.03±0.03 | −8.47±0.02 | −7.62±0.04 | – | −7.99±0.00 | −6.50±0.02 |
| C–(N)–A–S–H (Al/Si = 0.125, Pore size ~2 nm) | −6.01±0.00 | −8.39±0.08 | −7.54±0.11 | – | −7.96±0.05 | −6.50±0.03 |
| C–(N)–A–S–H (Al/Si = 0.125, Pore size ~8 nm) | −6.03±0.01 | −8.46±0.07 | −7.71±0.02 | – | −7.85±0.02 | −6.52±0.01 |

**Table S19.** Summary of the coordination numbers ($\mathrm{CN_{Cs-j}}$) for Cs–j pairs in the first coordination shell of $Cs^{+}$ ions for bulk and nanoconfined CsCl solutions within various gel environments. Coordination numbers were obtained by integrating the corresponding partial radial distribution functions (RDFs) shown in **Figure S21** up to the cutoff radius ($r_c$) listed in the table. The standard deviations reported in the table were calculated based on three independent simulations.

| **Gel type** | **Coordination number: $\mathrm{CN_{Cs-j}}$** | | | | | |
|---|---|---|---|---|---|---|
| | **Ow** | **Obts** | **Ob** | **Oh** | **Ohs** | **Cl** |
| | ($r_c$ = 3.95Å) | | | | | ($r_c$ = 4.55Å) |
| Bulk | 7.58±0.01 | – | – | – | – | 0.88±0.01 |
| C–(N)–A–S–H (Al/Si = 0) | 7.37±0.01 | – | 0.03±0.00 | 0.31±0.00 | – | 0.85±0.00 |
| C–(N)–A–S–H (Al/Si = 0.59) | 7.18±0.04 | 0.16±0.01 | 0.22±0.01 | 0.22±0.01 | 0.32±0.01 | 0.81±0.00 |
| C–(N)–A–S–H (Al/Si = 0.125) | 6.90±0.03 | 0.61±0.02 | 0.33±0.01 | – | 0.74±0.01 | 0.73±0.01 |
| C–S–H (Ca/Si = 1.67) | 7.59±0.02 | – | 0.04±0.01 | 0.04±0.01 | – | 0.96±0.01 |
| N–A–S–H (Al/Si = 0.33) | 7.19±0.00 | 0.22±0.01 | 0.25±0.02 | 0.60±0.01 | 0.02±0.00 | 0.80±0.00 |
| C–(N)–A–S–H (Al/Si = 0.125, Pore size ~1 nm) | 6.25±0.13 | 1.03±0.09 | 0.59±0.06 | – | 1.73±0.07 | 0.70±0.05 |
| C–(N)–A–S–H (Al/Si = 0.125, Pore size ~2 nm) | 6.92±0.07 | 0.57±0.07 | 0.34±0.04 | – | 0.98±0.07 | 0.73±0.02 |
| C–(N)–A–S–H (Al/Si = 0.125, Pore size ~8 nm) | 7.19±0.01 | 0.44±0.00 | 0.24±0.00 | – | 0.50±0.00 | 0.75±0.00 |

**Table S20.** Total binding strength (TBS) and its individual Cs–j components in the first coordination shell of $Cs^{+}$ ions for bulk and nanoconfined CsCl solutions within various gel environments. Each component of the total binding strength ($-\mathrm{CN_{Cs-j}}E_{\mathrm{Cs-j}}$) was obtained by multiplying the individual bond energies (**Table S18**) with corresponding coordination numbers (**Table S19**). Since attractive Cs–j interactions have negative energies, the leading negative sign converts them into positive binding-strength magnitudes for easier comparison. The total TBS values ($-\sum(\mathrm{CN}_{\mathrm{Cs-j}}E_{\mathrm{Cs-j}})$, where $j$ = Obts, Ob, Oh, Ohs) represent the collective contribution of all Cs–surface interactions. The standard deviations reported in the table were calculated based on three independent simulations.

| Gel type | Individual components of TBS: $-\mathrm{CN}_{\mathrm{Cs-j}}E_{\mathrm{Cs-j}}$ ($10^{-19}$ J) | | | | | | Total binding strength (TBS): $-\sum(\mathrm{CN}_{\mathrm{Cs-j}}E_{\mathrm{Cs-j}})$ | Diffusion coefficient: $D$ ($10^{-9}\mathrm{m}^2/\mathrm{s}$) | Immobilization extent (%) |
|---|---|---|---|---|---|---|---|---|---|
| | Ow | Obts | Ob | Oh | Ohs | Cl | | | |
| Bulk | 45.76 ±0.20 | 0.00 ±0.00 | 0.00 ±0.00 | 0.00 ±0.00 | 0.00 ±0.00 | 5.76 ±0.03 | 0.00±0.00 | 1.84±0.09 | 0.00±0.00 |
| C–(N)–A–S–H (Al/Si = 0) | 44.54 ±0.22 | 0.00 ±0.00 | 0.22 ±0.02 | 2.08 ±0.03 | 0.00 ±0.00 | 5.54 ±0.03 | 2.30±0.05 | 1.31±0.02 | 28.88±1.33 |
| C–(N)–A–S–H (Al/Si = 0.59) | 43.32 ±0.31 | 1.41 ±0.11 | 1.53 ±0.10 | 1.46 ±0.09 | 2.56 ±0.09 | 5.24 ±0.01 | 6.96±0.34 | 0.99±0.02 | 46.30±1.00 |
| C–(N)–A–S–H (Al/Si = 0.125) | 41.61 ±0.28 | 5.16 ±0.17 | 2.57 ±0.09 | 0.00 ±0.00 | 5.83 ±0.02 | 4.72 ±0.07 | 13.55±0.27 | 0.83±0.01 | 54.86±0.44 |
| C–S–H (Ca/Si = 1.67) | 45.86 ±0.18 | 0.00 ±0.00 | 0.39 ±0.09 | 0.45 ±0.06 | 0.00 ±0.00 | 6.25 ±0.06 | 3.18±0.22 | 0.97±0.07 | 47.49±3.71 |
| N–A–S–H (Al/Si = 0.33) | 43.29 ±0.29 | 1.86 ±0.13 | 1.91 ±0.13 | 4.19 ±0.04 | 0.18 ±0.02 | 5.17 ±0.04 | 8.14±0.08 | 0.94±0.04 | 48.74±2.27 |
| C–(N)–A–S–H (Al/Si = 0.125, Pore size ~1 nm) | 37.67 ±0.82 | 8.70 ±0.73 | 4.47 ±0.49 | 0.00 ±0.00 | 13.85 ±0.59 | 4.53 ±0.36 | 27.02±1.39 | 0.04±0.00 | 97.64±0.18 |
| C–(N)–A–S–H (Al/Si = 0.125, Pore size ~2 nm) | 41.59 ±0.43 | 4.81 ±0.60 | 2.57 ±0.31 | 0.00 ±0.00 | 7.80 ±0.64 | 4.75 ±0.15 | 15.18±1.53 | 0.42±0.02 | 77.20±0.93 |
| C–(N)–A–S–H (Al/Si = 0.125, Pore size ~8 nm) | 43.34 ±0.11 | 3.75 ±0.06 | 1.88 ±0.03 | 0.00 ±0.00 | 3.95 ±0.04 | 4.91 ±0.03 | 9.57±0.08 | 1.23±0.04 | 32.89±1.90 |

*The secondary Ca–Ow–Cs interaction for the C–S–H gel was also included in the calculation of the TBS due to its relatively high strength. The calculated amount of such interactions per $Cs^+$ ion within the gel system is provided in **Table S21**. According to our previous study [12] and the present

results, the Ca–Ow bond ($-15.14 \times 10^{-19}$J) is stronger than other interactions, such as Cs–Ow ($-6.01 \times 10^{-19}$J) and Na–Ow ($-7.84 \times 10^{-19}$J). In addition, interlayer $Ca^{2+}$ ions remain strongly bound to the gel surface throughout the simulation process. Therefore, the secondary immobilization pathway of $Cs^+$ ions via the Cs–Ow–Ca linkage should be taken into consideration for the C–S–H gel, whereas other gels do not exhibit comparably strong secondary interactions.

**Table S21.** The total binding strength (TBS) of $Cs^+$ ions within different gel nanopores, refined by correcting for multiple ion–surface atom interactions per ion, with cutoff radii of 3.4 Å for Ca/Cai–Ow and 3.95 Å for Cs–Obts/Oh/Ohs/Ob/Ow interactions. The standard deviations reported in the table were calculated based on three independent simulations.

| **Gel type** | **Surface-bound ion fraction: $f_{Cs-j}$** | | **Total binding strength (TBS): $-\sum f_{Cs-j}E_{Cs-j}$ (j=Obts/Ob/Oh/Ohs, and Ow−Ca/Cai)**** | **Diffusion coefficient: $D$ ($10^{-9}m^2/s$)** | **Immobilization extent (%)** |
|---|---|---|---|---|---|
| | **Ow−Ca/Cai*** | **Obts/Oh/Ohs/Ob** | | | |
| C–(N)–A–S–H (Al/Si = 0) | – | (14.73±0.14)/94=0.16±0.00 | 1.07±0.01 | 1.31±0.02 | 28.88±1.33 |
| C–(N)–A–S–H (Al/Si = 0.59) | – | (24.76±1.24)/94=0.26±0.01 | 2.27±0.11 | 0.99±0.02 | 46.30±1.00 |
| C–(N)–A–S–H (Al/Si = 0.125) | – | (32.64±0.13)/94=0.35±0.00 | 2.93±0.01 | 0.83±0.01 | 54.86±0.44 |
| C–S–H (Ca/Si = 1.67) | (37.17±1.57)/96=0.39±0.02 | (4.93±0.52)/96=0.05±0.01 | 2.88±0.16 | 0.97±0.07 | 47.49±3.71 |
| N–A–S–H (Al/Si = 0.33) | – | (21.08±0.40)/76=0.28±0.01 | 2.37±0.04 | 0.94±0.04 | 48.74±2.27 |
| C–(N)–A–S–H (Al/Si = 0.125, Pore size ~1 nm) | – | (19.86±0.05)/23=0.86±0.00 | 7.31±0.02 | 0.04±0.00 | 97.64±0.18 |
| C–(N)–A–S–H (Al/Si = 0.125, Pore size ~2 nm) | – | (24.57±1.62)/47=0.52±0.03 | 4.44±0.29 | 0.42±0.02 | 77.20±0.93 |
| C–(N)–A–S–H (Al/Si = 0.125, Pore size ~8 nm) | – | (43.16±0.65)/188=0.23±0.00 | 1.95±0.03 | 1.23±0.04 | 32.89±1.90 |

*The ion binds indirectly to the gel surface by coordinating with water oxygens that are themselves bonded to surface calcium atoms.

**$E_{\mathrm{Cs-Obts/Oh/Ohs/Ob}} = \min(E_{\mathrm{Cs-Obts}}, E_{\mathrm{Cs-Oh}}, E_{\mathrm{Cs-Ohs}}, E_{\mathrm{Cs-Ob}})$ and $E_{\mathrm{Cs-Ow-Ca/Cai}} = E_{\mathrm{Cs-Ow}}$

**Section S15 Effect of pore size on heavy-metal immobilization behaviors and mechanisms**

In this section, we examine pore-size effects on heavy-metal immobilization in C–(N)–A–S–H nanochannels (Al/Si = 0.125) containing 2 M CsCl aqueous solutions as a representative case, focusing on pore widths spanning ~1–8 nm, which cover the typical confinement regime of gel pores [17, 18].

Our results show that, within this range, the apparent pore-size dependence primarily arises from variations in the relative proportion of interfacial and bulk-like regions, rather than changes in the underlying interaction mechanisms. **Figure S22** shows the apparent pore-scale averaged diffusivity of various aqueous species and the calculated extent of immobilization (if present). Results show that with the pore size increasing, the diffusivity of all species increases and approached to their bulk values. As the channel width decreases, the two solid–solution interfacial regions increasingly overlap, leading to a higher fraction of confined species being influenced by surface effects. This interpretation is supported by the spatially resolved diffusivity and number density profiles (**Figure S23** and **Figure S24**). As shown in **Figure S23**, diffusivities progressively approach bulk values with increasing distance from the surface, with a nearly constant, bulk-like region observed in the center of the ~8 nm channel. In contrast, in ultra-narrow pores (e.g., ~1–2 nm), species transport is simultaneously affected by both confining surfaces due to interfacial overlap. Consistently, **Figure S24** shows that the number densities of all solution species gradually recover to their corresponding bulk values away from the gel surface. Therefore, although decreasing pore size enhances the overall extent of immobilization by increasing the relative contribution of interfacial regions, the nature and hierarchy of the underlying binding sites remain essentially unchanged.

Importantly, this mechanistic consistency is further supported by the spatial collapse of near-surface profiles across different pore sizes. Heavy-metal cations consistently exhibit near-surface enrichment and suppressed diffusivity, indicating that immobilization is governed by strong interfacial interactions. Notably, across all examined pore widths, the near-surface portions of the profiles (from $z$ = 0 nm outward) follow nearly identical trends until the influence of the opposite surface becomes significant. This behavior further confirms that pore size primarily modulates the degree of interfacial overlap, rather than introducing a distinct immobilization pathway.

The corresponding partial RDF profiles are presented in **Figure S25**, while the bond lengths,

bond energies, coordination numbers, and binding energies for Cs–X atom pairs are summarized in **Tables S17–S21 (Section S14**) for comparison. The binding analysis shows that the characteristic bond lengths and bond energies for each Cs–X pair remain essentially unchanged across all pore sizes, indicating that the interfacial immobilization mechanisms are not altered under different nanoconfinement conditions.

With decreasing pore size, the coordination number between heavy-metal ions and surface oxygen atoms increases, reflecting the enhanced proportion of interfacial regions. Correspondingly, the coordination number of Cs–Ow decreases. These changes are therefore attributed to geometric confinement effects rather than modifications of the underlying interaction mechanisms.

Overall, the dominant ion–surface interactions remain consistent across the pore sizes examined, confirming that the binding mechanisms are intrinsically preserved. These results collectively indicate that pore size primarily modulates the degree of confinement, without altering the fundamental immobilization mechanisms governed by local ion–surface interactions.

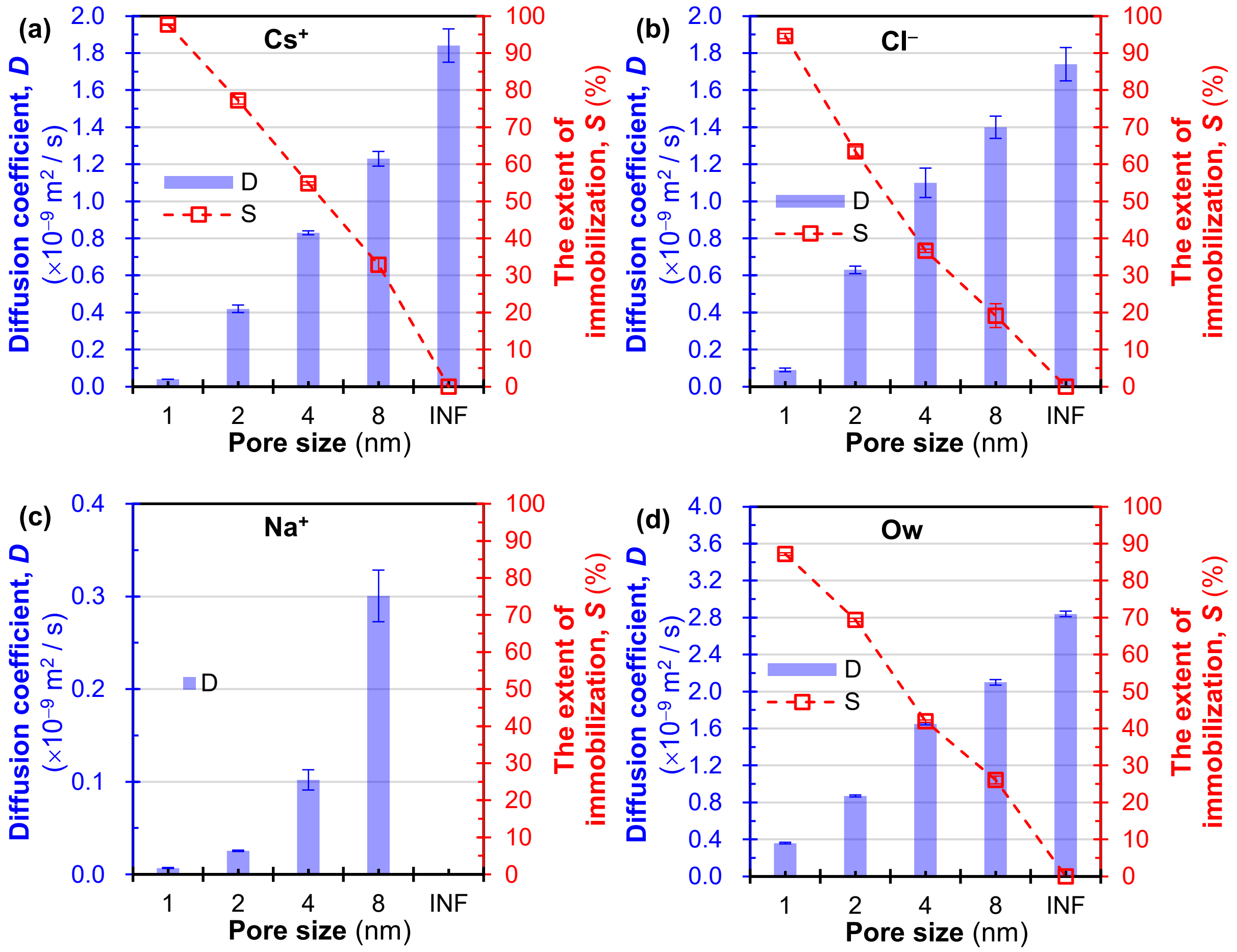

**Figure S22.** The pore-scale diffusivity for (a) heavy-metal cation ($Cs^+$) diffusivity, (b) anion ($Cl^-$) diffusivity, (c) counter-ion ($Na^+$) diffusivity, and (d) water (Ow) diffusivity for CsCl solutions confined in C–(N)–A–S–H gel nanochannels (Al/Si = 0.125) with pore sizes of ~1, ~2, ~4, ~8, and INF nm. The infinite pore size (INF) represents bulk solution conditions. Error bars represent standard deviations from three independent simulations.

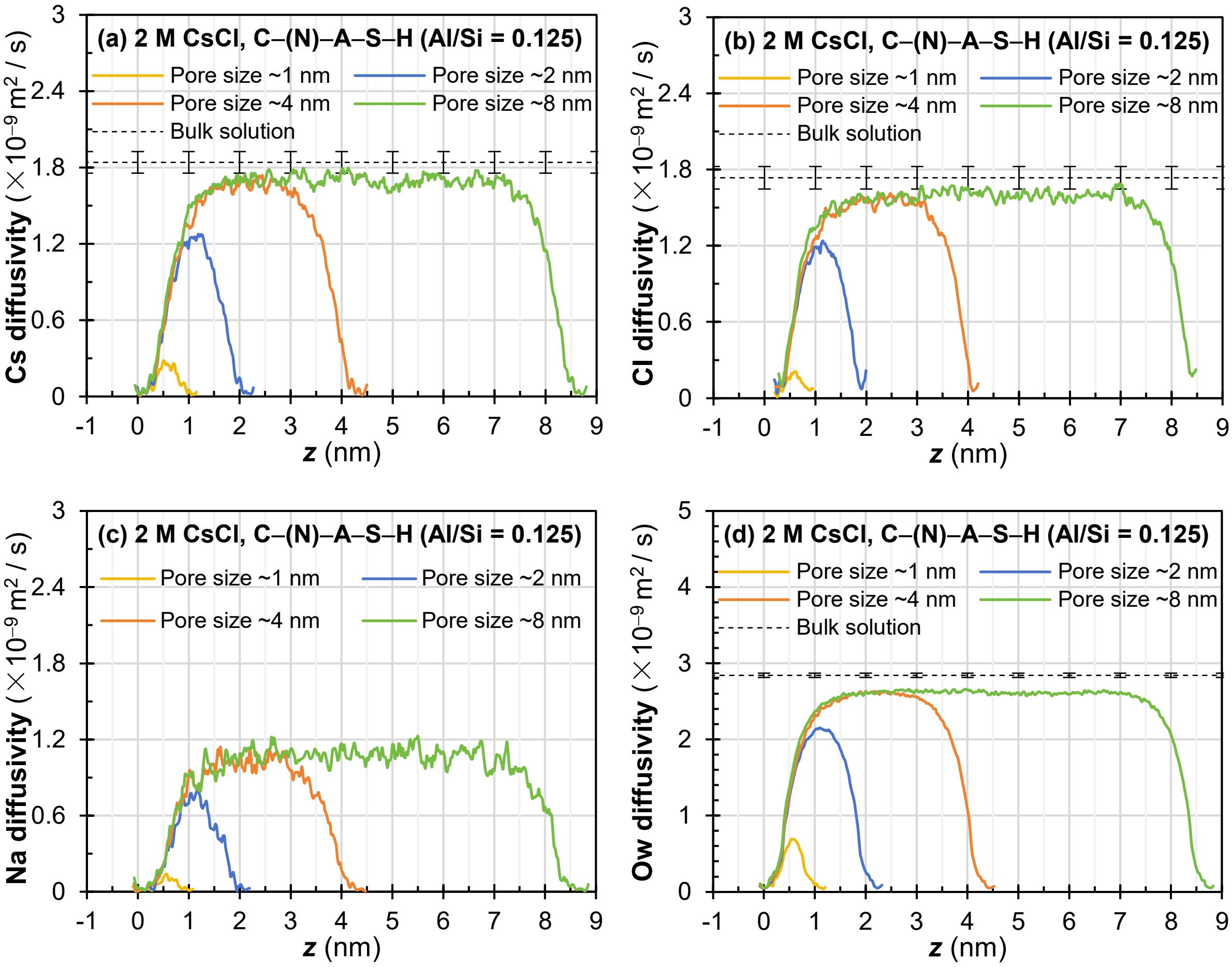

**Figure S23.** The spatial distributions of (a) heavy-metal cation (Cs) diffusivity, (b) anion (Cl) diffusivity, (c) counter-ion ($Na^+$) diffusivity, and (d) water (Ow) diffusivity for CsCl solutions confined in C–(N)–A–S–H gel nanochannels (Al/Si = 0.125) with pore sizes of ~1, ~2, ~4, and ~8 nm. Profiles were computed along the *z* direction using a bin size of 1 Å. The coordinate origin (z ≈ 0 nm) is defined as the averaged position of the Al/Si substitution sites to enable direct comparison across pore sizes. Dashed lines with error bars denote the corresponding bulk diffusivities.

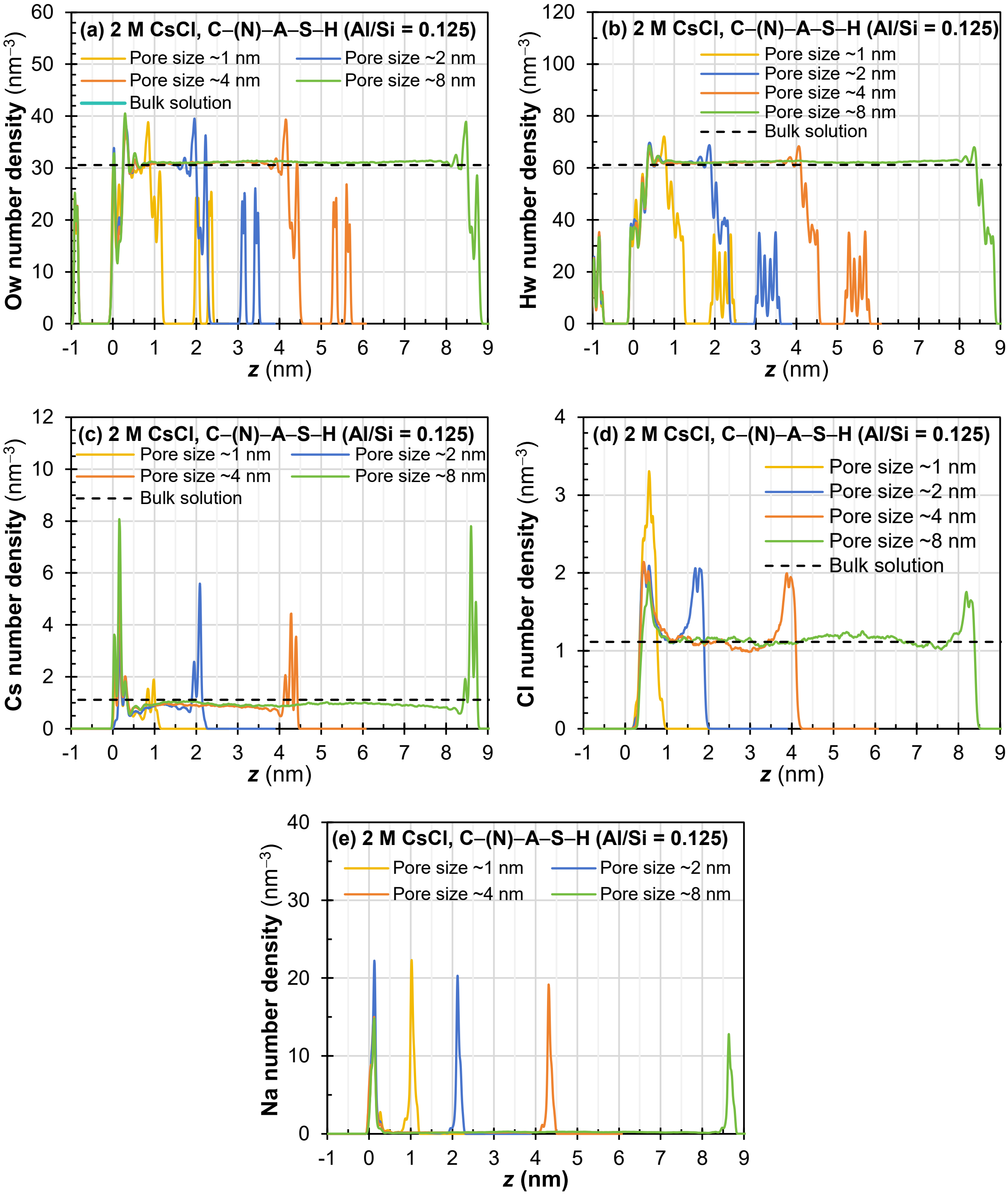

**Figure S24.** One-dimensional number density profiles of various solution species for nanoconfined 2 M CsCl within C–(N)–A–S–H gel nanochannels (Al/Si = 0.125) with pore sizes of ~1, ~2, ~4, and ~8 nm. The coordinate origin ($z \approx 0$ nm) is defined as the averaged position of the Al/Si substitution sites to enable direct comparison across pore sizes. Dashed lines denote the corresponding bulk number densities.

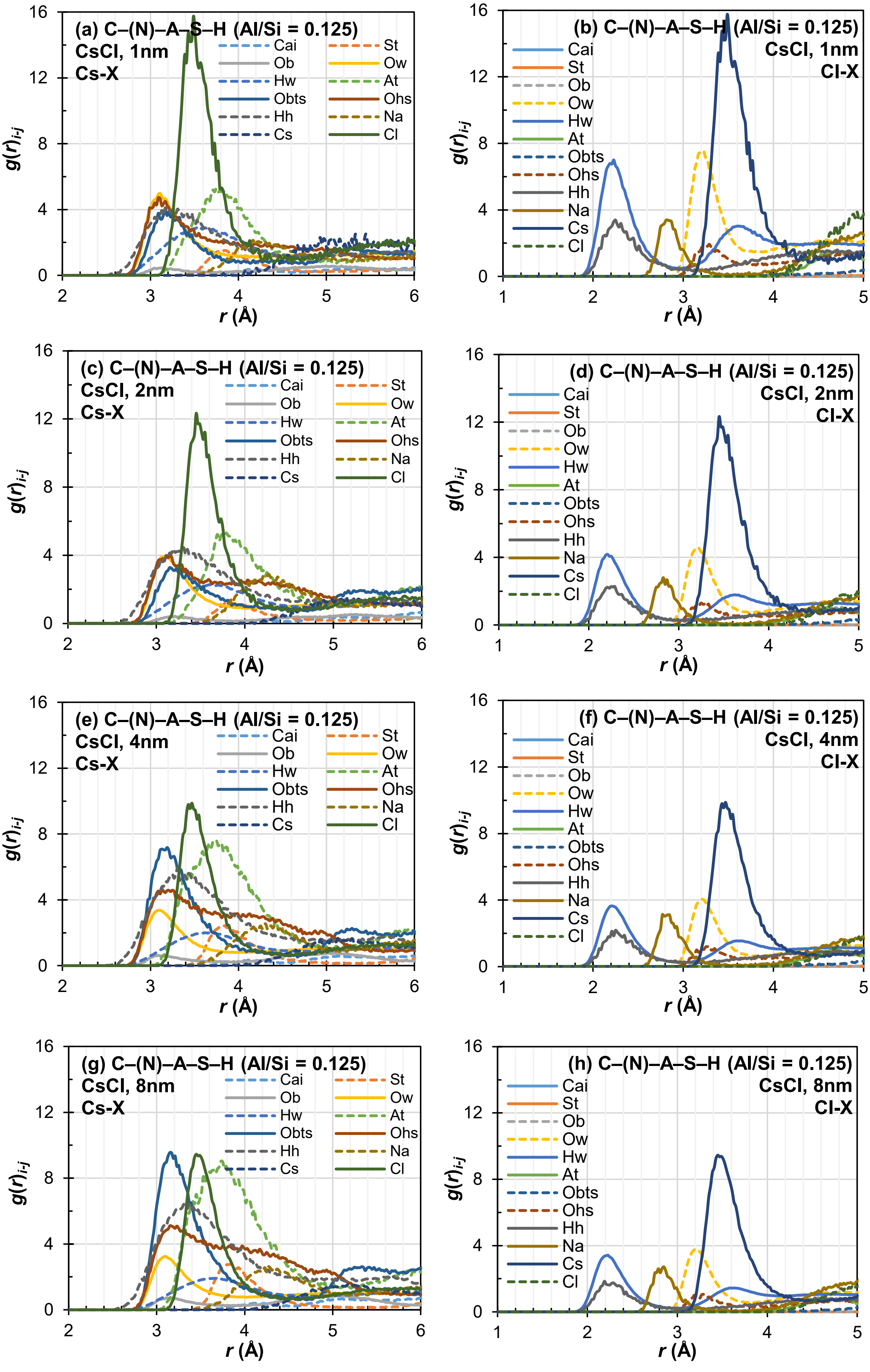
(a) C–(N)–A–S–H (Al/Si = 0.125)
CsCl, 1nm
Cs-X
Cai
St
Ob
Ow
Hw
At
Obts
Ohs
Hh
Na
Cs
Cl
g(r)i-j
r (Å)
(b) C–(N)–A–S–H (Al/Si = 0.125)
CsCl, 1nm
Cl-X
(c) C–(N)–A–S–H (Al/Si = 0.125)
CsCl, 2nm
Cs-X
(d) C–(N)–A–S–H (Al/Si = 0.125)
CsCl, 2nm
Cl-X
(e) C–(N)–A–S–H (Al/Si = 0.125)
CsCl, 4nm
Cs-X
(f) C–(N)–A–S–H (Al/Si = 0.125)
CsCl, 4nm
Cl-X
(g) C–(N)–A–S–H (Al/Si = 0.125)
CsCl, 8nm
Cs-X
(h) C–(N)–A–S–H (Al/Si = 0.125)
CsCl, 8nm
Cl-X

**Figure S25.** (a, c, e, g) Partial RDFs in the nanoconfined 2 M CsCl aqueous solution for Cs–X pairs inside the C–(N)–A–S–H gel with an Al/Si ratio of 0.125 and a pore size of (a) 1 nm, (c) 2 nm, (e) 4 nm, and (g) 8 nm, with their respective nearest neighbours. (b, d, f, h) The corresponding partial RDFs for Cl-X pairs. Atom labels are as follows: St = tetrahedral silicon atoms; At = tetrahedral aluminum atoms; Ca = aqueous interlayer calcium ion; Ow = oxygen atoms in water molecules; Hw = hydrogen atoms in water molecules; Ob = bridging oxygen atoms connected to tetrahedral Si atoms; Obts = bridging oxygen atoms connecting tetrahedral Si and Al; Ohs = hydroxyl oxygen atoms bonded to Al tetrahedral sites; Hh = hydrogen atoms in surface hydroxyl groups.

## Section S16 Correlation analysis of TBS components across cation types and pore sizes

Similar to the coordination-number correlation in the main text (**Figure 11**), the TBS correlation in **Figure S26** supports the same mechanistic interpretation. Correlation of individual TBS components for $Ba^{2+}$, $Pb^{2+}$, and $Cs^{+}$ systems (**Table S10**, **Table S15**, and **Table S20**) shows a strong linear relationship between $BaCl_2$ and $PbCl_2$ systems ($R^2 \approx 0.97$; **Figure S26**a), indicating similar interfacial binding characteristics. In contrast, the $BaCl_2$–CsCl comparison shows a clear deviation ($R^2 \approx 0.59$; **Figure S26**b), mainly due to the pronounced Cs–Ob and Cs–Oh contributions associated with inner-sphere $Cs^{+}$ adsorption. With increasing Al incorporation in C–(N)–A–S–H gels, the dominant Cs-binding environments progressively shift from Ob/Oh to Ohs/Obts sites. These results demonstrate that Al incorporation enhances overall adsorption strength while also differentiating the key interfacial oxygen environments controlling mobility suppression across gel families and cation types.

As shown in **Figure S26**c, Cs–O coordination numbers obtained for the 1, 2, and 8 nm pores correlate strongly with those from the 4 nm reference pore, with $R^2$ values of 0.97–1.00. This suggests that the relative hierarchy of $Cs^{+}$ binding sites is largely preserved as pore width changes. Although decreasing pore size increases the relative contribution of interfacial regions and therefore enhances the overall extent of immobilization (**Figures S22–S24**), it does not substantially alter the local surface coordination environments responsible for $Cs^{+}$ retention. These results indicate that, for the representative system examined here, pore size primarily modulates the proportion of ions affected by interfacial confinement, while the underlying adsorption-controlled immobilization mechanism remains governed by the same local ion–surface interactions.

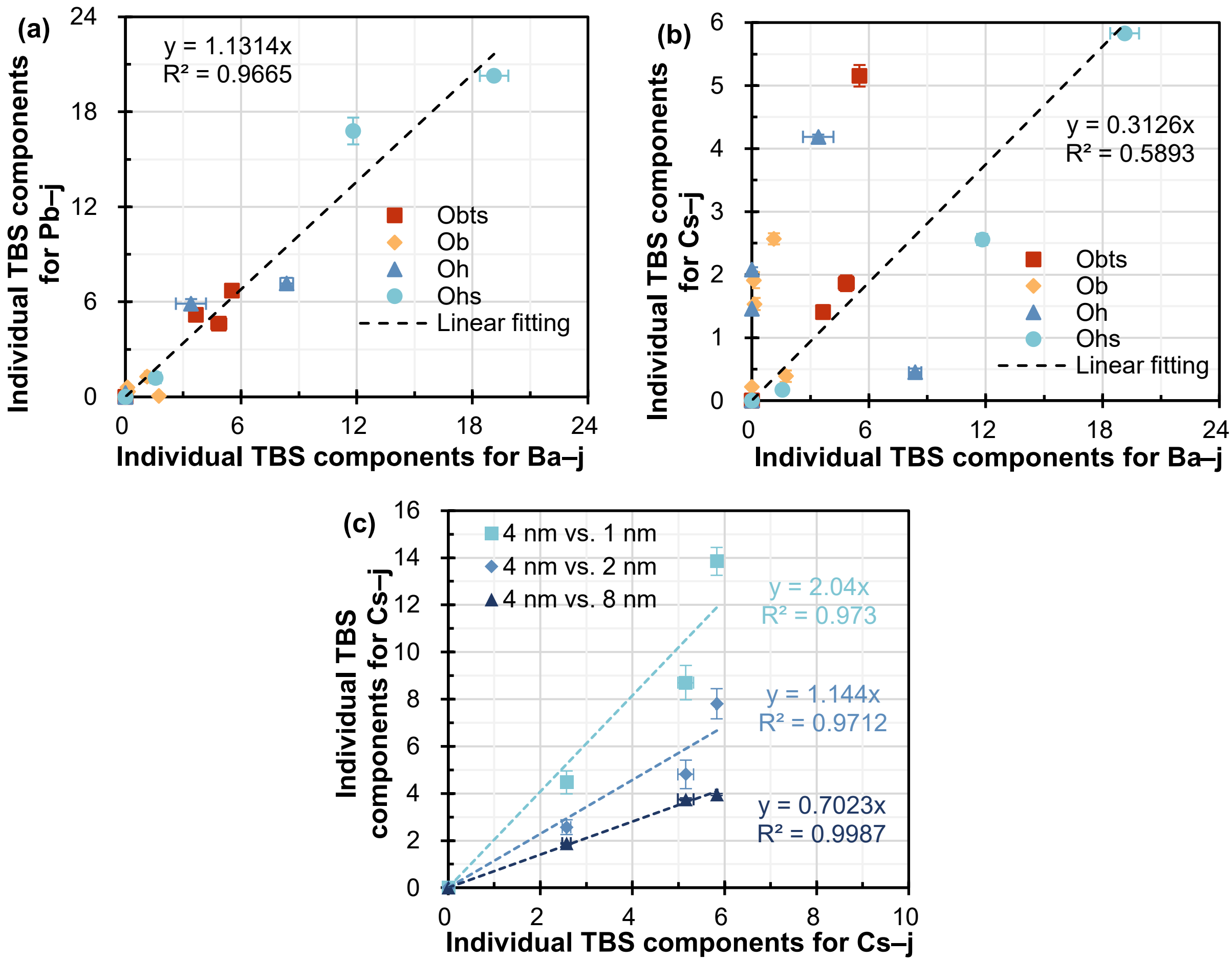

**Figure S26.** Correlations of individual TBS components: (a) $Ba^{2+}$ vs. $Pb^{2+}$, (b) $Ba^{2+}$ vs. $Cs^{+}$, and (c) $Cs^{+}$ systems with varying pore sizes, where each data point corresponds to the same TBS component evaluated at two different pore sizes. Error bars represent standard deviations from three independent simulations.

## Section S17 References

1. Pedone A, Malavasi G, Menziani MC, Cormack AN, Segre U. A new self-consistent empirical interatomic potential model for oxides, silicates, and silica-based glasses. J Phys Chem B. 2006;110(24):11780–95. https://doi.org/10.1021/jp0611018.
2. White CE, Page K, Henson NJ, Provis JL. In situ synchrotron X-ray pair distribution function analysis of the early stages of gel formation in metakaolin-based geopolymers. Appl Clay Sci. 2013;73:17–25. https://doi.org/10.1016/j.clay.2012.09.009.
3. Farrow C, Juhas P, Liu J, Bryndin D, Božin E, Bloch J, Proffen T, Billinge S. PDFfit2 and PDFgui: computer programs for studying nanostructure in crystals. J Phys Condens Matter. 2007;19(33):335219. https://doi.org/10.1088/0953-8984/19/33/335219.
4. Ihaddadene T, Claverie J, Bignonnet F, Amiri O. Diffusion and physicochemical behavior of chloride and sodium ions in CSH gel pores: A molecular dynamics investigation. Cem Concr Res. 2026;199:108053. https://doi.org/10.1016/j.cemconres.2025.108053.
5. Cong X, Kirkpatrick RJ. $^{29}$Si MAS NMR study of the structure of calcium silicate hydrate. Adv Cem Based Mater. 1996;3(3-4):144–56. https://doi.org/10.1016/S1065-7355(96)90046-2.
6. Cygan RT, Liang J-J, Kalinichev AG. Molecular models of hydroxide, oxyhydroxide, and clay phases and the development of a general force field. J Phys Chem B. 2004;108(4):1255–66. https://doi.org/10.1021/jp0363287.
7. Koneshan S, Rasaiah JC, Lynden-Bell R, Lee S. Solvent structure, dynamics, and ion mobility in aqueous solutions at 25 °C. J Phys Chem B. 1998;102(21):4193–204. https://doi.org/10.1021/jp980642x.
8. Smith DE, Dang LX. Computer simulations of NaCl association in polarizable water. J Chem Phys. 1994;100(5):3757–66. https://doi.org/10.1063/1.466363.
9. Aqvist J. Ion-water interaction potentials derived from free energy perturbation simulations. J Phys Chem. 1990;94(21):8021–24. https://doi.org/10.1021/j100384a009.
10. Smith DE, Dang LX. Computer simulations of cesium–water clusters: Do ion–water clusters form gas-phase clathrates? J Chem Phys. 1994;101(9):7873–81. https://doi.org/10.1063/1.468213.
11. Honorio T, Carasek H, Cascudo O. Water self-diffusion in CSH: Effect of confinement and temperature studied by molecular dynamics. Cem Concr Res. 2022;155:106775. https://doi.org/10.1016/j.cemconres.2022.106775.
12. Chen W, Gong K. Insights into ionic diffusion in C–S–H gel pore from molecular dynamics simulations: spatial distributions, energy barriers, and structural descriptor. J Phys Chem B. 2025;129(40):10550–67. https://doi.org/10.1021/acs.jpcb.5c04534.
13. The MathWorks Inc. MATLAB version: 9.14.0 (R2023a). https://www.mathworks.com. Accessed Aug 15, 2025.
14. Metzler R, Jeon J-H, Cherstvy AG, Barkai E. Anomalous diffusion models and their properties: non-stationarity, non-ergodicity, and ageing at the centenary of single particle tracking. Phys Chem Chem Phys. 2014;16(44):24128–64. https://doi.org/10.1039/c4cp03465a.
15. Rode B, Schwenk C, Hofer T, Randolf B. Coordination and ligand exchange dynamics of solvated metal ions. Coord Chem Rev. 2005;249(24):2993–3006. https://doi.org/10.1016/j.ccr.2005.03.032.
16. Bhattacharjee A, Hofer TS, Pribil AB, Randolf BR, Lim LHV, Lichtenberger AF, Rode BM. Revisiting the hydration of Pb (II): a QMCF MD approach. J Phys Chem B.

2009;113(39):13007–13. https://doi.org/10.1021/jp905848x.
17. Mehta PK, Monteiro P. Concrete: microstructure, properties, and materials. 4th ed. New York: McGraw-Hill Education; 2014.
18. Jiang Z-L, Pan Y-J, Lu J-F, Wang Y-C. Pore structure characterization of cement paste by different experimental methods and its influence on permeability evaluation. Cem Concr Res. 2022;159:106892. https://doi.org/10.1016/j.cemconres.2022.106892.